\documentclass[a4paper,fleqn,usenatbib]{mnras}

\usepackage{newtxtext,newtxmath}
\DeclareSymbolFont{CMletters}{OML}{cmm}{m}{it}
\DeclareMathSymbol{\nu}{\mathord}{CMletters}{23}
\DeclareMathSymbol{v}{\mathord}{CMletters}{`v}
\newcommand{\angstrom}{\mbox{\normalfont\AA}}

\usepackage[T1]{fontenc}
\usepackage{ae,aecompl}

\usepackage{graphicx}	
\usepackage{amsmath}	
\usepackage{amssymb}	

\title[The \ion{Mg}{ii} Forest]{Probing Reionization and Early Cosmic Enrichment with the Mg\,II Forest}

\newcommand{\beq}{\begin{equation}}
\newcommand{\eeq}{\end{equation}}

\def\be{\begin{equation}}
\def\ee{\end{equation}}
\def\bea{\begin{eqnarray}}
\def\eea{\end{eqnarray}}



\def \logTd6 {\hbox{log$( T/6 \kev)$} }

\def\myputfigure#1#2#3#4#5%
{\vskip#5pt\makebox[0pt]{\hskip#2in
\includegraphics[width=#3\textwidth]{#1}}\vskip#4pt\hfill}

\def \kev       {{\rm\ keV}}

\author[J. F. Hennawi et al.]{
  Joseph F. Hennawi,$^{1}$\thanks{E-mail: joe@physics.ucsb.edu (JFH)}
  Frederick B. Davies,$^{1,2}$
  Feige Wang,$^{1,3,\ast}$  
  and Jose O\~{n}orbe$^{4}$  
\\
$^{1}$ Department of Physics, Broida Hall, University of California, Santa Barbara Santa Barbara, CA 93106-9530, USA\\
$^{2}$ Lawrence Berkeley National Laboratory, CA 94720-8139, USA\\
$^{3}$ Steward Observatory, University of Arizona, 933 North Cherry Avenue, Tucson, AZ 85721, USA\\
$^{\ast}$ NHFP Hubble Fellow\\
$^{4}$ Facultad de F\'{i}sicas, Universidad de Sevilla, Avda. Reina Mercedes s/n, Campus de Reina Mercedes, 41012 Sevilla, Spain\\
}

\date{Accepted XXX. Received YYY; in original form ZZZ}

\pubyear{2020}

\hypersetup{draft} 

\begin{document}
\label{firstpage}
\pagerange{\pageref{firstpage}--\pageref{lastpage}}
\maketitle

\begin{abstract}
  Because the same massive stars that reionized the
  intergalactic
  medium (IGM)
  inevitably exploded as supernovae that polluted the Universe
  with metals, the history of cosmic reionization and
  enrichment are intimately intertwined.  While the overly sensitive
  Ly$\alpha$ transition completely saturates in a neutral IGM, strong
  low-ionization metal lines like the \ion{Mg}{ii} $\lambda
  2796,\lambda 2804$ doublet will give rise to a detectable
  `metal-line forest' if the metals produced during reionization
  ($Z\sim 10^{-3}Z_{\odot}$) permeate the neutral IGM.  We simulate
  the \ion{Mg}{ii} forest for the first time by combining a large
  hydrodynamical simulation with a semi-numerical reionization
  topology, assuming a simple enrichment model where the IGM is
  uniformly suffused with metals. In contrast to the traditional
  approach of identifying discrete absorbers, we treat the absorption
  as a continuous random field and measure its two-point correlation
  function, leveraging techniques from precision cosmology. We show
  that a realistic mock dataset of 10 JWST spectra can simultaneously
  determine the Mg abundance, $[{\rm Mg}\slash {\rm H}]$, with a
  $1\sigma$ precision of 0.02~dex and measure the global neutral
  fraction $\langle x_{\ion{H}{i}}\rangle$ to 5\% for a Universe with
  $\langle x_{\ion{H}{i}}\rangle = 0.74$ and $[{\rm Mg}\slash {\rm H}]
  = -3.7$. Alternatively, if the IGM is pristine, a null-detection of
  the \ion{Mg}{ii} forest would set a stringent upper limit on the IGM
  metallicity of $[{\rm Mg}\slash {\rm H}] < -4.4$ at 95\%
  credibility, assuming $\langle x_{\ion{H}{i}}\rangle > 0.5$ from
  another probe. Concentrations of metals in the circumgalactic
  environs of galaxies can significantly contaminate the IGM signal,
  but we demonstrate how these discrete absorbers can be easily
  identified and masked such that their impact on the correlation function is negligible.
  The \ion{Mg}{ii} forest thus has tremendous
  potential to precisely constrain the reionization and enrichment
  history of the Universe.
\end{abstract}




\section{Introduction}

The process of converting primordial hydrogen and helium into heavier
elements underlies the entire history of star and galaxy formation in
the Universe.  During the Epoch of Reionization (EoR) primeval
galaxies and accreting black holes ionized the hydrogen in the
intergalactic medium (IGM) ending the preceding cosmic `dark ages'.
Current Planck constraints suggest reionization took place in the
range $z_{\rm reion}\simeq 5.9-8.0$ \citep[$2\sigma$;][]{Planck18},
consistent with other astrophysical constraints from IGM damping wings
towards the highest redshift quasars 
\citep{Mortlock11,Greig17b,Banados18,Davies18b,Greig19,Wang20a,Yang20b}
and the disappearance of strong Ly$\alpha$ emission from galaxies
\citep{Mason18,Mason19,Hoag19}.  In the same way that the reionization
history provides a global census of ionizing photons emitted by all
galaxies and quasars, the metal content of the IGM provides a fossil
record of the Universe's integrated star-formation history.  Indeed,
the production of ionizing photons and metals go hand-in-hand because
the same massive stars emitting the ionizing photons explode as
supernovae ejecting metals into their surroundings. Considerations
based solely on the lifetimes and yields of massive stars generically
predict that in the process of producing the $\simeq 3$ photons per
hydrogen atom required to reionize the Universe the average
metallicity will reach $Z\sim 10^{-3}Z_{\odot}$
\citep{MadauShull96,MR97,GnedinOstriker97,Ferrara16}, largely
insensitive to the shape of the stellar IMF.

While the inevitable production of these metals at early times is
uncontroversial, their abundance, distribution, and ionization state
are far less clear. Models can be found where the pre-reionization IGM
is polluted to $Z\sim 10^{-3}Z_\odot$ already at $z\sim 8-9$
\citep{MFR01,Pallottini14,Jaacks18,Jaacks19,Doughty19,Kirihara20}, possibly explaining the background
metallicity of the IGM measured at $z\simeq 3$ \citep{Schaye03};
whereas in other studies metals remain highly concentrated around the
galaxies producing them \citep{Oppenheimer09,Pawlik17} while the IGM remains
pristine. There is even less consensus about whether these enriched
regions are necessarily simultaneously reionized, either by ionization
fronts from the galaxies responsible for the pollution, or the
`enrichment-front' powered by galactic outflows, which can shock heat
and collisionally ionize the gas \citep{MFR01,Ferrara16}. Alternatively, neutral enriched
material could exist in the pre-reionization IGM if it recombines
due to the stochasticity of reionization \citep{Oh02}, or if galactic outflows were not
fast enough to collisionally ionize it, or if these outflows do
not open up the necessary channels for ionizing photons to penetrate outflowing enriched gas. 

Metal absorption lines at the highest redshifts provide an additional
window into the physics of reionization.  At $z \lesssim 5$ after
reionization is complete, heavy elements are ubiquitously detected in
the IGM as extremely weak absorption in high-ionization states like
\ion{C}{iv} and \ion{O}{vi}
\citep[e.g.][]{Ellison00,Bergeron02,Simcoe11a,Dodorico16}, owing to low IGM
densities $n_{\rm H}\sim 10^{-5}-10^{-4}~{\rm cm^{-3}}$ and the
relatively hard UV background.  The smoking gun of reionization as
probed by metal absorbers would be a transition from these weak
high-ionization lines to a forest of low-ionization absorbers from
transitions like \ion{O}{i} $\lambda$1302~\AA, \ion{Si}{ii}
$\lambda$1260~\AA, \ion{C}{ii} $\lambda$1334~\AA, and \ion{Mg}{ii}
$\lambda$2796~\AA,$\lambda$2804~\AA~as the IGM becomes progressively
more neutral \citep{Oh02}, provided the pre-reionization IGM is
sufficiently enriched.

To date all of our knowledge of the so-called background metallicity
of the IGM comes from $z\sim 2-4$, where the sensitive high-resolution
and high ${\rm S\slash N}$ ratio ($\sim 50-200$) echelle absorption
line spectra required to measure this quantity are easiest to
obtain. This work has revealed that the IGM is enriched to a level
$Z\simeq 10^{-3}Z_{\odot}$ at $z\sim 2-4$
\citep{Ellison00,Bergeron02,Schaye03,Aguirre04,Aguirre08,Simcoe11a} down to as
low as the cosmic mean density, indicating that $> 50\%$ of the
baryonic mass in the universe has been polluted with metals
\citep{Booth12,Dodorico16}.  Interestingly, \citet{Schaye03} found no
evidence for evolution over the redshift range $1.8 < z < 4.1$, during
which the cosmic star-formation decreases by $\sim 0.5$\,dex,
suggesting that a significant amount of the IGM enrichment could have
occurred at early times.  The highest redshift IGM metallicity
measurement comes from \citet{Simcoe11a} who measured ${\rm C\slash H} \simeq
-3.55$ at $z\sim 4.3$,
and argued for a mild
($0.3-0.5$ dex) decrease from $z\sim 2.4$. 
Whether this decrease is at odds with the lack of evolution observed by
\citet{Schaye03} is unclear, and as emphasized by \citet{Simcoe11a}
could partly result from methodological differences between the
\citet{Schaye03} pixel optical depth technique and \citet{Simcoe11a}'s
traditional Voigt profile fitting approach.

Notwithstanding significant observational efforts, it has proven too
difficult to detect the extremely weak absorption lines that would
herald the transition to a neutral IGM at $5\lesssim z \lesssim 7$
because the relevant transitions are redshifted into the near-IR.  Due
to the higher sky background, increased detector noise, lower
resolution spectrographs, and exacerbated by the paucity of
sufficiently bright high-redshift quasars, the resulting spectra lack the
required sensitivity to probe the diffuse IGM.  For example, the vast
majority of $z > 5$ \ion{C}{iv} detections published in the literature to date
have $N_{\ion{C}{iv}} > 14\,{\rm cm^{-2}}$ -- a direct result of the
limiting column densities (i.e. $50\%$ completeness) of the respective
surveys \citep{RyanWeber09,Becker09,Simcoe11b,Dodorico13a}.  For
comparison, at $z\simeq 6$ if the IGM has a metallicity of
$Z=10^{-3.5}\,Z_\odot$ a CIV absorber arising from gas at the cosmic
mean density would have\footnote{This estimate is determined from a CLOUDY
  photoionization model for an absorber at $z=6$ at the cosmic mean
  density with a stopping column density of $\log_{10}N_{\ion{H}{i}} =
  15.3$ derived using the scaling relations from
  \citep{Schaye03}.} $N_{\ion{C}{iv}}= 10^{11.8}\,{\rm cm^{-2}}$. This
implies current $z\gtrsim 5$ \ion{C}{iv} searches are probing
overdense ($\rho\slash \langle \rho\rangle\sim 100$) gas in the
circumgalactic
medium (CGM) of galaxies which is at much higher metallicity than the background
IGM. Similarly, the majority of the \ion{O}{i} absorbers published by
recent surveys \citep{Becker06,Becker11a,Becker19}  have $N_{\ion{O}{i}}
> 10^{14}\,{\rm cm^{-2}}$ corresponding to rest-frame equivalent widths
of $W_\lambda\gtrsim 0.1$\AA, whereas the hydrodynamical simulations
performed by \citet{Keating14} demonstrate that such absorbers arise
from CGM gas at overdensities of $\rho\slash \langle \rho\rangle
\gtrsim 80$ characteristic of so-called sub-damped Ly$\alpha$ systems.
This is in line with the conclusions of \citet{Simcoe12}'s search for a forest of \ion{Mg}{ii} absorbers in the
neutral region surrounding a quasar at  $z\simeq 7$ --- 
with current sensitivity only significantly overdense
or chemically enriched absorbers would be detectable as discrete lines, but the diffuse IGM at metallicity
$[{\rm Mg\slash H}] < -3$ is presently beyond reach.  
Thus, for both high and low-ions, current metal absorption line
searches probing into the EoR can only individually identify the
strongest absorbers arising from dense gas in the CGM of galaxies.

It has been observed that the cosmic mass density of \ion{C}{iv}
absorbers $\Omega_{\ion{C}{iv}}$ drops by a factor of $\sim2$ at $z > 5.3$
\citep{RyanWeber09,Becker09,Simcoe11b,Dodorico13a}, at around the same
redshift that an upturn in the abundance of \ion{O}{i} absorbers is
observed \citep{Becker06,Becker11a,Becker19}, which \citet{Becker19}
attributes to the `reionization of the CGM'. This behavior is
not unexpected given the strong redshift evolution of the UVB towards
$z\gtrsim 6$ \citep{Calverley11,Wyithe11,Davies18ABC,Anson18}, possibly augmented by the
presence of strong UVB fluctuations \citep{davies16b} or a later than expected
end to reionization \citep{Kulkarni19b,Nasir20,Keating20}, and
offers a preview of the transition from high high-ions to low-ions
that one might observe in the IGM writ large. But unfortunately, if
one adheres to the traditional approach of detecting individual lines,
probing IGM absorbers into the reionization epoch is currently beyond
current sensitivity limits.
As emphasized by \citet{Simcoe12} and \citet{Dodorico13a},
progress would require either a major observational effort or waiting
for the extremely large telescopes.

In this paper we propose a novel statistical approach leveraging
methods from precision cosmological studies of the Ly$\alpha$ forest
which overcomes this limitation.  Whereas early studies of the
Ly$\alpha$ forest first treated it as a collection of discrete
absorption lines, what precipitated the breakthrough in our
understanding of the IGM --- that the forest naturally results from
hierarchical structure formation in a cold dark matter (CDM) Universe --- was the insight
that it can instead be analyzed as a continuous cosmological random
field. Adopting tools from large-scale structure analysis,
cosmologists measured statistics like the transmission probability distribution function (PDF)
\citep[e.g.][]{McDonald01,Lee15}, and the power spectrum 
\citep[e.g.][]{McDonald06,Walther17,Walther19} enabling quantitative
analysis and statistical inference that make the Ly$\alpha$ forest a
precision probe of cosmological parameters and the nature of dark
matter \citep{Seljak03,McDonald05,Viel13,Nathalie15,Irsic17}.

As opposed to the traditional approach of identifying discrete metal
lines in high-redshift quasar spectra, we will treat metal line forests
during the EoR analogously, as a continuous field.  Actually,
measurements along these lines have already been carried out in the
context of $z\sim 2$ baryon acoustic oscillation (BAO) measurements
using the Ly$\alpha$ forest as a density tracer.  Whereas the first
such measurements identified the BAO peak at $z\sim 2$ in the 3D
Ly$\alpha$ forest correlation function \citep{Busca13,Slosar13}, it
was quickly realized \citep{Pieri14} that complementary BAO
constraints can be obtained by measuring the correlation functions of
low-redshift metal-line forests, which has already led to competitive BAO
measurements using the \ion{C}{iv} forest \citep{Blomqvist18} at $z
\simeq 2$ and the \ion{Mg}{ii} forest \citep{Bourboux19} at $z\simeq
0.6$.

To illustrate the power of this new approach, we focus our attention
on the \ion{Mg}{ii} forest \citep[see also][]{Simcoe12} during the
EoR\footnote{Although \ion{Mg}{ii} is singly ionized, because the
  \ion{Mg}{i} ionization edge (0.56 Rydberg) lies below 1 Rydberg, low
  IGM densities and the star-formation powered UVB imply Mg will
  entirely populate the \ion{Mg}{ii} state in a neutral IGM. See the
  next section for additional details.}.  There are several reasons to
prefer \ion{Mg}{ii} over the other low-ions that have been discussed
\citep{Oh02}.  First and foremost, \ion{Mg}{ii} is a doublet absorbing
at $\lambda$2796\AA\, and $\lambda$2804\AA\, in an approximate
$2\!:\!1$ ratio dictated by the ratio of oscillator strengths. This
gives rise to a strong correlated absorption feature at a velocity lag
$\Delta v_{\ion{Mg}{ii}} = 768\,{\rm km\,s^{-1}}$ set by the doublet
separation, which provides definitive confirmation that one is
actually detecting the \ion{Mg}{ii} forest and not noise or
systematics. Second, the bluer aforementioned ions probe less
pathlength because Ly$\alpha$ Gunn-Peterson absorption from the IGM
wipes out rest-frame wavelengths blueward of $1216$\AA, whereas the
redder \ion{Mg}{ii} transition can probe a much longer line-of-sight
pathlength extending from the quasar redshift down to the redshift at
which reionization is complete.  Finally, the redder \ion{Mg}{ii}
forest region is expected to be completely uncontaminated by
foreground absorption\footnote{The only strong resonant lines that
  could potentially contaminate the \ion{Mg}{ii} forest are the
  \ion{Ca}{ii} H+K doublet $\lambda 3969$, $\lambda 3934$ and
  \ion{Na}{i}$\lambda 5897$. But these absorbers are extremely weak
  \citep{ZhuMenard13}, because they are not the dominant ionization
  state of either element in the presence of a radiation field that
  cuts off at energies exceeding the Lyman limit, as is the case in
  the neutral IGM.  These lines are thus only observable in rare
  extremely strong absorbers where dust can attenuate the ionizing
  continuum redward of the Lyman limit.}, whereas for
e.g. \ion{O}{i} $\lambda$1302, lower-$z$ absorbers from redder ionic
transitions must be identified and masked.
\footnote{For example for a
  quasar at $z_{\rm QSO} = 7.5$ there would be contamination from \ion{Mg}{ii}
  at $z\sim 2.9$, \ion{Al}{ii} $\lambda1670$ at $z\sim 5.5$, \ion{C}{iv}$ \lambda 1548$
  and \ion{Si}{ii} $\lambda 1526$ at $z=6.1$, etc.}

This paper explores the detectability of the \ion{Mg}{ii}
forest during the EoR. Our method for simulating the \ion{Mg}{ii} forest is described
in \S~\ref{sec:simulate}.  The dependence of the \ion{Mg}{ii} forest correlation function on our model
parameters and spectral resolution is studied in \S~\ref{sec:corrfunc}, and a method for
statistical inference is presented in \S~\ref{sec:inference}, along with sensitivity estimates
that would result from a hypothetical observing program with the
\emph{James Webb Space Telescope} (JWST). CGM absorption associated with galactic metal reservoirs can contaminate the IGM \ion{Mg}{ii} forest. A model for CGM absorbers is implemented in
\S~\ref{sec:CGM}, we show how this CGM absorption alters the flux PDF of the \ion{Mg}{ii} forest in \S~\ref{sec:PDF}, and implement a procedure for identifying
and masking CGM absorbers in \S~\ref{sec:mask}. We summarize and conclude in 
\S~\ref{sec:summary}.

Throughout this work we adopt a $\Lambda$CDM cosmology with the
following parameters: $\Omega_{\rm m}=0.3192$,
$\Omega_{\Lambda}=0.6808$, $\Omega_{\rm b}=0.04964$, $h=0.67038$,
which agree with latest cosmological constrains from the CMB
\citep{Planck18} within one sigma. All distances are quoted in
comoving units denoted as cMpc or ckpc.  In this cosmology, a
line-of-sight velocity of $v_\parallel = 100~{\rm km~s^{-1}}$
corresponds to $r_\parallel = \frac{v_\parallel}{aH(z)} = 0.90~{\rm
  cMpc}$ in the Hubble flow at $z=7.5$. All equivalent widths are in
the rest-frame in units of \AA~and are denoted by the symbol
$W_\lambda$. For doublet transitions the equivalent width of the
stronger transition in the doublet are quoted.

\section{Simulating the \ion{Mg}{ii} Forest}
\label{sec:simulate}

\subsection{General Considerations}
\label{sec:general}
To simulate the \ion{Mg}{ii} forest we need to model the distribution
of metals in the IGM during the reionization phase transition. This is
clearly an extreme challenge to simulate, as one must capture not only
the physical state of the reionizing IGM, but also the production and
dispersal of metals, and their ionization state.  While progress
on simulating all of this complex physics has been made in recent
years \citep[e.g.][]{Oppenheimer09,Pallottini14,Pawlik17,Jaacks18,Jaacks19,Doughty19,Kirihara20},
our goal here is to investigate
detectability and perform a sensitivity analysis. To this end we adopt
a highly simplistic toy model whereby the entire IGM is suffused with
metals at a fixed metallicity $Z$ with solar relative
abundances.

While in principle ionization corrections would be required to
determine the fraction of Mg in the \ion{Mg}{ii} state, this can
be trivially simplified. Note that ionization edge of \ion{Mg}{i} is at
0.56 Rydberg which lies below the 1 Rydberg ionizing edge for hydrogen,
whereas the \ion{Mg}{ii} edge is at 1.11 Rydberg blueward of the Lyman
edge.  The IGM is thus essentially completely transparent at the
energies that ionize neutral Mg, and it is expected that even prior to
reionization, the ultraviolet radiation field sourced by cosmic
star-formation will produce a metagalactic UV background sufficiently
intense at the \ion{Mg}{i} edge to ionize all Mg into the \ion{Mg}{ii}
state. We explicitly checked this by running a CLOUDY \citep{Cloudy17} model with gas at the mean
density of the IGM at $z=7.5$ subjected to a \citet{HM12} UV background truncated at energies
greater than 1 Rydberg. We find negligible $\lesssim 10^{-8}$ abundance of Mg in any ionization state
besides \ion{Mg}{ii} irrespective of the total $N_{\rm H}$ used to set the so-called
stopping criterion. Thus for the purposes of the present study it is
an excellent approximation to assume that the ionization state of Mg is
simply tied to that of hydrogen, and neutral regions of the Universe
are in the \ion{Mg}{ii} state.

Having described our model of the metallicity and ionization state of
Mg, we now turn to modeling the IGM during reionization. It is well
known that reionization photoheating modifies the small-scale
structure of the IGM. Although baryons trace dark matter fluctuations
on large scales, on smaller scales gas is supported against
gravitational collapse by thermal pressure.  Analogous to the classic
Jeans argument, baryonic fluctuations are suppressed relative to the
pressureless dark matter, and gas is `pressure smoothed' or `filtered'
on small scales \citep{GH98,Kulkarni15,Rorai17}. As a result, the
small-scale structure or clumpiness of the pre-reionization IGM is
intimately related to its thermal evolution. It is currently unknown whether the IGM adiabatically cooled to extremely low temperatures $T\lesssim 1\,{\rm K}$ just
before reionization at $z\sim 7-8$, or if a metagalactic X-ray
background, sourced by faint AGN \citep{Madau04,Ricotti04a}, X-ray
binaries \citep{Madau17}, or PopIII stellar remnants \citep{Xu16},
photoelectrically heated it to much higher temperatures $T \sim
1000\,{\rm K}$ \citep[e.g.][but see \citet{Fialkov14}]{Furlanetto06}.
During reionization ionization fronts propagate supersonically through
the IGM, impulsively heating reionized gas to $\sim 10^4~{\rm K}$.
This rapid temperature increase drives the Jeans scale from $\sim
1~{\rm ckpc}$ (for $T\lesssim 1~{\rm K}$) up to $\sim 100~{\rm ckpc}$,
dissipating pre-reionization IGM small-scale structure on a timescale
of $\sim 1~{\rm Gyr}$ \citep{Anson20,Davies20_smallscale}. 

Since the \ion{Mg}{ii} forest absorption arises from neutral regions
of the IGM, in principle predicting its clustering strength depends on
the small-scale structure of the IGM, and hence on the
pre-reionization IGM temperature evolution. Accurately simulating the
pre-reionization IGM is a daunting numerical problem \citep{Anson20,Davies20_smallscale}.  Ideally, the
simulation domain would be sufficiently large ($\gtrsim 10\,{\rm
  cMpc}$) to remain linear in its fundamental mode, and encompass the
$768\,{\rm km\,s^{-1}}$ ($6.9~{\rm cMpc}$) doublet separation of
\ion{Mg}{ii} and typical size ($\sim 10~{\rm cMpc}$) of ionized
bubbles, while simultaneously resolving the extremely small $\sim
1~{\rm ckpc}$ Jeans scale corresponding to the potentially very low
temperatures $T\lesssim 1~{\rm K}$ prevailing in the pre-reionization
IGM. Naively, this would require a $\sim 20,000^3$ grid or comparable
number of SPH particles, unattainable even with the world's largest
supercomputers. Given our lack of knowledge of pre-reionization IGM
temperature evolution, and the numerical challenge, we will utilize a
snapshot of a hydrodynamical simulation at $z = 7.5$ prior to
reionization photoheating. This simulation, of a
$L_{\rm box} = 40\,{h^{-1}\,{\rm cMpc}}$ ($v_{\rm box} \equiv  aH(z)L_{\rm box} = 6600~{\rm km~s^{-1}}$ at $z=7.5$)
domain on a $2048^3$ grid, has a grid scale  of $29\,{\rm ckpc}$ ($3.2~{\rm km~s^{-1}}$), which would fail to
resolve the $\sim 1~{\rm ckpc}$ Jeans scale for pre-reionization IGM temperatures
$T\lesssim 1~{\rm K}$, but marginally resolve the $\sim 30~{\rm ckpc}$
Jeans scale if an X-ray background preheated the IGM to $T \sim
1000\,{\rm K}$.

As this paper focuses on the clustering of the \ion{Mg}{ii} forest
which is tied to the clustering of the IGM  via our simple enrichment
model, a discussion of the impact of this unresolved structure on our
results is in order. Unlike the Gunn-Peterson optical depth for
\ion{H}{i} Ly$\alpha$ absorption, which probes neutral gas in an
ionized medium and hence scales  $\tau_{\rm Ly\alpha} \propto n_{\ion{H}{i}} \propto n_{\rm H}^2$ quadratically with density,
the analogous optical depth for the \ion{Mg}{ii} forest $\tau_{\ion{Mg}{ii}}\propto n_{\ion{Mg}{ii}} \propto n_{\ion{H}{i}} \propto n_{\rm H}$
scales only linearly with density, because the medium is predominantly neutral.
Thus if $F = e^{-\tau_{\ion{Mg}{ii}}}$, $\delta_f \equiv (F-\langle F\rangle)\slash \langle F\rangle \approx \tau_{\ion{Mg}{ii}}\propto n_{\rm H}$ is always
a good approximation for the small optical depths we consider, the  \ion{Mg}{ii} forest
flux correlation function, $\xi_{\ion{Mg}{ii}} \equiv \langle \delta_f \delta_f\rangle \propto \xi_{\rm \rho,1D}$,
is simply proportional
to the correlation function of the overdensity projected along skewers, which can
in turn be written as the Fourier transform
\begin{equation}
  \xi_{\rm \rho,1D}(r) = \int_0^\infty \Delta_{\rm \rho,1D}^2(k) \cos(kr)d\ln k\label{eqn:xiden}. 
\end{equation}
of the analogous 1D density dimensionless power spectrum $\Delta_{\rm
  \rho, 1D}^2(k) \equiv kP_{\rm \rho, 1D}(k)\slash \pi$.  The
dimensionless power $\Delta_{\rho, 1D}^2(k)$ is a smoothly rising
function that traces the underlying clustering of the CDM, but is truncated
by a sharp cutoff at the Jeans scale of the pre-reionization
IGM. Failure to resolve this Jeans scale would then effectively
truncate $\Delta_{\rm \rho, 1D}^2(k)$ at the simulation grid scale
$k_{\rm trunc}\simeq 2\pi\slash r_{\rm grid}$. Note that the
$\cos(kr)$ factor in eqn.~(\ref{eqn:xiden}) implies that only
wavenumbers with $kr \lesssim 1$ for which $\cos(kr) \sim 1$
contribute significantly to the integral, whereas wavenumbers
$kr \gg 1$ contribute negligibly because of cancellations induced by the
highly oscillatory $\cos(kr)$ term. 
In other words, only Fourier modes with wavelengths
$\lambda \gtrsim r$ ($k \lesssim 2\pi \slash \lambda$) larger than the scale of interest
contribute to the correlation function. Since the velocity
separations (length scales) that we would realistically probe with
real data ($\gtrsim 30~{\rm km~s^{-1}}$) correspond to spatial scales
a least an order of magnitude larger than the pre-reionization IGM
Jeans scale ($\sim 1-30~{\rm ckpc}$ or $\sim 0.1-3~{\rm km~s^{-1}}$),
we do not expect this missing small-scale power to have a 
significant impact on our results.

\subsection{Hydrodynamical Simulations of the Pre-Reionization IGM}

We simulate the \ion{Mg}{ii} forest using \texttt{Nyx}, a massively
parallel N-body gravity + grid hydrodynamics code specifically
designed for simulating the IGM \citep{Almgren13,Lukic15}.  Initial
conditions were generated using the \textsc{music} code \citep{Hahn11}
using a transfer function generated by \textsc{camb}
\citep{CAMB,Howlett12}.  We assumed a $\Lambda$CDM cosmology with the
following parameters: $\Omega_{\rm m}=0.3192$,
$\Omega_{\Lambda}=0.6808$, $\Omega_{\rm b}=0.04964$, $h=0.6704$,
$\sigma_{8}=0.826$ and $n_{\rm s}=0.9655$ which agree with latest
cosmological constrains from the CMB \citep{Planck18} within one
sigma.  We adopted hydrogen and helium mass abundances ($X_{\rm
  p}=0.76$ and $Y_{\rm p}=0.24$) in agreement
with the recent CMB observations and Big Bang nucleosynthesis
\citep{Coc14}. We simulated a domain with a box size of $40~{\rm
  cMpc}~h^{-1}$ using a $2048^3$ grid, and the simulation was started
at $z_{\rm ini}=159$.

We analyze a snapshot at $z = 7.5$, which is prior to the redshift of
reionization in this simulation. Specifically, hydrodynamical
simulations like \texttt{Nyx} which do not attempt to model
reionization \citep[but see][]{Onorbe17a,Onorbe19} typically treat reionization
by assuming a spatially uniform, time-varying metagalactic UVB
radiation field, input to the code as a list of photoionization and
photoheating rates that vary with redshift. The simulation we analyze
reionizes at $z = 6.0$, which is to say that the UVB is turned on
at this redshift, which occurs at a later time than the snapshot we analyze at $z =
7.5$. For additional details about the numerical approach see
\citet{Onorbe19}.  The simulation analyzed here is similar to the
`flash' reionization simulations in that paper, where flash refers to the fact
that the UVB is abruptly turned on at $z = 6.0$ causing the
simulation to instantaneously reionize.

\subsection{Semi-Numerical Reionization Topology}

Our \texttt{Nyx} hydrodynamical simulation yields the baryon density,
peculiar velocity, and temperature at each grid cell. The only other
quantity required to simulate the \ion{Mg}{ii} forest is the hydrogen
neutral fraction $x_{\ion{H}{i}}$.  For this we must create a model of the
global reionization topology in the simulation domain.
To generate the \ion{H}{i} neutral fraction $x_{\ion{H}{i}}$ throughout
the simulation  we use a modified version of the semi-numerical
reionization code
\texttt{21cmFAST}\footnote{\url{https://github.com/andreimesinger/21cmFAST}}
\citep{Mesinger11}, to be presented in further detail in Davies \&
Furlanetto (in prep.). The semi-numerical approach computes the
fraction of material that has collapsed into dark matter halos,
$f_{\rm coll}$, following conditional Press-Schechter \citep{LaceyCole93}
applied to an approximate non-linear density field computed by applying the Zel'dovich
approximation \citep{Zeldovich70} to the initial conditions of the
\texttt{Nyx} simulation, and evolving it to $z=7.5$. 
A region is considered ionized if
$f_{\rm coll} > \zeta^{-1}$ on \emph{any} scale, where $\zeta$ is the
``ionizing efficiency," which combines 
several parameters governing the efficiency of star formation and the production and escape of
ionizing photons from galaxies into a single parameter that
corresponds to the total number of ionizing photons emitted per
collapsed baryon. An ionizing mean free path $\lambda_{\rm mfp}$ is implemented by suppressing the contribution
of ionizing photons with a smooth exponential decline rather than the traditional hard cutoff ($R_{\rm max}$). 
The neutral fraction $x_{\ion{H}{i}}$ is computed on a $256^3$ grid encompassing the
\texttt{Nyx} simulation domain, which is chosen to be sufficiently fine to sample the ionization topology
but also coarse enough to guarantee sufficient numbers of dark matter halos (which source the ionizing photons)
in each cell.

We generated a sequence of 51 different reionization topologies,
parameterized by the volume filling fraction of neutral hydrogen
$\langle x_{\ion{H}{i}}\rangle$ at $z = 7.5$, spanning the range $\langle
x_{\ion{H}{i}}\rangle$ = 0.0 to 1.0 in steps of 0.02. The models used here
adopted an ionizing photon mean free path $\lambda_{\rm mfp} = 20~{\rm cMpc}$,
and the ionizing efficiency was then adjusted to get the full range
of $\langle x_{\ion{H}{i}}\rangle$ required. 
For additional details about the
semi-numerical reionization model see \citet{Davies18b}, \citet{Onorbe19}, and Davies \&
  Furlanetto, in prep.

\subsection{Creating \ion{Mg}{ii} Forest Skewers}

To simulate the \ion{Mg}{ii} forest we generate skewers by drawing random locations along one face of the simulation
cube, and record the baryon density, the line-of-sight component of
the peculiar velocity field, the temperature, and neutral fraction at
each line-of-sight location in the grid, where the latter comes from the
semi-numerical reionization topology just described.
The doublet nature of the \ion{Mg}{ii} ion requires that we deal with
resonant absorption at two wavelengths $\lambda2796,\lambda 2804$\AA~,
which corresponds to a velocity difference of $\Delta v=768~{\rm
  km~s^{-1}}$. In practice we compute the optical depth
$\tau_{2796}$ for the stronger $\lambda2796$ transition,
and then rescale this array by the oscillator
strength ratio of $f_{2804}/f_{2796} = 0.497$ to obtain $\tau_{2804}$,
which we shift by the doublet separation $\Delta v$, allowing us to
compute the total optical depth $\tau_{\ion{Mg}{ii}} = \tau_{2796}
+ \tau_{2804}$. Below we describe our approach for generating skewers of
$\tau_{2796}$, the optical depth in the $\lambda$2796\AA\, resonance.

The optical depth for resonant absorption for an ionic transition
$X^{n+}$ of an element $X$ in its $n$th ionization state is
\be
\tau_\nu = \int n_X x^{n+}\sigma_\nu dr \label{eqn:tau_int},
\ee
where $\nu$ is the frequency, $n_X$ is the number density of the element
$X$, $x^{n+}$ is the fraction of this element populating the $n$th ionization state, and
\be \sigma_\nu = \frac{\pi e^2}{m_e c}f_{\ell
  u}\phi_\nu \label{eqn:sigma}
\ee
is the frequency specific cross-section. Here $e$ and $m_e$ are the electron charge and mass, 
respectively, $c$ is the speed of light, $f_{\ell u}$ is the oscillator strength
for the resonant transition $\ell \rightarrow u$ at rest-frame frequency  $\nu_{\ell
  u}$, and $\phi_\nu$ is the
line profile. Whereas in general $\phi_\nu$ is described by the Voigt
profile, for the extremely low optical depths characterizing
metal-line forests it is a very good approximation to represent
$\phi_\nu$ with a Gaussian form \be \phi_\nu =
\frac{1}{\sqrt{\pi}}{\Delta \nu_{\rm D}}\exp{\left[-\left(\frac{\nu -
      \nu_{\ell u}}{\Delta \nu_{\rm
        D}}\right)^2\right]} \label{eqn:phi}, \ee where $\Delta
\nu_{D} = \left(b\slash c\right)\nu_{\ell u}$ is the Doppler frequency
width of the Gaussian profile determined by the Doppler parameter $b =
\sqrt{2 k_{\rm B} T\slash m_X}$ characterizing thermal broadening of
the absorption lines, where $k_{\rm B}$ is Boltzmann's constant and
$m_{\rm X}$ is the mass of element $X$.

If we assume, as we will throughout this work, that the metal line
species $X$ is uniformly mixed with the baryons in the IGM, then we
can write $n_X = \langle n_{\rm X}\rangle \Delta$, where $\Delta
\equiv \rho_{\rm B}\slash\langle \rho_{\rm B}\rangle$, $\rho_{\rm B}$
is baryon mass density, angle brackets represent an average over the
volume of the Universe, and $\langle n_{\rm X}\rangle = Z\slash
Z_\odot\left(n_{\rm X}\slash n_{\rm H}\right)_{\odot} \langle n_{\rm
  H}\rangle$, where $Z\slash Z_\odot$ is the metallicity in solar
units, and $\left(n_{\rm X}\slash n_{\rm H}\right)_{\odot}$ is the
abundance of element $X$ in the sun.

Combining
eqns.~\ref{eqn:tau_int}-\ref{eqn:phi}, and transforming to velocity
coordinates using the Doppler formula $\nu^\prime = \nu_{\ell u}[1 -
  (v^\prime - v)\slash c]$ and the Hubble relation $dr = dv\slash
H(z)$, where $H(z)$ is the Hubble expansion rate at redshift $z$, we
finally arrive at
\be \tau_{v} = \tau_{X,0} \int
\frac{x^{n+}\Delta}{\sqrt{\pi}}\exp{\left[-\left(\frac{v^\prime -
      v}{b}\right)^2\right]} \frac{dv^\prime}{b}\label{eqn:tauv},
\ee
where we have defined $\tau_{X,0}$ the metal-line forest analog of the
Gunn-Peterson optical depth \citep{Oh02},
\be
\tau_{X,0} = 
\frac{\pi e^2 f_{\ell u} \lambda_{\ell u} \langle n_{\rm X}\rangle}{m_e c H(z)} \label{eqn:tau0}
\ee
or plugging in numbers for the \ion{Mg}{ii} forest
\be
\tau_{X,0} = 0.06\left(\frac{Z\slash Z_\odot}{10^{-3}}\right)\left(\frac{\,\left(n_X\slash n_{\rm H}\right)_{\odot}}{3.4\times 10^{-5}}\right)\left(\frac{f_{\ell u}}{0.62}\right)\left(\frac{\lambda_{\ell u}}{2796~{\text \AA}}\right)\left(\frac{1 + z}{8.5}\right)^{3\slash 2}\label{eqn:mgiitau0},  
\ee
where we used the Mg abundance $\left(n_{\rm Mg}\slash n_{\rm H}\right)_{\odot} = 3.4\times 10^{-5}$ determined from the solar photosphere \citep{Asplung09}, and the oscillator strength $f_{2796}$ and wavelength $\lambda_{\ell u}$ of the stronger \ion{Mg}{ii} $\lambda 2796~$\AA~ transition.

Special attention must be paid to the construction of metal-line
forest skewers because of the extremely small Doppler parameters
$b$. This results from both cold pre-reionization IGM temperatures and
the fact that metals are much heavier than hydrogen. The temperature
of the pre-reionization IGM is currently unknown and could be anywhere
in the range $T\sim 1-1000\,{\rm K}$, as discussed in \S~\ref{sec:general}. 
Plugging in numbers for the Doppler parameter for an ion $X$
\be
b = 0.26\,{\rm km\,s^{-1}}\left(\frac{T}{100\,{\rm K}}\right)^{-1\slash
  2}\left(\frac{m_X\slash m_H}{24}\right)^{-1\slash 2}, \label{eqn:bval}
\ee
where we normalized using the atomic weight of Mg. 
The \texttt{Nyx} 
hydrodynamical simulations employed in this work have a grid scale
of  $dv = 3.2\,{\rm km\,s^{-1}}$ at $z=7.5$, thus the thermal
broadening of \ion{Mg}{ii} forest absorption lines are far from being
resolved by our native velocity grid.  One option would be to
interpolate the simulated density and temperature fields onto a much
finer grid before performing the convolution in eqn.~(\ref{eqn:tauv})
required to construct simulated \ion{Mg}{ii} forest skewers, but this
is computationally intensive given the small $b$ parameters one would
need to resolve. Instead, we adopt the clever approach described in
Appendix B of \citet{Lukic15}, which is far faster because it enables
one to work on the native grid, but nevertheless explicitly conserves
optical depth. Specifically, we discretize the integral in eqn.~(\ref{eqn:tauv}),
taking the overdensity and temperature as constant across each grid cell.
For the $k$th pixel at velocity $v_k = kdv$ in the Hubble expansion,
the optical depth is
\be
\tau_{v_j} = \tau_{X,0} \sum_i \frac{x^{n+}_i\Delta_i}{2}\left[{\rm erf}(y_{i-1\slash 2}) - {\rm erf}(y_{i + 1/2})\right]\label{eqn:tau}, 
\ee
where the error function\footnote{Note that equivalent eqn. B5 in \citet{Lukic15} is missing a factor of $1\slash 2$} results from the integral of the Gaussian
profile across the $i$th pixel (from $v_{i-1\slash 2}$ to $v_{i+1\slash 2}$),
and $y_{i - 1/2} = (v_j - v_{\parallel, i} - v_{i-1\slash 2})\slash b$, where
$v_{\parallel,i}$ is the component of the gas peculiar velocity parallel to the
sightline, $v_{i-1\slash 2} = (i-1\slash 2)dv$, and an analogous expression
holds for $y_{i + 1/2}$.

\subsection{Generating the Model Parameter Grid}

Our goal is to construct a large set of \ion{Mg}{ii} forest skewers
for a model grid governed by two parameters, $[{\rm Mg\slash H}]$ and
$\langle x_{\ion{H}{i}}\rangle$.  We start with 10,000 skewers of
$\Delta$, $v_{\parallel}$, and $T$, extracted from the \texttt{Nyx}
simulation  at random locations along one face of the cube.  For
the $x_{\ion{H}{i}}$ field, we use the set of 51 reionization
topologies in the range $\langle x_{\ion{H}{i}}\rangle = 0.0$ to 1.0. 
Given
that the \ion{Mg}{ii} forest optical depth $\tau_{\ion{Mg}{ii}}$
is linear in $Z$ (see eqn.~\ref{eqn:tau0}), which we define as
$Z\slash Z_{\odot} \equiv 10^{[{\rm Mg\slash H}]}$, we perform the
convolutions in eqn.~(\ref{eqn:tau}) for each value of $\langle x_{\ion{H}{i}}\rangle$, but a
single metallicity, and scale the resulting optical depth to the desired metallicity.
For the metallicity grid we use 201 models spanning the range $[{\rm Mg\slash H}] = [-6.0, -2.0]$.
The result of this procedure is a set of 10,000 \ion{Mg}{ii} forest skewers generated
for a grid of $51\times 201 = 10,251$ models.

\subsection{Forward Modeling Observed Data}
\label{sec:observations}

For the purpose of visualizing real observational data and performing
statistical inference we create mock spectra with smearing induced by
finite spectral resolution and add noise consistent with a realistic
${\rm S\slash N}$ ratio.  We parameterize the data quality with the
FWHM of the spectral resolution assuming a Gaussian line spread function, and the ${\rm S\slash N}$ per pixel, where
the spectral sampling is assumed to be $n_{\rm samp} = 3$ pixels per
spectral resolution element of width the FWHM. Our simulated spectra
are convolved with a Gaussian consistent with the spectral resolution, interpolated
onto a velocity grid set by the spectral sampling, and then 
Gaussian random noise is added with a standard deviation $\sigma_{\rm S\slash N} = ({\rm S\slash N})^{-1}$. 

For a quasar at $z = 7.5$, and considering a redshift
interval of $\Delta z = 0.6$ from  $z = 6.9-7.5$ where we expect the Universe
to be significantly neutral, the \ion{Mg}{ii} forest
is redshifted to observed frame wavelengths of $\lambda =
2.21-2.38\mu{\rm m}$ in the $K$-band. 
In this
work we model spectra from JWST/NIRSpec for which the sky background
at the relevant wavelengths comes from zodiacal light and is
relatively smooth. For this case, assuming spectral noise that is
constant with wavelength is a reasonable approximation. On the other
hand, ground-based observations of this spectral region
would  have heteroscedastic
noise owing to the forest of atmospheric OH airglow lines.
As we will aim to measure the two-point correlation function
of the \ion{Mg}{ii} forest (see \S~\ref{sec:corrfunc}), formally the
noise in the estimated correlation function averages to zero at all
non-zero velocity lags irrespective of whether or not it is
heteroscedastic. As such, even for ground-based observations, we do not expect the heteroscedasticity of the
noise to be a significant issue\footnote{This assumes that correlated
  noise resulting from sky subtraction systematics are insignificant,
  which we expect to be the case.}, provided that one take our assumed
${\rm S\slash N}$ ratio to be a suitable average over the spectral region in
question.

We primarily focus on mock data with ${\rm FWHM}=100~{\rm km~s^{-1}}$ and
${\rm S\slash N}=100$, representative of what can be achieved with
JWST/NIRSpec in a 10hr integration for a typical $z \gtrsim 7$ quasar
with an AB apparent magnitude of $m_{\rm 1450} = 20.5$, chosen to correspond to
the observed frame $J$-band. This estimate is based on calculations performed with the
JWST/NIRSpec exposure time calculator\footnote{https://jwst.etc.stsci.edu/}.

\subsection{Results}
\begin{figure}
  \hskip -0.09cm
  \includegraphics[trim=20 4 25 0,clip,width=0.49\textwidth]{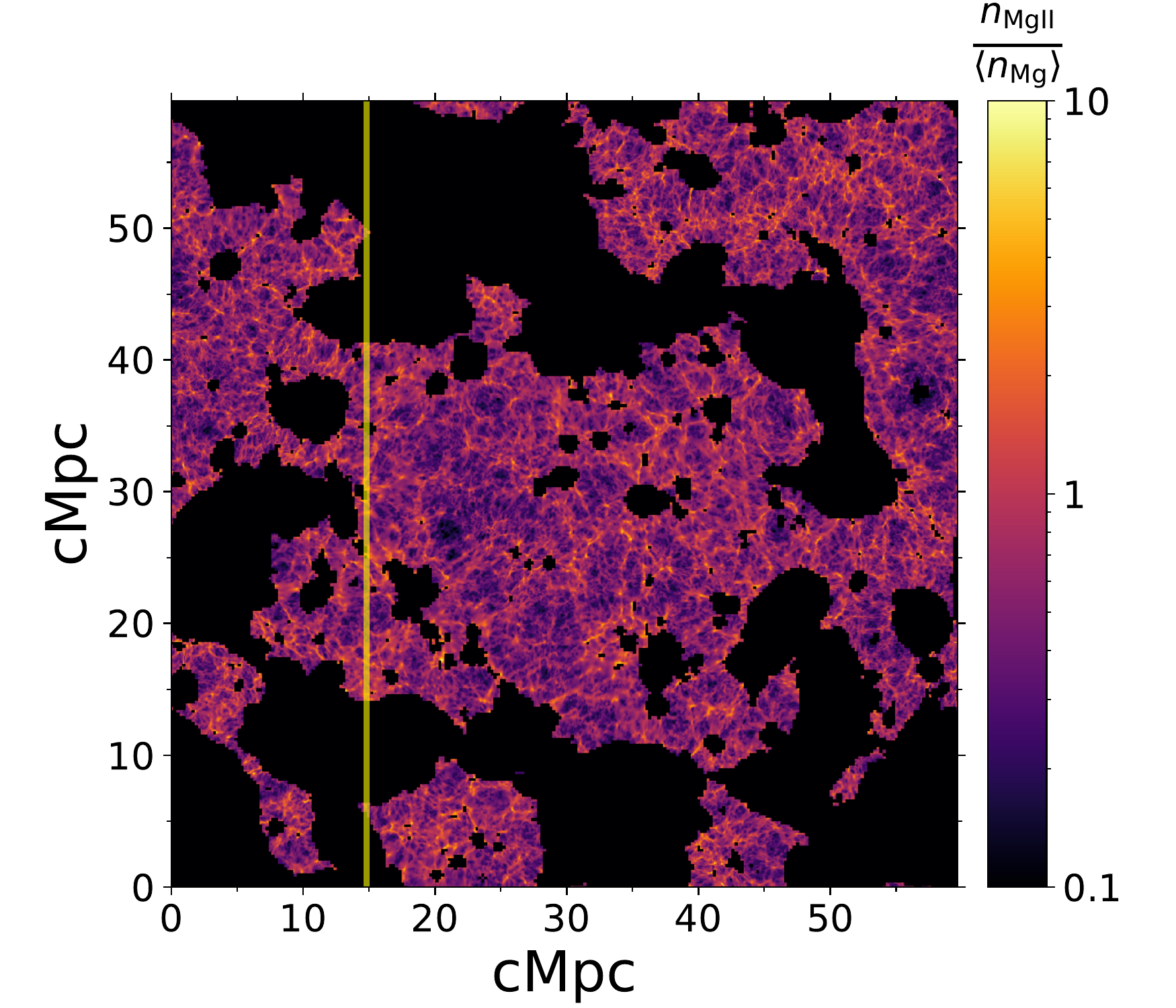}
  \vskip -0.1cm
  \caption{ The topology of \ion{Mg}{ii} absorbing gas. The image is a
    single pixel wide ($29~{\rm ckpc}$) slice through our simulation
    of the distribution of \ion{Mg}{ii} for a universe with a volume
    averaged neutral fraction $\langle x_{\ion{H}{i}}\rangle = 0.50$ at $z
    = 7.5$.  The quantity plotted is $x_{\ion{H}{i}}\Delta$ where $\Delta
    = \rho\slash \langle \rho_{B}\rangle$ which is equivalent to
    $n_{\ion{Mg}{ii}}\slash \langle n_{\rm Mg}\rangle$ given our
    assumption of a uniform metallicity distribution and $x_{\ion{H}{i}}=x_{\ion{Mg}{ii}}$ (see \S~\ref{sec:general}).  The color scale
    is logarithmic, as indicated by the colorbar at right. Black
    regions have been reionized, whereas colored regions are still
    neutral. The vertical yellow line shows a skewer through the
    volume for which \ion{Mg}{ii} forest absorption spectra are shown
    in Fig.~\ref{fig:skewers}\label{fig:slice}}
\end{figure}

\begin{figure*}
  \includegraphics[trim=48 50 83 53,clip,width=\textwidth]{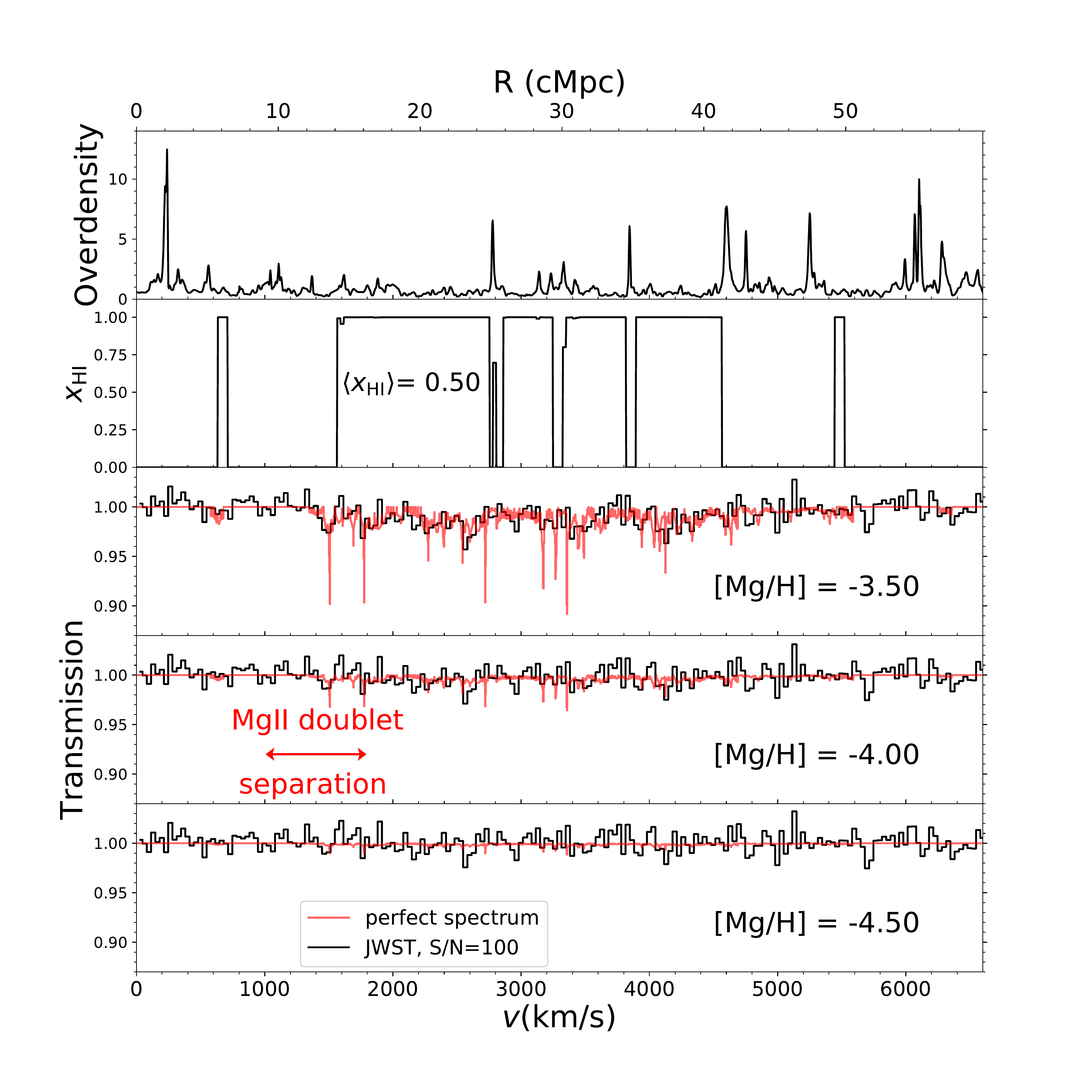}
  \vskip -0.2cm
  \caption{Simulations of the \ion{Mg}{ii} forest for a skewer
    indicated by the yellow vertical line in
    Fig.~\ref{fig:slice}. \emph{Top:} Baryon overdensity along the
    skewer obtained from a hydrodynamical simulation of the $z = 7.5$
    \emph{pre-reionization} IGM. \emph{Second from Top:} The IGM
    neutral fraction $x_{\ion{H}{i}}$ along the skewer determined from a
    semi-numerical reionization model with $\langle x_{\ion{H}{i}}\rangle=0.5$. \emph{Lower Panels:} Simulated \ion{Mg}{ii}
    forests for various values of $[{\rm Mg}\slash {\rm H}]$ (relative
    to solar), for perfect spectra (red curves) and realistic mock
    JWST/NIRSpec spectra (black histograms; FWHM=$100~{\rm
      km~s^{-1}}$, ${\rm S\slash N}=100$ per pixel) for $\sim 10\,{\rm
      hr}$ of exposure per source for quasars comparably bright as the
    two known at $z = 7.5$.\label{fig:skewers}}
\end{figure*}

The topology of \ion{Mg}{ii} absorbing gas predicted by our model is shown in
Fig.~\ref{fig:slice} for a universe
with a volume averaged neutral fraction $\langle x_{\ion{H}{i}}\rangle = 0.50$ at $z=7.5$. 
The figure shows a single pixel ($29~{\rm ckpc}$) slice through $x_{\ion{H}{i}}\Delta$,
which is equivalent to $n_{\ion{Mg}{ii}}\slash \langle n_{\rm Mg}\rangle$ given our assumption of a
uniform metallicity distribution and $x_{\ion{H}{i}}=x_{\ion{Mg}{ii}}$ (see
\S~\ref{sec:general}). The vertical yellow line shows a skewer through this
volume taken to be the line-of-sight direction towards the observer.  
Fig.~\ref{fig:skewers} shows our predictions for the
\ion{Mg}{ii} forest spectra along this sightline.  The top panel shows
the baryon overdensity, which traces the clumpy
structure of the pre-reionization IGM determined by the underlying
CDM as discussed in \S~\ref{sec:general}. 
The second panel from the top shows the IGM neutral fraction,
$x_{\ion{H}{i}}$, from our semi-numerical reionization
topology.  The lower panels show the simulated \ion{Mg}{ii} forest for
various values of $[{\rm Mg}\slash {\rm H}]$, where the red curves are
perfect spectra and black histograms show realistic mock JWST/NIRSpec
spectra with a resolution ${\rm FWHM}=100~{\rm km~s^{-1}}$ and
${\rm S\slash N}=100$ per pixel (see \S~\ref{sec:observations} for
details).

\section{The Correlation Function of the \ion{Mg}{ii} Forest}
\label{sec:corrfunc}

It is clear from the mock observations in Fig.~\ref{fig:skewers} that even with the
exquisite spectra delivered by JWST, the traditional approach of
detecting individual \ion{Mg}{ii} absorption systems appears
hopeless. Nevertheless, the \ion{Mg}{ii} forest is still detectable by
statistically averaging down the noise to reveal the correlated
structure present. To this end, we compute the two-point correlation
function of the
transmission, which has two important
advantages.
First, since the spectrograph and background noise are white,  the noise covariance averages
down to be consistent with zero at
all non-zero lags $\Delta v > 0$. Second, the weak absorption field is
highly correlated at the doublet separation $\Delta v_{\ion{Mg}{ii}} =
768\,{\rm km\,s^{-1}}$, which will give rise to a pronounced peak in the
correlation function at this velocity lag.

Specifically, if $F$ is the continuum normalized flux, we
define the relative flux fluctuation
\be
\delta_f \equiv \frac{F - \langle F\rangle}{\langle F\rangle}\label{eqn:delta},
\ee
where  $\langle F\rangle$ is the mean flux. We then compute
the correlation function
\be
\xi(\Delta v) = \langle \delta_f(v)\delta_f(v + \Delta v)\rangle
\label{eqn:xi}
\ee
by averaging over all pairs of pixels separated by
velocity lag  $\Delta v$. 

\subsection{Dependence on Model Parameters}
\label{sec:model}
\begin{figure}
  \hskip -0.09cm
  \includegraphics[trim=25 10 0 0,clip,width=0.49\textwidth]{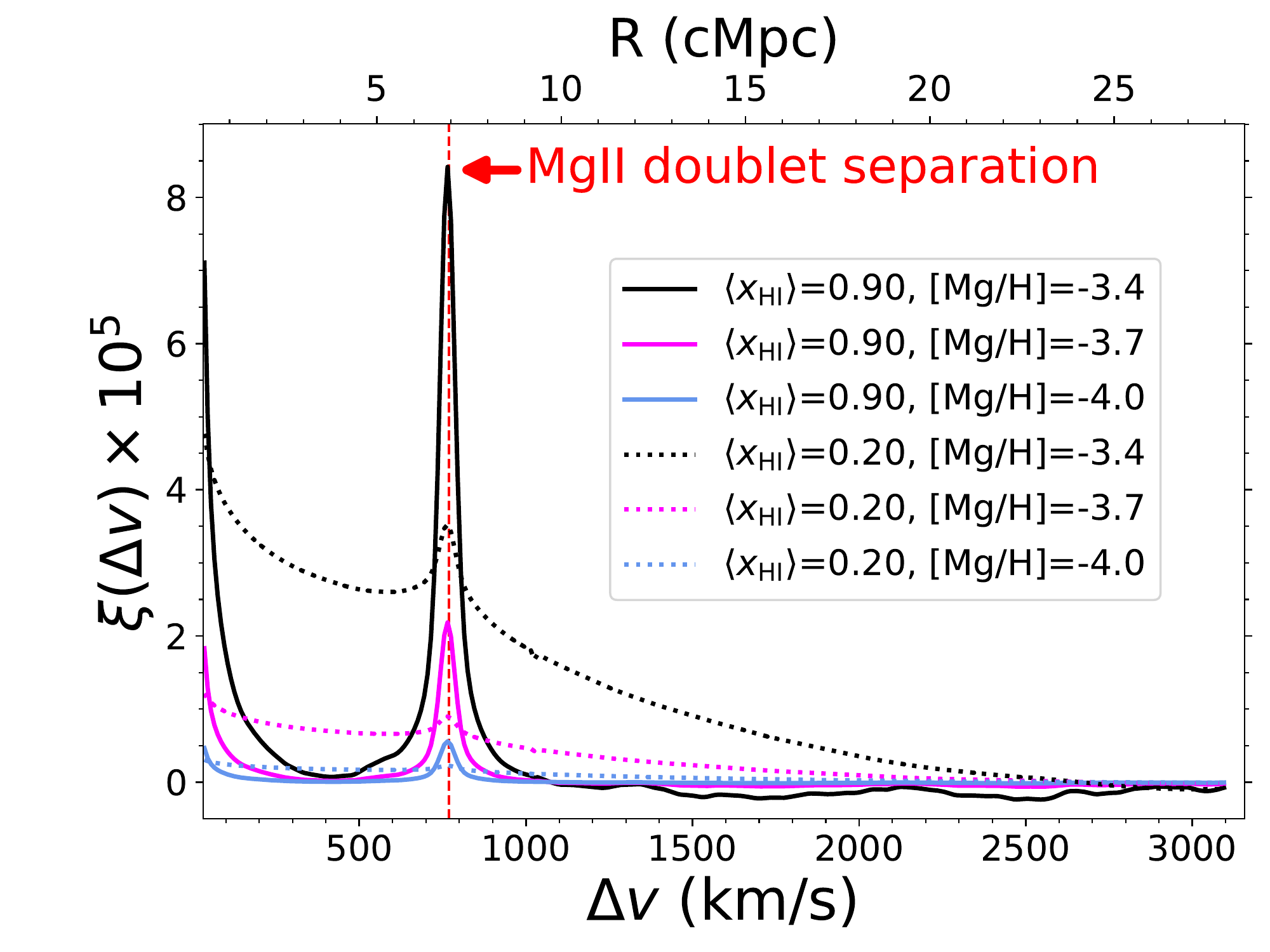}
  \vskip -0.1cm
   \caption{Predicted \ion{Mg}{ii} forest correlation function. Curves
     show the correlation function $\xi(\Delta v)$ of the relative flux fluctuation
     $\delta_f$ (see eqn.~\ref{eqn:delta})
     where colors show different IGM enrichment $[{\rm Mg\slash H}]$,
     whereas solid and dotted curves show average IGM neutral
     fractions of $\langle x_{\ion{H}{i}}\rangle =0.9$ and $\langle x_{\ion{H}{i}}\rangle=0.2$, respectively.  The correlation function
     exhibits a conspicuous peak at $\Delta v=768~{\rm km\,s^{-1}}$
     corresponding to the \ion{Mg}{ii} doublet separation, indicated
     by the vertical red dashed line. Because the correlation function
     shape has a distinct dependence on each of these parameters they
     can both be independently measured. A spectral resolution of
     ${\rm FWHM}=30~{\rm km~s^{-1}}$ typical of a ground based echelle
     spectrograph has been assumed.}
   \label{fig:corrfunc}
\end{figure}

The correlation function $\xi(\Delta v)$ is shown in
Fig.~\ref{fig:corrfunc} for several different combinations of
metallicity, $[{\rm Mg}\slash {\rm H}]$, and volume averaged neutral
fraction, $\langle x_{\ion{H}{i}}\rangle$. For this
computation we have assumed noiseless data, and a resolution of ${\rm
  FWHM}=30~{\rm km~s^{-1}}$ typical of a ground based echelle
spectrograph.

There are several important features of the correlation function shown in Fig.~\ref{fig:corrfunc} which
we now describe.  First, because pre-reionization IGM baryons trace
the clumpy small-scale structure set by the underlying CDM, there is
significant variance on small scales and as a result the 
\ion{Mg}{ii} forest correlation function exhibits a precipitous rise
towards small velocity lags. Second, there is a strong peak at the doublet
separation $\Delta v_{\ion{Mg}{ii}} = 768\,{\rm km\,s^{-1}}$, indicated by the red vertical dashed line,
which arises from the doublet nature of the \ion{Mg}{ii} transition. The height of this peak is a result of
the significant small-scale power, since pixels separated
by around $\Delta v_{\ion{Mg}{ii}} = 768\,{\rm km\,s^{-1}}$ are in reality probing correlated fluctuations in the
gas at much smaller velocity lags. Finally, at intermediate to large velocity lags
$\xi(\Delta v)$ exhibits an overall power-law dependence on $\Delta v_{\ion{Mg}{ii}}$,
which appears to be highly sensitive to the value of $\langle x_{\ion{H}{i}}\rangle$. 
This occurs because line-of-sight fluctuations in
neutral fraction $x_{\ion{H}{i}}$ (see Fig.~\ref{fig:skewers} second
panel from top) modulate the \ion{Mg}{ii} forest sourcing fluctuations
on a hierarchy of scales set by the topology of the neutral regions
during reionization (see Fig.~\ref{fig:slice}). 

Naively one might expect a perfect degeneracy between the global
neutral fraction $\langle x_{\ion{H}{i}}\rangle$ and Mg abundance $[{\rm
    Mg}\slash {\rm H}]$, since the optical depth for
the \ion{Mg}{ii} forest in eqn.~(\ref{eqn:tauv}) depends on the
degenerate product of $x_{\ion{H}{i}}$ and $[{\rm Mg}\slash {\rm H}]$.
However, this naive intuition proves incorrect, as is clear by
comparing the solid, $\langle x_{\ion{H}{i}}\rangle = 0.9$, and dotted,
$\langle x_{\ion{H}{i}}\rangle = 0.2$, curves in Fig.~\ref{fig:corrfunc}.
At fixed $\langle x_{\ion{H}{i}}\rangle$, the amplitude of the correlation
function scales as the square of the Mg abundance $[{\rm Mg}\slash
  {\rm H}]$, as expected from eqn.~(\ref{eqn:tauv}) and
eqn.~(\ref{eqn:xi}) --- when $F$ is small $|\delta_f| \approx |\tau -
\langle \tau \rangle|$, and thus $\xi \propto \delta_f^2 \propto
\tau_{\ion{Mg}{ii}}^2$. But the dependence of $\xi(\Delta v)$ on $x_{\ion{H}{i}}$ is more complex, which probes large-scale $\sim 1-30~{\rm cMpc}$
fluctuations arising from the global topology of reionization.

\subsection{Dependence on Resolution}
\begin{figure}
  \hskip -0.09cm \includegraphics[trim=25 10 0
    0,clip,width=0.49\textwidth]{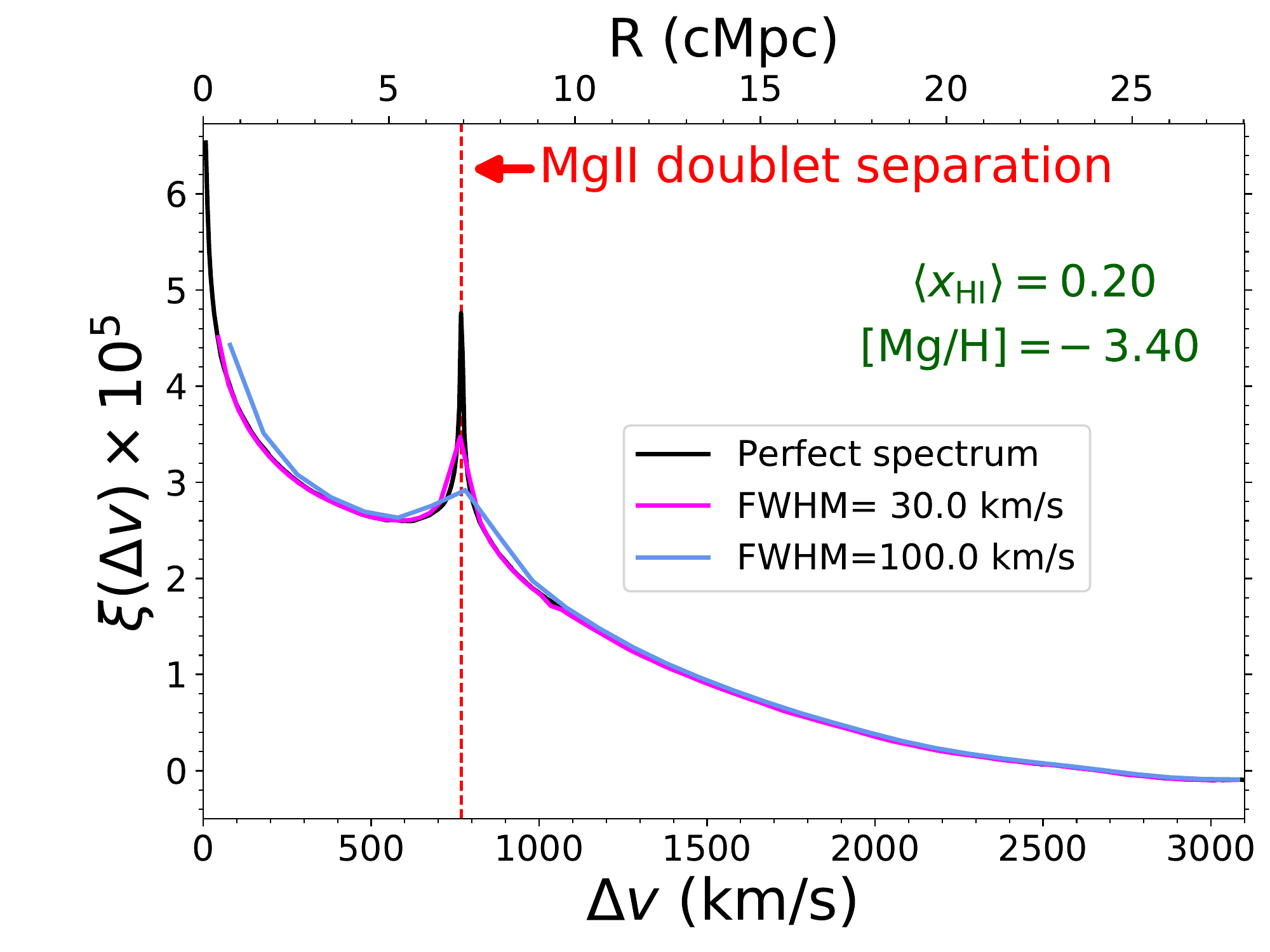}
  \vskip -0.1cm
   \caption{Impact of spectral resolution on the \ion{Mg}{ii} forest correlation function. Curves
     show the correlation function $\xi(\Delta v)$ of the transmission observed
     at different spectral resolutions. Whereas the black curve represents perfect resolution
     (i.e. the highest resolution we can model with our current simulations), the
     magenta (${\rm FWHM}=100~{\rm km~s^{-1}}$; e.g. JWST) and blue (${\rm FWHM}=30~{\rm km~s^{-1}}$; ground-based)
     curves illustrate what can be achieved at moderate and echelle 
     spectral resolution, respectively. The impact of finite spectral resolution is to smooth out the
     rise in the small-scale power at velocity lags smaller and comparable to the FWHM, and to
     broaden the peak in the correlation function at the doublet separation.}
   \label{fig:resolution}
\end{figure}

In Fig.~\ref{fig:resolution} we illustrate the impact of spectral
resolution on $\xi(\Delta v)$ for a model with $[{\rm Mg}\slash {\rm
    H}] = -3.4$ and $\langle x_{\ion{H}{i}}\rangle = 0.2$. Finite spectral
resolution smears out the small-scale structure in the \ion{Mg}{ii}
forest reducing the amount of variance at velocity lags smaller than
the FWHM of the spectrograph. This is readily apparent from the curves
in Fig.~\ref{fig:resolution}, where one sees that the rise in the
correlation function has been smoothed out at velocity lags comparable
to the spectral resolution. Because the peak in the correlation
function at the doublet separation $\Delta v=768~{\rm km~s^{-1}}$ also
results from small-scale correlations (see discussion in
\S~\ref{sec:model}), we observe that this peak is broadened by the
spectral smearing and its resulting width is effectively determined by
the spectral resolution. The black curves (labeled Perfect Spectrum)
have not been explicitly smoothed to model spectral resolution,
however the small-scale power is nevertheless be smoothed by the
finite spatial resolution of our simulation, which has a grid scale of
$29~{\rm ckpc}$ corresponding to $3.2 {\rm km~s^{-1}}$ in the Hubble
flow. As discussed in \S~\ref{sec:general}, our simulations will not
resolve the small-scale structure of the IGM if pre-reionization
baryons are at temperature $T\lesssim 1000\,{\rm K}$ resulting in a
Jeans scale smaller than our grid scale. However, we expect this
smoothing to only impact the correlation function at velocity lags
smaller than the grid scale $3.2~{\rm km~s^{-1}}$ (see
eqn.~\ref{eqn:xiden}). But measuring lags this small would require high-resolution
near-IR spectra which we do not consider here.

\section{Statistical Inference}
\label{sec:inference}

To assess the precision with which model parameters can be measured
from real observational data we construct mock observations and
perform statistical inference.

\subsection{The Mock Dataset}
We consider a realistic mock dataset
of $n_{\rm QSO}=10$ quasar spectra, each covering a pathlength of $\Delta
z=0.6$ of the \ion{Mg}{ii} forest, resulting in a total pathlength of
$\Delta z_{\rm tot} = n_{\rm QSO}\times \Delta z = 6.0$. Our
forward modeled spectra have a velocity extent of $6567~{\rm km~s^{-1}}$ set
by how our simulation box fits onto the spectral velocity grid, which is far smaller than the $21,936~{\rm
  km~s^{-1}}$ corresponding to
\footnote{For $\Delta z = 0.6$ and $z = 7.5$, we assume spectra
  covering the range $z = 6.9-7.5$ centered at $z_{\rm eff} = 7.2$,
  such that velocity interval covered is $c\Delta z/(1 +z_{\rm eff}) =
  21,936~{\rm km~s^{-1}}$} $\Delta z = 0.6$. We thus create a mock
dataset with the same effective pathlength by aggregating the
equivalent integer number of shorter $6567~{\rm km~s^{-1}}$
skewers. In other words, we model our mock dataset comprising of
$n_{\rm QSO} = 10$ quasars with $\Delta z=0.6$ and desired pathlength
of $\Delta z_{\rm tot} = 6.0$, with an integer number of $n_{\rm
  path}=33$ skewers, corresponding to a slightly shorter pathlength of
$\Delta z_{\rm tot} = 5.93$. As described in
\S~\ref{sec:observations}, we parameterize the data quality with the
spectral resolution FWHM and the ${\rm S\slash N}$ ratio, and here
consider observations with JWST/NIRSpec and assume FWHM=$100~{\rm
  km~s^{-1}}$ and ${\rm S\slash N}=100$.

\subsection{The Likelihood}
\begin{figure}
  \hskip -0.09cm
  \includegraphics[trim=0 0 0 0,clip,width=0.49\textwidth]{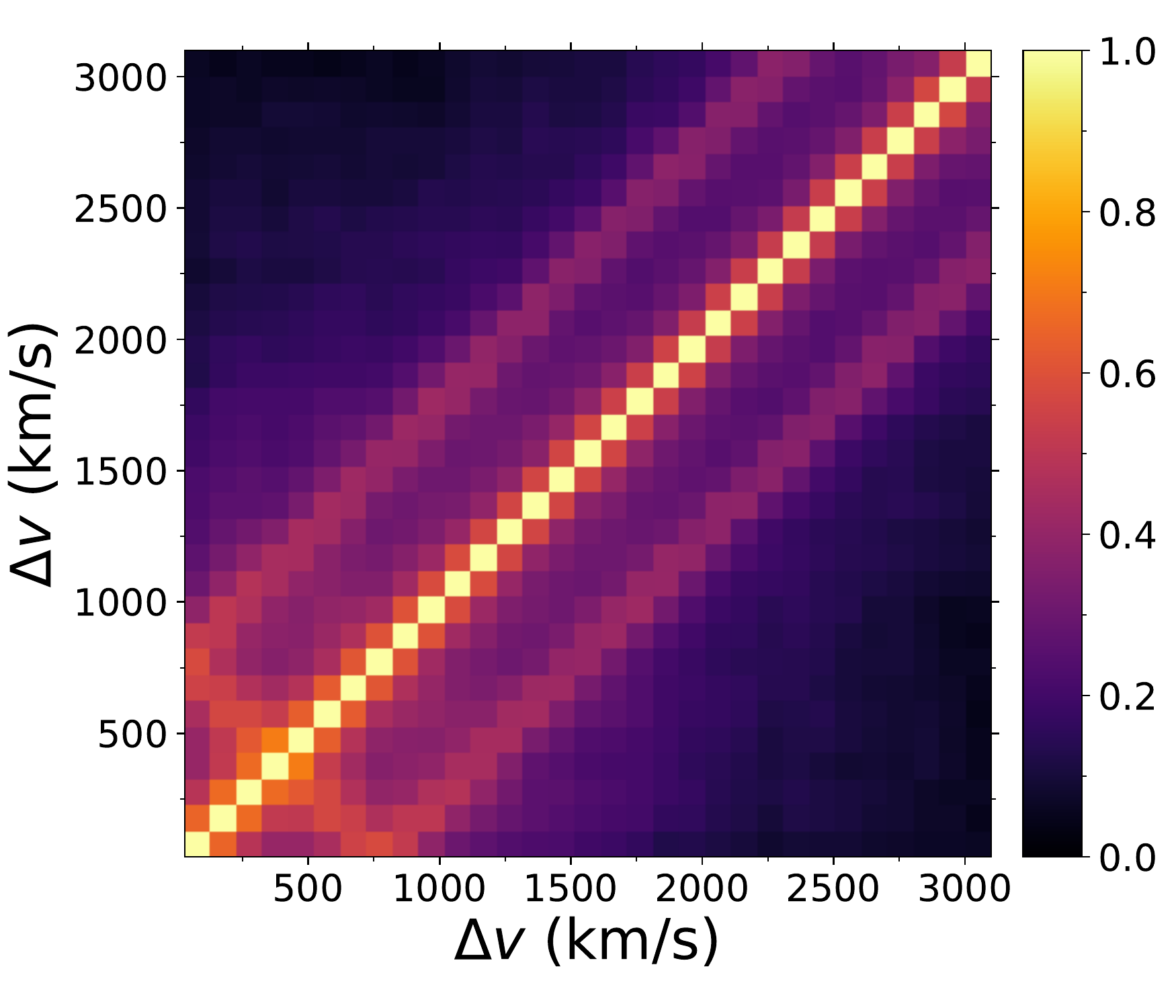}
  \vskip -0.1cm
  \caption{Correlation matrix of the \ion{Mg}{ii} forest correlation function.
    Correlation matrix for a model with $\langle x_{\ion{H}{i}}\rangle =
    0.74$ and $[{\rm Mg\slash H}] = -3.7$.  The diagonal band
    structure and the  high level of correlations along the `base' of the `trident'
    extending diagonally across the correlation matrix at small velocity lags
    result from correlations induced by the doublet nature
    of \ion{Mg}{ii}, which preferentially correlates velocity bins which
    probe the same underlying structures (see text for more details). 
    \label{fig:corr}}
\end{figure}

Following standard practice for correlation function measurements, we adopt a multivariate Gaussian likelihood for the \ion{Mg}{ii} forest correlation function
\be
L({\hat \xi}(\Delta v) \vert [{\rm Mg\slash H}], \langle x_{\ion{H}{i}}\rangle) = \frac{1}{\sqrt{(2\pi)^k \det{\mathbf{C}}}} \exp{\left(-\frac{1}{2} \mathbf{d}^\intercal \mathbf{C}^{-1} \mathbf{d}\right)}\label{eqn:lhood}, 
\ee
where $\mathbf{C}$ is the covariance matrix and
$\mathbf{d} \equiv {\hat \xi}(\Delta v) - \xi(\Delta v \vert [{\rm Mg\slash H}], \langle x_{\ion{H}{i}}\rangle)$,
where  ${\hat \xi}(\Delta v)$ is the correlation function estimated from the data and $\xi(\Delta v \vert [{\rm Mg\slash H}], \langle x_{\ion{H}{i}}\rangle)$ is the parameter dependent model correlation function,
which  we will henceforth simply denote by $\xi(\Delta v)$. 
We compute the
correlation function in $k=31$ linearly spaced velocity bins of equal width, which determines the dimensionality of $\mathbf{d}$ and
$\mathbf{C}$. We choose the bin width to match our resolution of $100~{\rm km~s^{-1}}$, and the bin centers extend from velocity
lags $80~{\rm km~s^{-1}}$ to $3080~{\rm km~s^{-1}}$.
From our ensemble of 10,000 skewers we compute the average value of
$\xi(\Delta v)$ at each location on our 2D grid ($51\times 201$)
of $\langle x_{\ion{H}{i}}\rangle$ and  $[{\rm Mg\slash H}]$ models. 

Typically the covariance matrix is either determined from the data itself, via i.e. a bootstrap procedure, or synthesized
from forward models. Given the relatively small mock dataset that we consider $n_{\rm QSO} = 10$ or $\Delta z_{\rm tot} = 6.0$, the covariance estimated from the data would be too noisy so 
 we adopt the latter approach. The covariance matrix is defined via
\be
  C_{ij} \equiv \langle [{\hat \xi}(\Delta v) - \xi(\Delta v)]_i [{\hat \xi}(\Delta v) - \xi(\Delta v)]_j\rangle\label{eqn:covar}
\ee
where the indices $i$ and $j$ denote bins of velocity lag $\Delta v$, and the angle brackets denote the average
over an ensemble of mock realizations of the dataset in question, which in this case is a correlation function
${\hat \xi}(\Delta v)$ computed from a set of  $n_{\rm path}=33$ skewers. Note that this covariance matrix
depends on the model parameters ($\langle x_{\ion{H}{i}}\rangle,[{\rm Mg\slash H}]$).
For each model in our $51\times 201$ grid, we generate $10^6$ mock datasets by grabbing $n_{\rm path} = 33$
random skewers from our sample of 10,000 without replacement, and computing their average 
correlation function  ${\hat \xi}(\Delta v)$,  allowing us to estimate the covariance from eqn.~(\ref{eqn:covar}).

A useful tool for visualizing the covariance structure is the
correlation matrix defined by \be {\rm Corr}_{ij} \equiv
\frac{C_{ij}}{\sqrt{C_{ii}C_{jj}}}.  \ee Fig.~\ref{fig:corr} shows an
example correlation matrix for a model with $\langle x_{\ion{H}{i}}\rangle
= 0.74$ and $[{\rm Mg\slash H}] = -3.7$. The diagonal band structure
of the correlation matrix can be easily understood. By definition the
correlation matrix is unity along the diagonal. The other two
prominent sidebands result from correlations induced by the doublet
nature of \ion{Mg}{ii}, i.e. a fluctuation in $\xi(\Delta v)$ in a bin
at $\Delta v = 1200~{\rm km~s^{-1}}$ will preferentially correlate
with fluctuations in velocity bins at $\Delta v = 1200 \pm 768\,{\rm
  km\,s^{-1}}$. Finally, the steep rise of the correlation function
towards zero-lag and its `mirror image' at $\Delta v=768\,{\rm
  km\,s^{-1}}$ (see Fig.~\ref{fig:corrfunc}), implies correlation function
estimates at small velocity lags correlate more strongly with each
other.  For example the correlation function bin at $\Delta v = 500~{\rm km~s^{-1}}$, which is
$268~{\rm km~s^{-1}}$ away from the doublet peak at  $\Delta v = 768\,{\rm
  km\,s^{-1}}$, actually contains contributions from the same structures producing the small-scale
rise of the correlation function towards zero lag at velocity $\Delta v = 268~{\rm km~s^{-1}}$, resulting
in a high value for the correlation matrix $\simeq 0.6$. These effects conspire
to produce the high level of correlations along the `base' of the `trident'
extending diagonally across the correlation matrix at small velocity lags.

Finally, we note that the `zero-lag' $\xi(\Delta v)$ bin is completely omitted
from our inference calculations. Although in principle this bin
contains information, utilizing it would require that one subtract off
the noise variance. Our $1\sigma$ noise level per spectral pixel is
$1\%$, whereas the standard deviation of the \ion{Mg}{ii} forest per
spectral pixel is $0.5\%$ for a model with $\langle x_{\ion{H}{i}}\rangle
= 0.74$ and $[{\rm Mg\slash H}] = -3.7$, and scales as roughly
metallicity squared (see discussion in \S~\ref{sec:model}). Thus using
the `zero-lag' bin would presume that one can determine the absolute noise level to
exquisite accuracy, whereas using non-zero lags assumes that the noise correlations are much smaller
than the signal correlations, which is a far weaker assumption given that the noise is expected to be white.

\subsection{Results}
\begin{figure*}
  \hskip -0.09cm
  \includegraphics[trim=3 0 20 10,clip,width=0.60\textwidth]{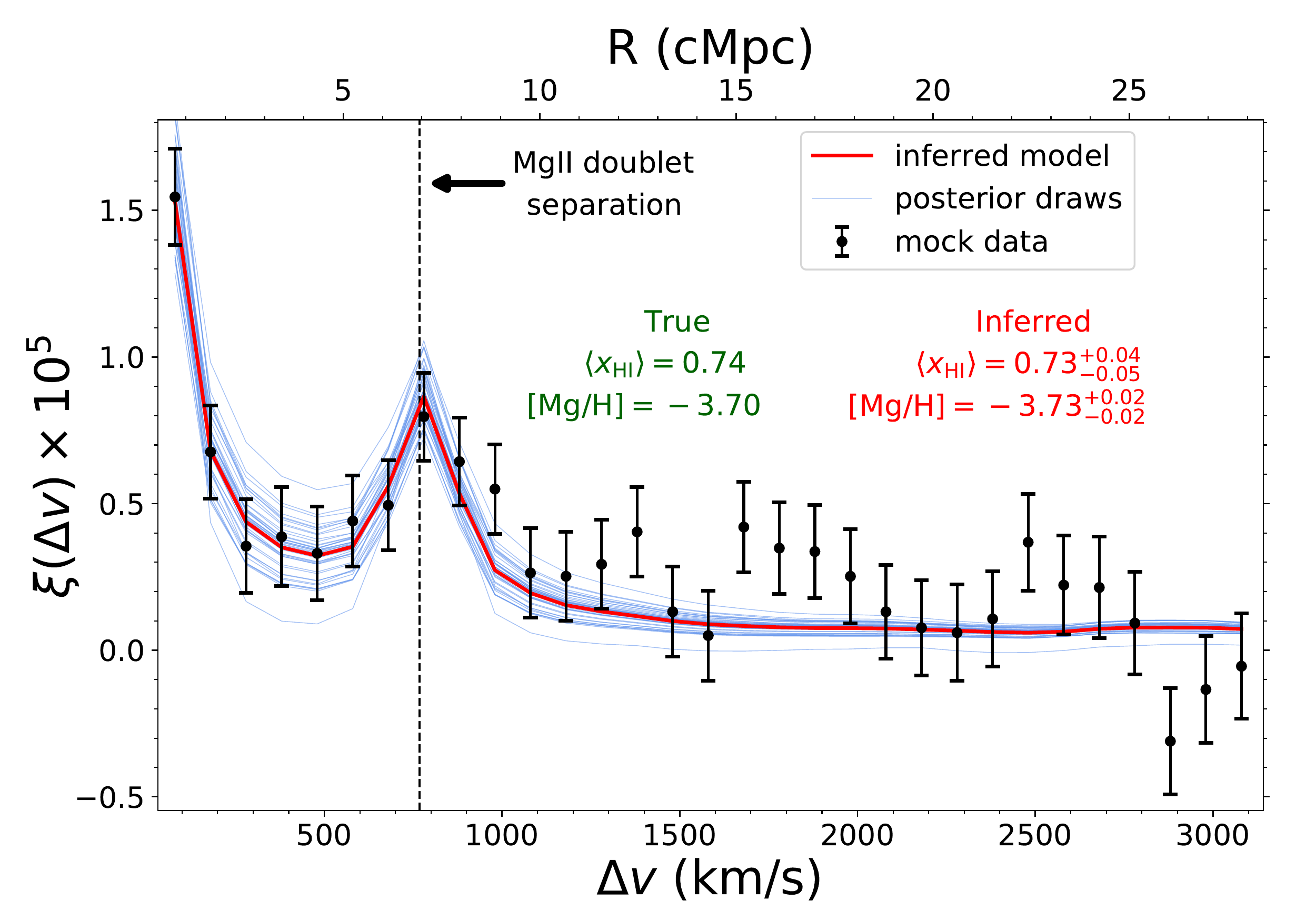}
  \raisebox{0.17cm}{\includegraphics[trim=7 10 15 8,clip,width=0.40\textwidth]{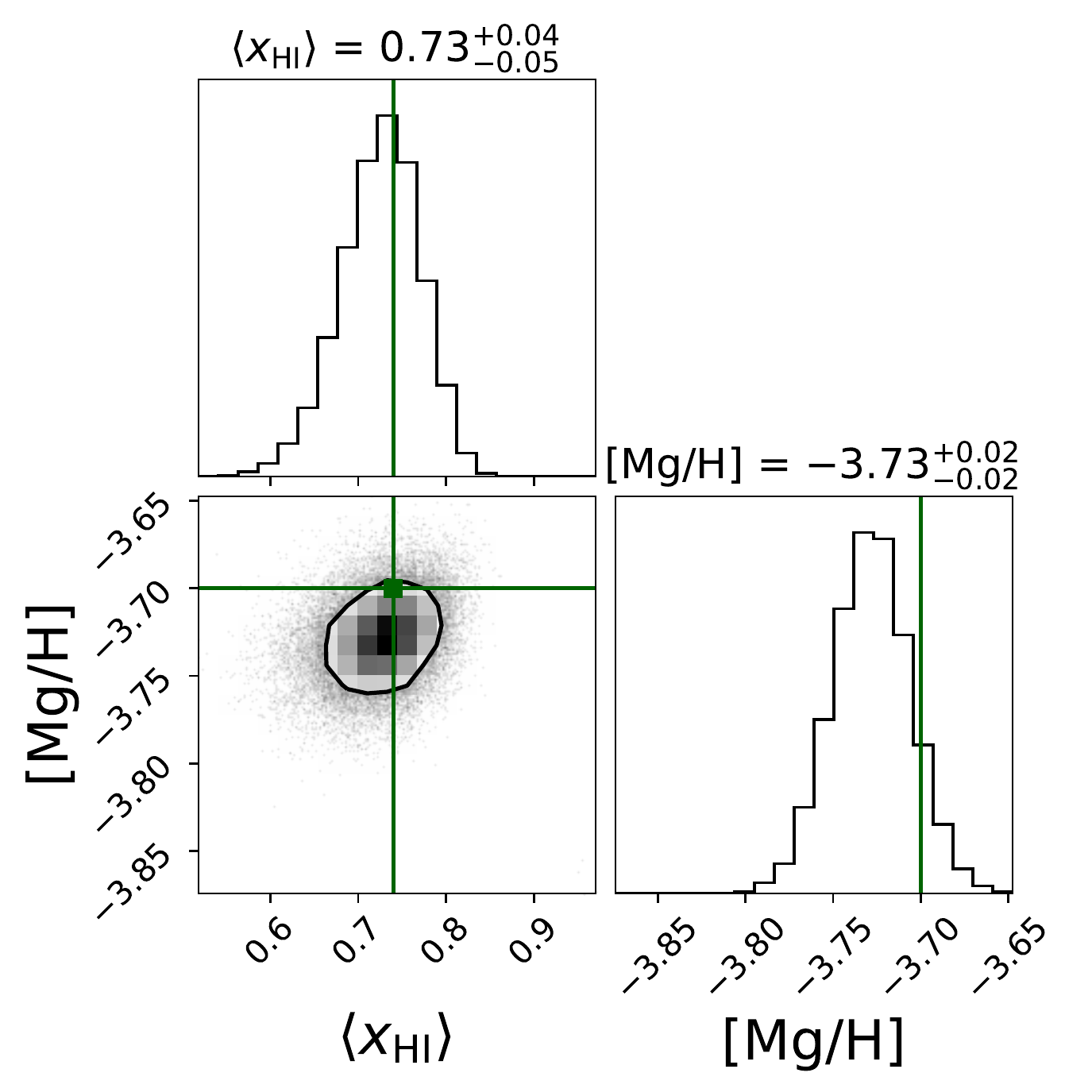}}
  \vskip -0.1cm
  \caption{Simulated correlation function measurement and
    inference for a mock \ion{Mg}{ii} forest dataset. \emph{Left:} Points
    with error bars show the measured correlation function for a mock
    JWST dataset of $n_{\rm QSO}=10$ quasars (FWHM=$100~{\rm
      km~s^{-1}}$ and ${\rm S\slash N}=100$, see
    \S~\ref{sec:observations} and Fig.~\ref{fig:skewers}) covering a
    total pathlength of $\Delta z_{\rm tot} = 6.0$. The thin blue
    lines are random draws from the posterior sampled via MCMC (shown
    in the right panel), and the red curve is the mean inferred IGM
    model.  \emph{Right}: Corner plot determined via MCMC sampling of
    the posterior distribution.  The true model had $[{\rm Mg\slash H}]
    = -3.7$ and $\langle x_{\ion{H}{i}}\rangle = 0.74$ as indicated by
    the dark green square and lines, whereas we recover values of $[{\rm Mg\slash H}] = -3.73 \pm 0.02$ and
    $\langle x_{\ion{H}{i}}\rangle = 0.73^{+0.04}_{-0.05}$, which are the median and 68\% credible
    intervals of the marginalized posterior distributions for each parameter. 
    At this location in
    parameter space we jointly determine the Mg abundance $[{\rm Mg \slash H}]$ with a
    $1\sigma$ precision of 0.02~dex and measure the volume averaged neutral
    fraction $\langle x_{\ion{H}{i}}\rangle$ to 5\%. \label{fig:inference}}
\end{figure*}

Given our mock JWST dataset of $n_{\rm QSO}=10$ quasars
(FWHM=$100~{\rm km~s^{-1}}$ and ${\rm S\slash N}=100$, see
\S~\ref{sec:observations} and Fig.~\ref{fig:skewers}) we can use the
likelihood in eqn.~(\ref{eqn:lhood}) to perform Markov Chain Monte
Carlo (MCMC) parameter inference.  We assume a flat linear prior on
the volume averaged neutral fraction extending from $x_{\ion{H}{i}} =
[0.0, 1.0]$, and a flat prior in the $\log_{10}$ of the Mg abundance
from $[{\rm Mg}\slash {\rm H}] = [-6.0, -2.0]$, i.e.  our prior is
uninformative and spans the parameter space covered by our model grid.
For the fiducial model, we choose $[{\rm Mg}\slash {\rm H}] = -3.7$
and $x_{\ion{H}{i}} = 0.74$, where the former is motivated by IGM
metallicity measurements at lower-$z$, and the latter by current
reionization constraints from the CMB \citep[e.g.][]{Planck18},
IGM damping wings towards $z \gtrsim 7$ quasars
\citep{Mortlock11,Greig17b,Banados18,Davies18b,Greig19,Wang20a,Yang20b}
and the disappearance of strong Ly$\alpha$ emission from galaxies
\citep{Mason18,Mason19,Hoag19}.  The resulting mock correlation
function and parameter constraints are shown in
Fig.~\ref{fig:inference}. Our analysis indicates that for this
combination of model parameters one can simultaneously determine the
Mg abundance $[{\rm Mg}\slash {\rm H}]$, with a $1\sigma$ precision of
0.02~dex, and measure the global neutral fraction $\langle x_{\ion{H}{i}}\rangle$ to 5\%.

\begin{figure}
  \hskip -0.09cm
  \includegraphics[trim=0 0 0 50,clip,width=0.49\textwidth]{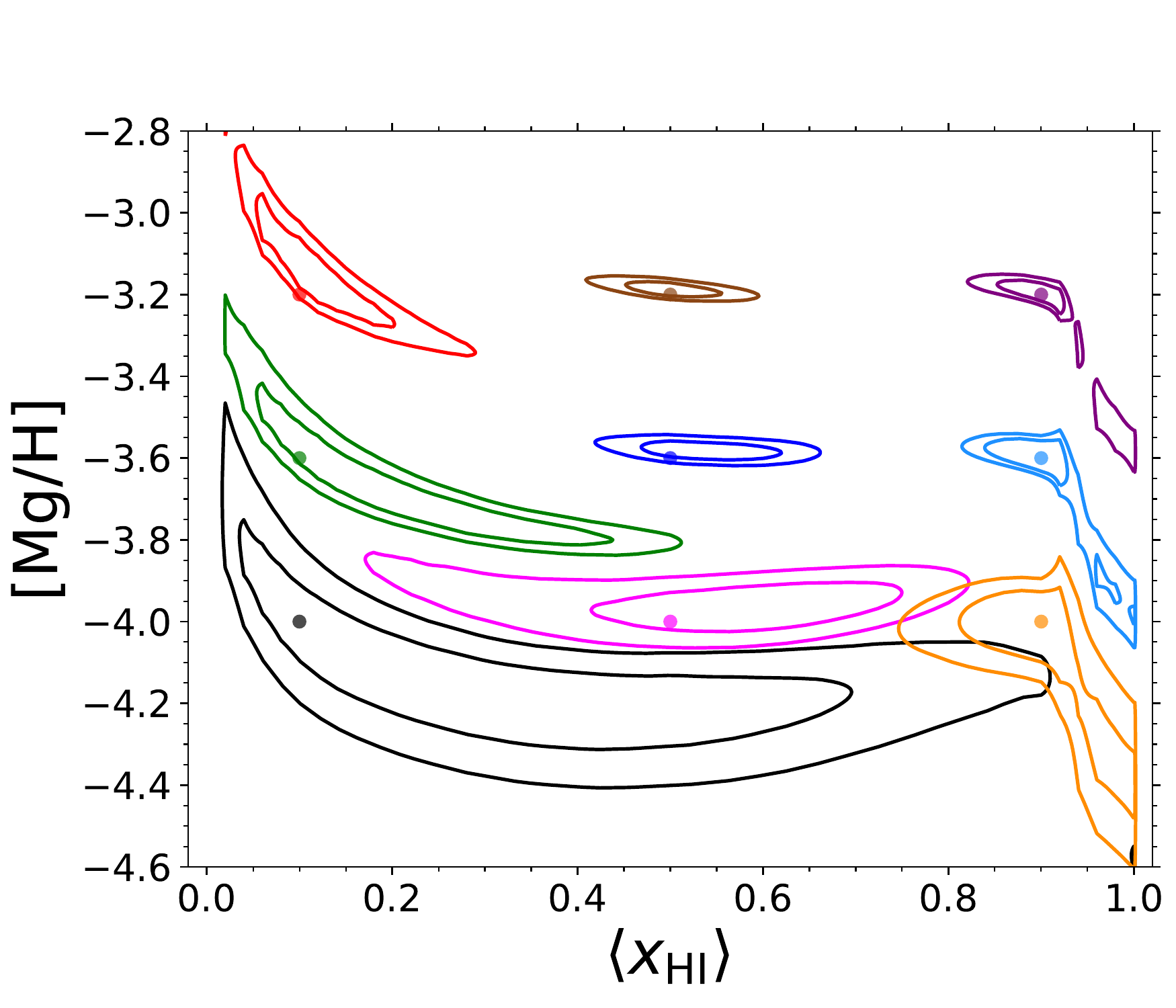}
  \vskip -0.2cm
  \caption{Constraining power of a mock \ion{Mg}{ii} forest dataset at
    different locations in parameter space.  Contours indicate the
  $68\%$ and $95\%$ confidence regions resulting from a mock JWST
  dataset of $n_{\rm QSO}=10$ quasars (FWHM=$100~{\rm km~s^{-1}}$ and
  ${\rm S\slash N}=100$, see \S~\ref{sec:observations} and
  Fig.~\ref{fig:skewers}) covering a total pathlength of $\Delta
  z_{\rm tot} = 6.0$. The filled circles indicate the true value of
  the model from which the mock dataset was drawn, which were chosen
  on a grid of locations in the 2D parameter space at $[{\rm Mg\slash
      H}] = [-3.2, -3.6 ,-4.0$] and $\langle x_{\ion{H}{i}}\rangle = [0.1,
    0.5, 0.9]$.\label{fig:contours}}
\end{figure}

The constraining power of the \ion{Mg}{ii} correlation function, or
more precisely the width and orientation of the contours in the
$\langle x_{\ion{H}{i}}\rangle$-$[{\rm Mg}\slash {\rm H}]$ plane in
the right panel of Fig.~\ref{fig:inference}, are a strong function of
the true value of the parameters. We illustrate this dependence in
Fig.~\ref{fig:contours}, which shows the resulting 68\% and 95\%
confidence intervals (colored lines) at a grid of values for the true
model (indicated by filled circles) in the $\langle
x_{\ion{H}{i}}\rangle$-$[{\rm Mg}\slash {\rm H}]$ plane.  For
intermediate values of the neutral fraction $\langle
x_{\ion{H}{i}}\rangle \simeq 0.5$ and Mg abundances $[{\rm Mg}\slash
  {\rm H}] \gtrsim -4$, there is no significant degeneracy between the
two parameters. At low $\langle x_{\ion{H}{i}}\rangle \lesssim 0.1$
and high $\langle x_{\ion{H}{i}}\rangle \gtrsim 0.9$ neutral fractions
a degeneracy between the parameters starts to emerge. This degenerate
behavior can be qualitatively understood as follows. For $\langle
x_{\ion{H}{i}}\rangle \lesssim 0.1$, the steepening power law shape of
the correlation function washes out the peaks arising from small-scale
structure toward both zero-lag and the doublet separation (see
Fig.~\ref{fig:corrfunc}), particularly at JWST resolution (${\rm
  FWHM}=100~{\rm km~s^{-1}}$) where these peaks are smeared (see
Fig.~\ref{fig:resolution}). In this regime the additional constraining
power provided by these peaks is washed out, and the $[{\rm Mg}\slash
  {\rm H}]$ increases the roughly power-law correlation function
amplitude, whereas $\langle x_{\ion{H}{i}}\rangle$ alters its
amplitude and slope, resulting in a degeneracy. The degeneracy at
$\langle x_{\ion{H}{i}}\rangle \gtrsim 0.9$ occurs for similar
reasons.  For these largely neutral models, the power-law behavior of
the correlation (left panel of Fig.~\ref{fig:corrfunc}) due to the
topology of reionization is suppressed, and all the signal is
concentrated at small velocity lags and at the doublet separation. In
this regime the naive degeneracy expected from the optical depth (see
eqn.~\ref{eqn:tauv} and the discussion at the end of
\S~\ref{sec:model}) sets in, since the degenerate product of
metallicity and neutral fraction determine the amplitude of
fluctuations and hence the correlation function. This degeneracy is
exacerbated by the fact that reionization occurs from the inside out,
and at high values of $\langle x_{\ion{H}{i}}\rangle$ the rare ionized
patches will be co-spatial with the highest density gas. As a fully
neutral Universe is approached $\langle x_{\ion{H}{i}}\rangle\rightarrow
1$ these dense regions become neutral, and the amplitude of the
correlation function will become hyper-sensitive to $\langle
x_{\ion{H}{i}}\rangle$ due to the outsize contribution of these dense
regions to the fluctuations.  Together we expect some degeneracy
between $[{\rm Mg}\slash {\rm H}]$ and $\langle x_{\ion{H}{i}}\rangle$
to emerge at high values $\langle x_{\ion{H}{i}}\rangle\gtrsim 0.9$,
and that a small change in $\langle x_{\ion{H}{i}}\rangle$ can
compensate for a relatively large change in $[{\rm Mg}\slash {\rm
    H}]$, which is exactly the behavior observed in
Fig.~\ref{fig:contours}.

It is conceivable that the neutral IGM is actually totally pristine,
as would be the case if the metals produced by the star-formation that
drives reionization remain highly concentrated around the galaxies
producing them. In this scenario, we would obtain a null detection of
the \ion{Mg}{ii} forest correlation function even if the IGM were
significantly neutral. Such a null detection would nevertheless
provide an upper limit on the enrichment of the IGM during the EoR,
providing an extremely interesting constraint on the enrichment
history of the Universe. To quantify this, we assume that an
independent constraint on the reionization history exists from other
probes (e.g. CMB, IGM damping wings, Ly$\alpha$ disappearance in galaxies, or 21cm observations)
such that $\langle x_{\ion{H}{i}}\rangle > 0.5$ at $z = 7.5$. 
For our fiducial model we choose $[{\rm
    Mg}\slash {\rm H}] = -6.0$ (the lowest metallicity in our grid)
and $x_{\ion{H}{i}} = 0.74$, which results in a correlation function
effectively consistent with zero.  We perform statistical inference
via MCMC as before, but now adjust the prior to have $\langle x_{\ion{H}{i}}\rangle > 0.5$, and importantly, we adopt a \emph{linear prior}
on the Mg abundance to be in the range $10^{[{\rm Mg}\slash {\rm H}]}
= [0.0, 1.0]$. The reasoning behind changing the abundance prior to be
linear, as opposed to the $\log_{10}$ prior adopted above, is that for a logarithmic
prior, the resulting upper limit on $[{\rm Mg}\slash {\rm H}]$ would
depend on the prior range adopted, whereas this is not the case with a
linear prior. Marginalizing over the unknown neutral fraction with the MCMC samples, we find
that a null correlation function detection from our mock dataset would place an upper limit on the Mg abundance of $[{\rm Mg}\slash {\rm H}] < -4.4$ at 95\% confidence. This stringent limit is nearly 0.5 dex more
sensitive than the most metal-poor Lyman Limit Systems (LLSs) and Damped Ly$\alpha$ Systems (DLAs) known \citep{Fumagalli_sci11,Crighton16,Cooke17,Robert19}
and is in the realm of the alpha element abundances of the most metal-poor stars known \citep{Frebel15}. 
But whereas these metal-poor absorbers and stars constitute the rarest outliers from their respective 
parent populations, the sensitive \ion{Mg}{ii} forest abundance constraint one would obtain is the average
for the IGM as a whole.

\section{Modeling Circumgalactic \ion{Mg}{ii} Absorbers}
\label{sec:CGM}

Up to this point our analysis has ignored the impact of \ion{Mg}{ii} absorption line systems associated
with galaxies. Specifically, our toy enrichment model assumes that all the gas in the Universe is suffused with
Mg parameterized by a uniform abundance $[{\rm Mg\slash H}]$. In reality, there will be a high concentration of
Mg in the circumgalactic environs of galaxies, and it is important to understand how contamination from these
CGM absorbers impacts the \ion{Mg}{ii} forest correlation function and our resulting parameter constraints. 

We now expand our toy enrichment model to have two components, the
uniform metallicity IGM that we considered previously plus additional
CGM absorbers arising from galaxies. To model the latter, we must consider
two quantities: their line density as a function of absorption line
strength and the spatial distribution distribution of metal
absorbers. 

\subsection{The Abundance of CGM Absorbers}

To populate our simulated skewers with CGM absorbers we require
the distribution function  $\frac{d^2N}{dz
  dW_\lambda}$  of rest-frame equivalent width. 
Whereas studies sensitive to the strongest absorbers typically
adopt an exponential form \citep[e.g.][]{ntr05,Chen17} for this distribution function, echelle
based searches sensitive enough to detect weak \ion{Mg}{ii} absorbers $W_\lambda \lesssim 0.1$~\AA~
find that a Schechter-like \citep{Schechter76} function  provides a better
fit to the data \citep{Kacp11,Vulture17}, 
\begin{equation}
  \frac{d^2N}{dz dW_\lambda} = \frac{N_\ast}{W_\ast}\left(\frac{W_\lambda}{W_\ast}\right)^{\alpha}\exp{\left(-\frac{W_\lambda}{W_\ast}\right)}\label{eqn:dNdzdW},
\end{equation}
which we will adopt here since, as we will see, only the weakest CGM absorbers are expected to significantly
contaminate the \ion{Mg}{ii} forest signal arising from the neutral IGM.

The abundance of weak absorbers at $z > 6$ is currently not well
constrained by observations. \citet{Chen17} measured $\frac{d^2N}{dz
  dW_\lambda}$ from the redshift range $z = 6.0-7.08$ from a sample of high-redshift
quasars, and \citet{Bosman17} surveyed absorbers over nearly the same
interval using a single high-quality spectrum of the ULAS J1120$+$0641
($z_{\rm em}=7.08$) quasar sightline.  These measurements are shown in
Fig.~\ref{fig:dNdzdW}.  Note that they are not independent since the
ULAS J1120$+$0641 quasar is also in the \citet{Chen17} sample, although their
spectrum is not as sensitive.  Clearly current data are too noisy to
independently constrain the three parameters $(\alpha, W_\ast,
N_\ast)$ governing the equivalent width distribution in
eqn.~(\ref{eqn:dNdzdW}), as also emphasized by \citet{Bosman17}.  We
thus adopt the following approach to set their values. The most
important parameter is the slope $\alpha$, since it has the largest
impact on the abundance of weak absorbers that dominate the
contamination of the \ion{Mg}{ii} forest. \citet{Vulture17} used a
large archival echelle dataset to measure $\frac{d^2N}{dz dW_\lambda}$
for  ($0.01\,{\text\AA} < W_\lambda < 10\,{\text \AA}$) over the redshift range
$0.14 < z < 2.64$, and found slopes in the range
$\alpha=-1.1$ to $-0.8$. We thus adopt the value to $\alpha=-0.8$
consistent with their measurement of $\alpha=-0.81\pm 0.12$ in the
highest redshift bin ($1.53 < z < 2.64$) that they studied. To set the
other parameters, we simply fix $W_\ast=1.0$~\AA, and then we
determine $N_\ast$ by requiring that our equivalent width distribution
reproduce the abundance of absorbers $dN\slash dz$ in the range
$0.6\,{\text \AA} < W_\lambda < 1.0\,{\text \AA}$ at $z =
7.5$. Specifically, \citet{Chen17} fit the $dN\slash dz$ with a
functional form $dN\slash dz = A(1 + z)^\beta$ with $A=0.09$ and
$\beta=0.82$ (see their Table 7 and Figure 10) implying $dN\slash dz =
0.52$ at $z=7.5$. This procedure finally yields $(\alpha, W_\ast,
N_\ast) = (-0.80, 1.0\,{\text \AA},2.34)$ which gives the equivalent
width distribution shown as the red curve in Fig.~\ref{fig:dNdzdW}.
We assume that CGM \ion{Mg}{ii} absorbers follow this distribution,
but we truncate it at the low and high equivalent widths of $W_{\lambda,{\rm min}}=0.01$\,\AA~ and
$W_{\lambda,{\rm max}} = 10$\,\AA, respectively, where these values correspond to roughly
the weakest and strongest absorbers that have been observed to date \citep[e.g.][]{Vulture17}.

\subsection{The Clustering of CGM Absorbers}
Because of their higher abundance, most of what we know about the
spatial distribution of metals in the IGM comes from \ion{C}{iv} absorbers.
Significant effort has been dedicated to understanding how strong \ion{C}{iv} (i.e. $N_{\ion{C}{iv}} \gtrsim 10^{13}~{\rm cm^{-2}}$ or $W_\lambda \gtrsim
0.20\,{\text \AA}$) systems cluster, including
both auto-correlation studies \citep{Quashnock99,Coppolani06,Martin10}, as well as
cross-correlation analyses with both Lyman Break Galaxies
\citep{adel03,ass+05} and quasars
\citep{Vikas13,QPQ6}. Similarly, the auto-correlation of strong
\ion{Mg}{ii} absorbers (i.e. $W_\lambda \gtrsim 1.0\,{\text \AA}$) has
been measured \citep{SteidelSargent92,QuashnockVB98,Tytler09} as well as the
cross-correlation with so-called luminous red galaxies 
\citep[LRGs; e.g.][]{BoucheLRG06,Lundgren09,Gauthier14}.
The qualtiative
picture that emerges from these studies is that these strong absorbers are clustered
similar to co-eval galaxies and reside in dark matter halos of $\sim 10^{12}~M_\odot$.
But as we will see, the strong absorbers that contaminate a \ion{Mg}{ii} forest measurement
will be easy to identify and mask in the JWST spectra that we envision obtaining, and it is the weak absorption line systems that cannot
be individually detected which will be our dominant contaminant. 

Much less is known about the clustering of weak absorption systems, as
these can only be identified in high ${\rm S\slash N}$ ratio echelle
resolution spectra
\citep[e.g.][]{Churchill99,Songaila05,Narayanan07,Dodorico10,Vulture17,MasRibas18},
and the relative paucity of such data inhibits the compilation of the
large absorber samples required to measure weak clustering signals.
The most comprehensive and sensitive study is the work by
\citet{Boksenberg15}, who measured the clustering of weak \ion{C}{iv}
absorbers 
($N\gtrsim 10^{12}~{\rm cm^{-2}}$ or $W_\lambda \gtrsim 0.02\,{\text
  \AA}$)
from a sample of $\sim 200$ systems over the redshift range
$1.6 \lesssim z \lesssim 4.4$.  They found significant clustering for
$\Delta v\lesssim 300\,{\rm km~s^{-1}}$ ($r_{\parallel} = 2.7~{\rm
  cMpc}$ if it were in the Hubble flow), but clustering is not detected on larger
scales. Furthermore, they argue rather convincingly that this
small-scale clustering signal likely arises from the complex
kinematics of individual \ion{C}{iv} components, which can be grouped
together into aggregate absorption `systems', and that this signal arises
primarily from the stronger absorbers in their sample.  Furthermore,
after grouping these systems into aggregate systems the clustering
signal measured is consistent with zero.
These results are in qualitative agreement with previous
work on weak \ion{C}{iv} based on smaller samples
\citep[][but see \cite{Scannapieco06}]{Sargent80,Sargent88,Petitjean94,Rauch96,Pichon03} as well as an
analogous analyses of weak
\ion{Mg}{ii} absorbers \citep[$W_\lambda \gtrsim
0.3~{\text \AA}$;][]{Petitjean90,Churchill03}. Given the
lack of convincing evidence for $\gtrsim 1~{\rm cMpc}$ clustering of weak absorption line systems, we
therefore neglect absorber clustering in our CGM model. This is a
reasonable assumption because the clustering of the \ion{Mg}{ii} forest
on large scales ($\Delta v \gtrsim 1000~{\rm km~s^{-1}}$ or
$r_{\parallel}\gtrsim 9~{\rm cMpc}$) results from large coherent
fluctuations in the IGM neutral faction, which should dominate over
any weak large scale clustering of CGM absorbers.

\subsection{The Final CGM Model}

\begin{figure}
  \includegraphics[trim=0 0 0 0,clip,width=0.49\textwidth]{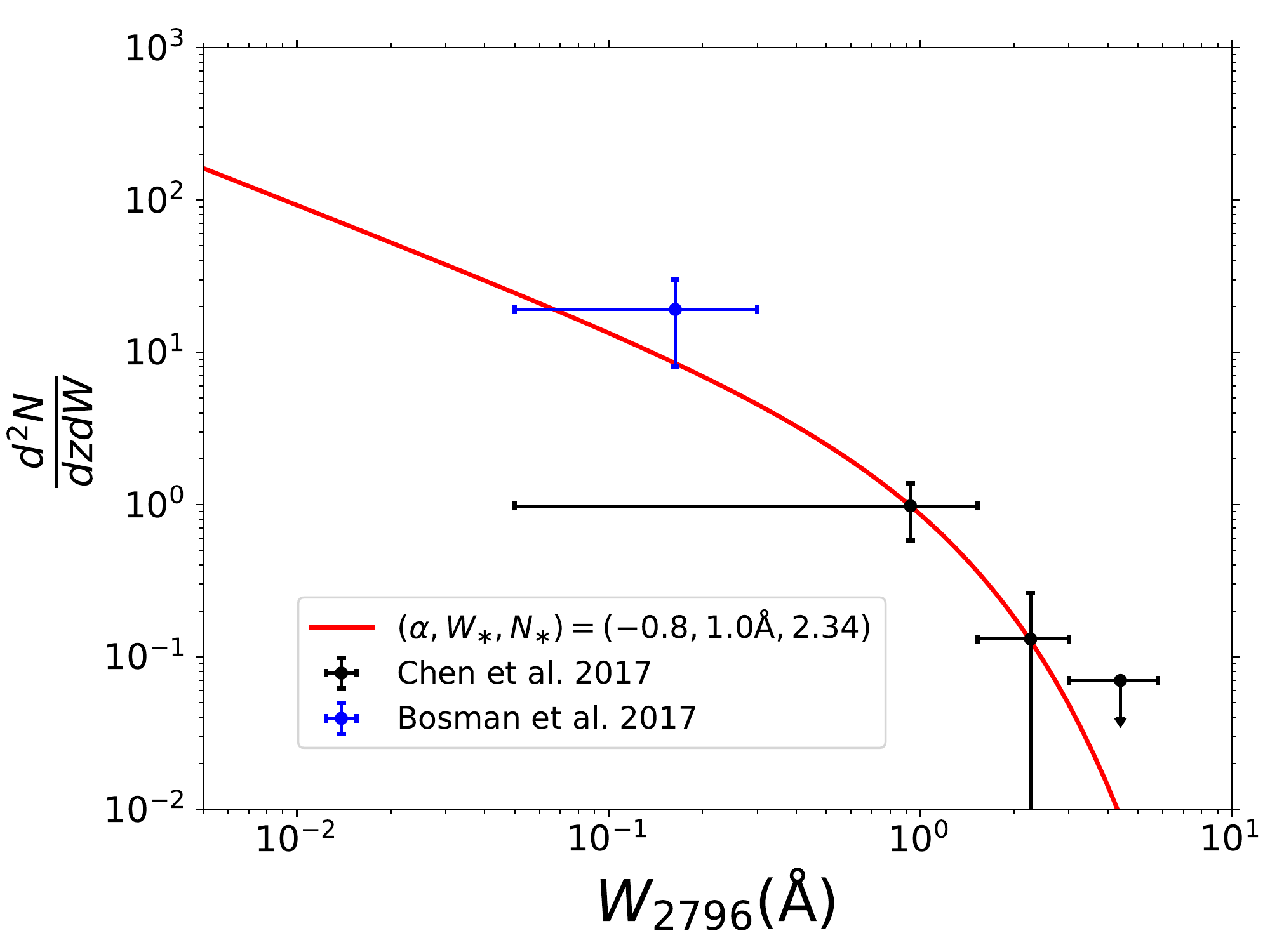}
  \caption{Rest-frame equivalent width distribution for CGM
      \ion{Mg}{ii} Absorbers. The red curve shows the Schechter
    function model (see eqn.~(\ref{eqn:dNdzdW})) used for the
    distribution of CGM equivalent widths. The black and blue points
    are measurements probing the redshift range $6.0 \lesssim z
    \lesssim 7.08$ from \citet{Chen17} and \citet{Bosman17},
    respectively.\label{fig:dNdzdW}}
  \vskip -0.2cm
\end{figure}

We populate our simulated \ion{Mg}{ii} forest spectra with CGM contaminants by drawing a number
of absorbers from the equivalent width distribution shown in Fig.~\ref{fig:dNdzdW} commensurate
with the pathlength $\Delta z$ probed by the skewer. Each absorber is randomly assigned a velocity along the skewer,
consistent with our assumption of no absorber clustering. The optical depth of each absorber is added to the skewer using
the full Voigt profile for an assumed  Gaussian velocity distribution. This requires a recipe for choosing a $N_{\ion{Mg}{ii}}$ and $b$-value
that gives the desired rest-frame equivalent width $W_\lambda$. On the linear part of the curve-of-growth (COG) the relationship between
$N_{\ion{Mg}{ii}}$ and $W_\lambda$ is
\begin{equation}
  W_\lambda = 0.43\,\angstrom\,\left(\frac{N_{\ion{Mg}{ii}}}{10^{13}~{\rm cm^{-2}}}\right)\label{eqn:Wr}, 
\end{equation}
and the optical depth at line center for the Gaussian core of the Voigt profile is
\begin{equation}
\tau_0 = 1.3\left(\frac{N_{\ion{Mg}{ii}}}{10^{13}~{\rm cm^{-2}}}\right)\left(\frac{b}{20~{\rm km~s^{-1}}}\right)^{-1}\label{eqn:tau0_cgm}, 
\end{equation}
For weak \ion{Mg}{ii} absorbers $W_\lambda \lesssim 0.1$~\AA~ the optical
depth weighted second moments are typically $\sim 20~{\rm km~s^{-1}}$,
whereas strong absorbers with $W_\lambda > 1$~\AA~ are significantly broader
$\sim 100~{\rm km~s^{-1}}$ \citep{Churchill01,Vulture17}. According to eqns.~(\ref{eqn:Wr}) and (\ref{eqn:tau0_cgm}), the COG saturates around $N_{\ion{Mg}{ii}} = 10^{13}~{\rm cm^{-2}}$ or a $W_\lambda\simeq 0.4$~\AA~ for $b = 20~{\rm km~s^{-1}}$. Thus, it will be impossible to
generate the strongest absorbers $W_\lambda > 1$~\AA~ with such a low $b$-value. To model changes in $b$ for stronger absorbers, we adopt
\begin{eqnarray}
  b &=& b_{\rm weak} + \left(b_{\rm strong} - b_{\rm weak}\right) \times\nonumber\\
  &&\left[1 + \exp{\left(-\frac{\log_{10} N_{\ion{Mg}{ii}} - \log_{10}N_{\rm strong}}{\Delta \log_{10} N}\right)}\right]^{-1}, 
\end{eqnarray}
where the second term is the `logistic sigmoid' function that
guarantees a smooth transition with column density between the value
of $b_{\rm weak}$ and $b_{\rm strong}$ around the transition column
density $\log_{10} N_{\rm strong}$, over a column density interval set by
 $\Delta \log_{10}N$.
We adopt $b_{\rm weak} = 20~{\rm
  km~s^{-1}}$, $b_{\rm strong}=200~{\rm km~s^{-1}}$, $\log_{10} N_{\rm
  strong} = 16$ and $\Delta \log_{10}N = 0.25$, where all column
densities are in units of ${\rm cm^{-2}}$. We will see that the contamination of the IGMs \ion{Mg}{ii} forest signal is insensitive to the
strongest absorbers, which are easily identified and masked, and is instead dominated by the weaker absorbers below the detection limit
of our simulated spectra, which is $W_\lambda \simeq 0.04$~\AA~ or $N_{\ion{Mg}{ii}} \simeq 10^{12}~{\rm cm^{-2}}$ for the mock JWST spectra. 
These weak absorbers have $b\approx 20~{\rm km~s^{-1}}$ much smaller than the resolution of the JWST spectra
(FWHM=$100~{\rm km~s^{-1}}$) that we simulate. For these reasons our results are insensitive to the details of the $b$ values assumed: the
weak absorbers are not resolved by our spectra and the strong absorbers are masked.

\section{The Flux Probability Distribution Function of the \ion{Mg}{ii} Forest}
\label{sec:PDF}
The goal of this section is to understand the contamination of the IGM \ion{Mg}{ii} forest absorption
by CGM absorption arising from the enriched halos of galaxies. To do so we must quantify, for a given absorption
level, the likelihood that it arises from the IGM versus the CGM.
Our modeling of CGM absorbers focused on the distribution of rest-frame equivalent widths, $W_\lambda$,  which is the obvious
choice for measurements of discrete \ion{Mg}{ii} absorbers, particularly in the regime where they are
not spectrally resolved. But the notion of equivalent
width loses its utility for a continuous absorption field,
analogous to the situation for the Ly$\alpha$ forest at lower redshift. 
In this regime, choosing the spectral regions for the equivalent
width integral would be arbitrary -- it is impossible to decide where the absorption starts and ends.
Indeed, the more appropriate description of a continuous random field is the flux PDF,
which will be the focus of this section.

\begin{figure}
  \includegraphics[trim=7 0 5 0,clip,width=0.49\textwidth]{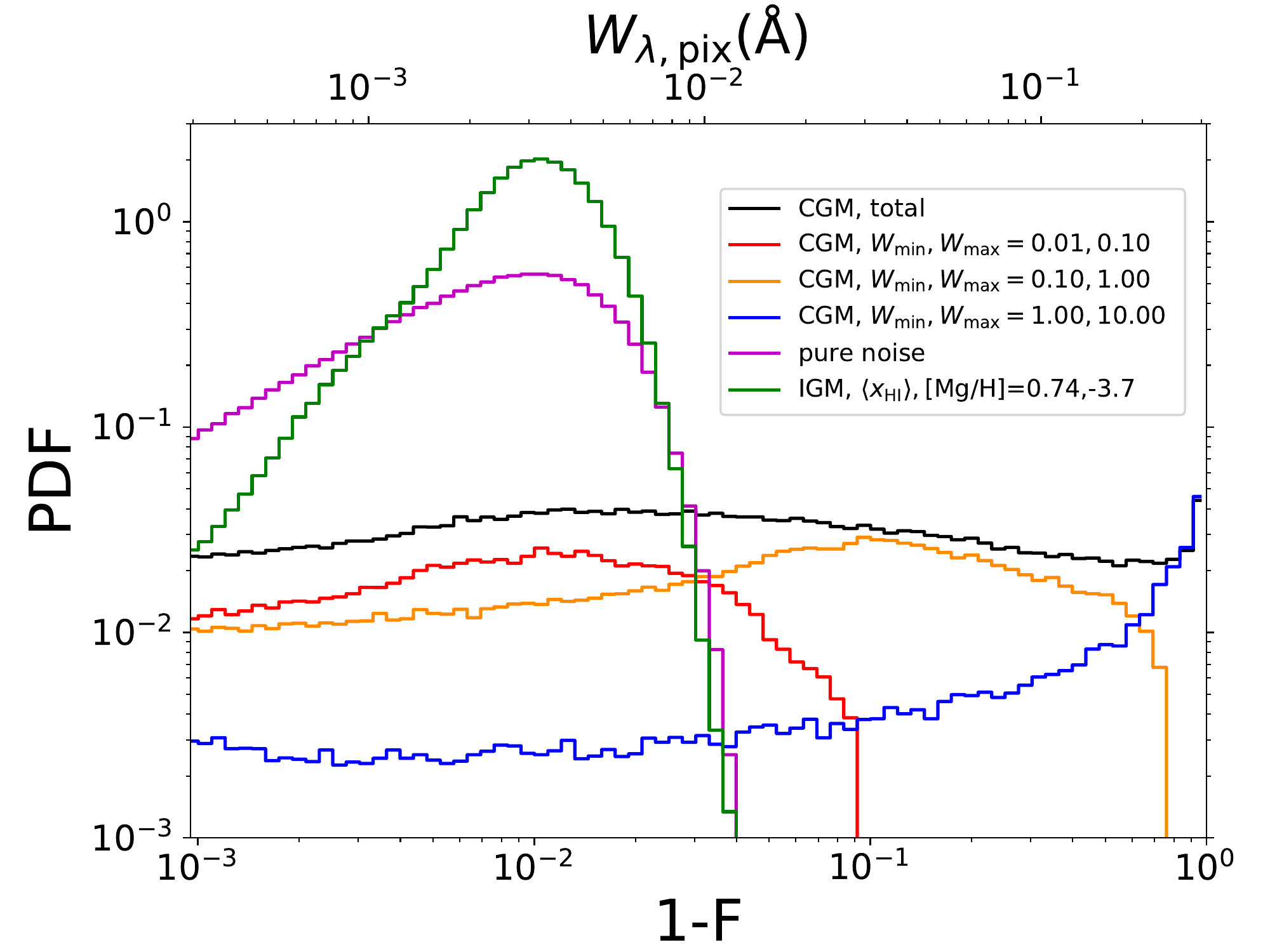}
  \vskip -0.2cm
  \caption{Flux PDF resulting from CGM absorbers compared to that of
    the IGM. The black histogram shows the PDF resulting from the
    entire population of CGM absorbers that we add to our spectra,
    whereas the red, yellow, and blue histograms show the
    contributions from each decade of $W_\lambda$ to the total CGM
    PDF. For comparison, the green histogram shows the PDF resulting
    from pure IGM \ion{Mg}{ii} forest absorption for our fiducial
    model $(\langle x_{\ion{H}{i}}\rangle, [{\rm Mg\slash H}]) = (0.74,
    -3.7)$.  Noise has not been added to the spectra used to construct
    these PDFs, but for comparison, the magenta histogram shows the
    PDF for pure Gaussian noise (${\rm S\slash N} = 100$). The
    counterintuitive appearance of these PDFs arises from the
    logarithmic scale and because we show only the positive
    fluctuations. \label{fig:PDF_CGM}}
    \vskip -0.2cm
\end{figure}

\begin{figure*}
  \includegraphics[trim=7 0 5 0,clip,width=0.49\textwidth]{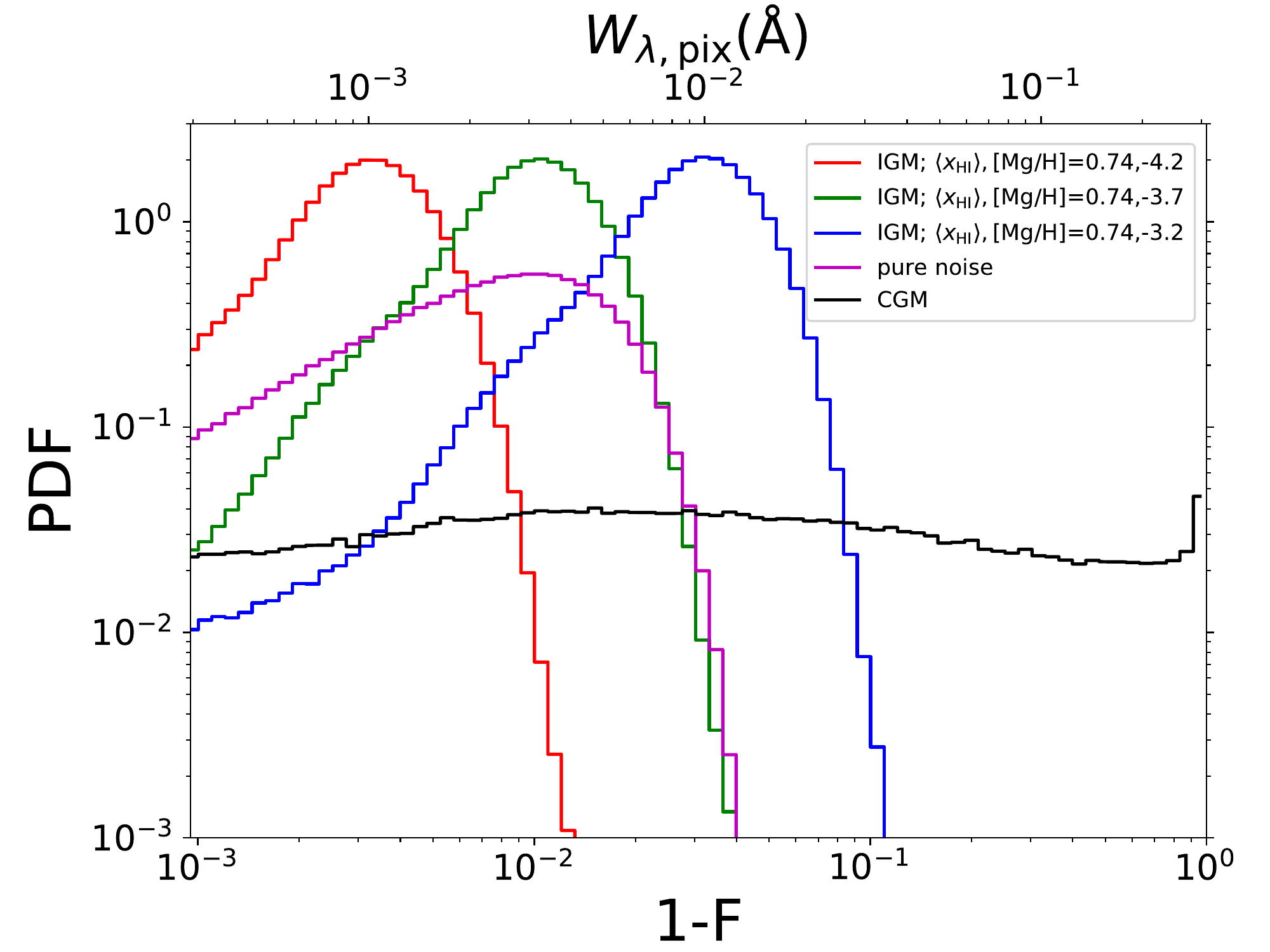}
  \includegraphics[trim=7 0 5 0,clip,width=0.49\textwidth]{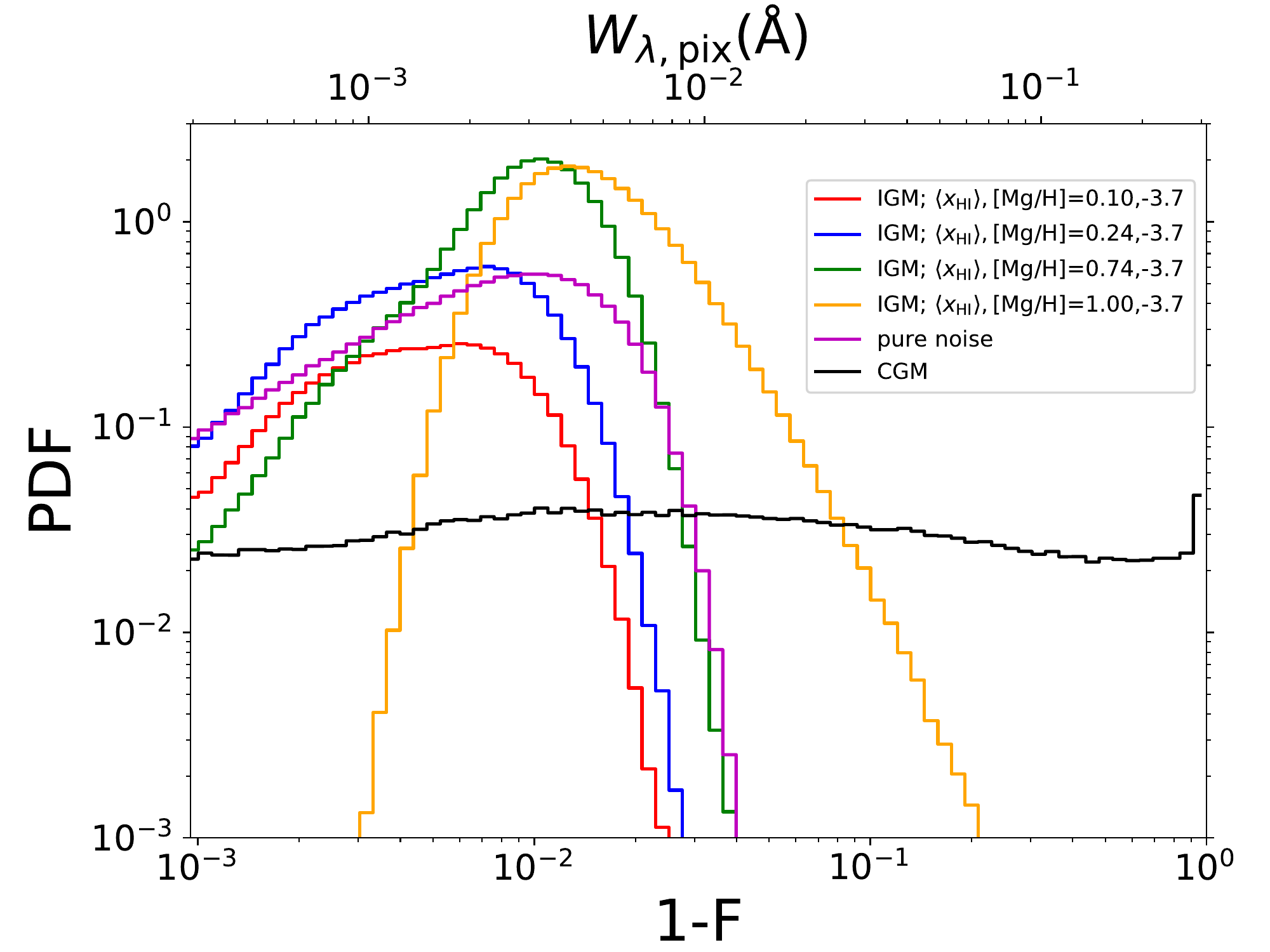}
  \vskip -0.2cm
  \caption{Dependence of IGM flux PDF on model parameters compared to that of the CGM.
    \emph{Left:} Red, green, and blue histograms show
    the effect of changing $[{\rm Mg\slash H}]$ at fixed $\langle
    x_{\ion{H}{i}}\rangle = 0.74$.  For comparison the black histogram
    shows the PDF resulting from CGM absorbers.  Noise has not been
    added to the spectra used to construct these PDFs, but 
    the magenta histogram shows the PDF for pure Gaussian
    noise (${\rm S\slash N} = 100$). \emph{Right:} Red, blue, green,
    and orange histograms show the effect of changing $\langle x_{\ion{H}{i}}\rangle$ at fixed $[{\rm Mg\slash H}]=-3.7$. Black and
    magenta histograms are the same as in the left
    panel. \label{fig:PDF_IGM}}
  \vskip -0.2cm
\end{figure*}

The equivalent width distribution and the flux probability distribution are however related to each other. Recall the definition
of equivalent width
\begin{equation}
  W_\lambda = \int (1-F) d\lambda. \label{eqn:EW}
\end{equation}
Defining the limits of integration to simply be the boundaries of a pixel in our spectrum gives rise to the concept of a pixel
equivalent width $W_{\lambda, {\rm pix}} \equiv (1-F)\Delta \lambda$, where $\Delta \lambda$ is the rest-frame width of a spectral
pixel, which is $0.31$\AA~for the JWST spectra that we model here. Note that unlike the conventional definition of equivalent width, the pixel equivalent width does depend on the spectral resolution if the `pixel' size is chosen to be
smaller than the resolution element of the spectrograph.

Consequently, we want to study the probability distribution of $1-F$, which is linearly related to the distribution of
$W_{\lambda, {\rm pix}}$. Given the large dynamic range of approximately three decades in $W_\lambda$ (see Fig.~\ref{fig:dNdzdW}) that we
model and the high ${\rm S\slash N} = 100$ of the mock spectra, it is preferable to work with $\log_{10}(1-F)$. We thus
define our PDF as
\begin{equation}
  \int_{a}^{b} \frac{dP}{d\log_{10}{\left(1-F\right)}} d\log_{10}{\left(1-F\right)} = {\rm frac}(a,b)
\end{equation}
where ${\rm frac}(a,b)$ is the fraction of the total pixels lying in the interval $\{a,b\}$. 

In Fig.~\ref{fig:PDF_CGM} we compare the flux PDF resulting from CGM
absorbers with that arising from the IGM and noise. These PDFs are
computed from our simulated IGM (see \S~\ref{sec:simulate}) and CGM
skewers (see \S~\ref{sec:CGM}), respectively. Specifically, the green
histogram shows pure IGM \ion{Mg}{ii} forest absorption for our
fiducial model $(\langle x_{\ion{H}{i}}\rangle, [{\rm Mg\slash H}]) =
(0.74, -3.7)$ with no CGM contamination, whereas the black histogram
shows the same skewer pathlength populated only by CGM
absorbers. Noise has not been added to the spectra used to construct
these PDFs, but for comparison, the magenta histogram in
Fig.~\ref{fig:PDF_CGM} shows the PDF for pure Gaussian noise (${\rm
  S\slash N} = 100$). The counterintuitive appearance of these PDFs
arises from the logarithmic scale and because we show only the
positive fluctuations. The other colored histograms illustrate the
contribution of CGM absorbers within a given decade of $W_\lambda$ to
the total CGM PDF (shown in black). The cutoffs at large values of
$1-F$ ($W_\lambda$) in these decade-specific PDFs can be easily
understood.  For example, the strongest absorbers in the range
$(W_{\rm min},W_{\rm max}) = (0.01\,{\text \AA},0.1\,{\text \AA})$
will be at the upper edge of the bin $W_\lambda \simeq 0.1$~\AA. For
such an absorber the highest value of $1-F \simeq 0.1$
($W_{\lambda,{\rm pix}} \simeq (1-F)\Delta \lambda = 0.1\times
0.31\,{\text\AA} = 0.03\,{\text \AA}$).  In other words, the most
absorbed pixel contributes about one third of the total $W_{\lambda}$,
which is the integral over the full absorption profile in
eqn.~(\ref{eqn:EW}).  The flat PDF shape at smaller values of $1-F$
results from both the range of $W_\lambda$ considered in each decadal
bin, as well as very weak absorption imprinted on a large number of
pixels by the Gaussian wings of the spectrograph line spread
function. Finally, the contribution from each decade of $W_\lambda$ to the
final CGM PDF (black), that is the relative normalization of each
histogram, results from the shape of the equivalent width
distribution, $\frac{d^2N}{dz dW_\lambda}$ (see
Fig.~\ref{fig:dNdzdW}).

The shape and amplitude of the IGM PDF, and thus relative importance of noise and CGM contamination,
depend on the structure of the \ion{Mg}{ii} forest as parameterized by $\langle x_{\ion{H}{i}}\rangle$ and $[{\rm Mg\slash H}]$.
The left panel of Fig.~\ref{fig:PDF_IGM} shows that increasing the Mg abundance at a fixed volume averaged neutral fraction ($\langle x_{\ion{H}{i}}\rangle = 0.74$)
simply shifts the PDF to the right. This is intuitive -- because the \ion{Mg}{ii} forest optical depth depends linearly on metallicity (see eqn.~(\ref{eqn:tau0})), and in the low optical depth limit $1-F\approx \tau \propto Z$, and thus a change in metallicity amounts to a simple rescaling of $1-F$, as is apparent in the left panel of Fig.~\ref{fig:PDF_IGM}. Changing $\langle x_{\ion{H}{i}}\rangle$ at fixed  $[{\rm Mg\slash H}]$ produces more complex
changes in PDF shape, as illustrated in the right panel of Fig.~\ref{fig:PDF_IGM}. As compared to a model with
$(\langle x_{\ion{H}{i}}\rangle,[{\rm Mg\slash H}])=(1.0,-3.7)$ (orange), a lower $\langle x_{\ion{H}{i}}\rangle = 0.1$ (red) reduces the abundance of percent level $1-F$ fluctuations by about an order of magnitude,
which is the naive expectation given the order of magnitude change in volume filling factor. But low values of $\langle x_{\ion{H}{i}}\rangle \lesssim 0.3$ also flatten out the peak in the PDF and shift it to lower $1-F$ values, as is apparent for the $\langle x_{\ion{H}{i}}\rangle = 0.1$ (red) and  $\langle x_{\ion{H}{i}}\rangle = 0.24$ (blue) histograms in Fig.~\ref{fig:PDF_IGM}.

In summary, for our fiducial IGM model $(\langle x_{\ion{H}{i}}\rangle,
[{\rm Mg\slash H}]) = (0.74, -3.7)$ (green histograms in
Figs.~\ref{fig:PDF_CGM} and \ref{fig:PDF_IGM}), the \ion{Mg}{ii}
forest produces a distribution of $1-F$ fluctuations peaking around a
percent, which are a factor of about four more abundant than noise
fluctuations at our assumed ${\rm S\slash N} = 100$. CGM absorbers
produce a flat distribution of $1-F$ fluctuations, which are almost
two orders of magnitude less abundant than IGM fluctuations for
$1-F\simeq 0.01$, but which overwhelmingly dominate at $1-F\gtrsim
0.03$, where both IGM fluctuations and noise fluctuations are
exponentially suppressed. As the parameters governing the IGM are
varied (see Fig.~\ref{fig:PDF_IGM}), the $1-F$ value at which the IGM
PDF peaks shifts, as does the location of the exponential
cutoff at high $1-F$. However, qualitatively the picture is unchanged.
At the $1-F$ values where the IGM PDF peaks, it exceeds the flat CGM
PDF by at least an order of magnitude for the majority of IGM
parameter space.  This indicates that CGM absorbers can simply be identified and masked
without significantly modifying the distribution of IGM fluctuations, and hence
preserving the information about enrichment and reionization encoded in
the \ion{Mg}{ii} forest. 

\section{Identifying and Masking CGM Absorption}
\label{sec:mask}

\begin{figure*}
  \vskip -0.4cm
  \includegraphics[trim=10 13 0 15,clip,width=\textwidth]{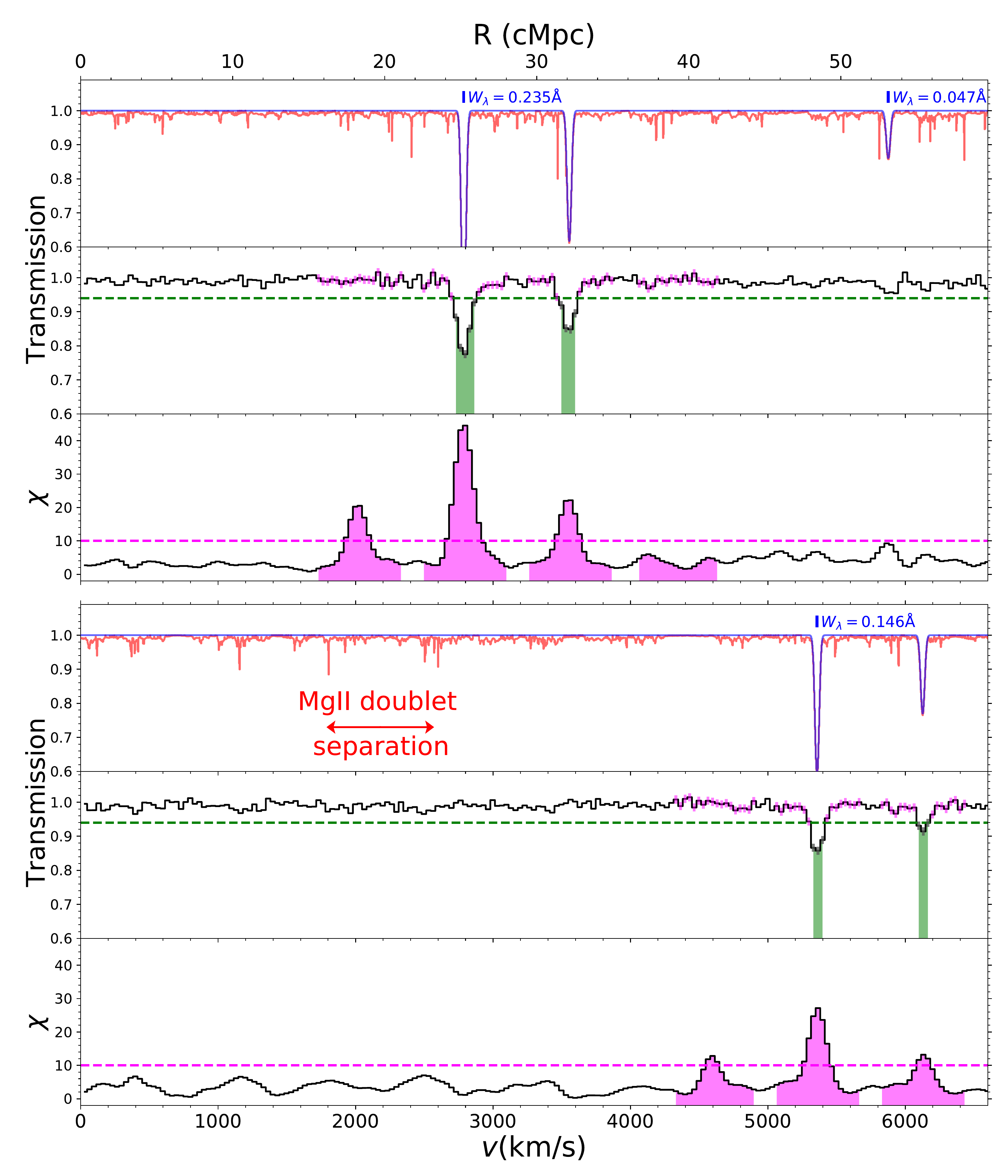}
  \vskip -0.3cm
  \caption{Illustration of our procedure for masking CGM
    absorbers. The two sets of three panel figures (top and bottom)
    show simulated \ion{Mg}{ii} forest spectra of the IGM contaminated
    by CGM absorbers for two absorption skewers. \emph{Top Panels:}
    The perfect input spectra (IGM in red, CGM in blue), with the
    individual CGM absorbers labeled by their rest-frame equivalent
    width $W_\lambda$. \emph{Middle Panels:} Forward modeled JWST
    spectra with resolution ${\rm FWHM}=100~{\rm km~s^{-1}}$ and ${\rm
      S\slash N}=100$.  Vertical rectangles indicate pixels that have
    been masked by our `flux-filtering' (green), `$\chi$-filtering'
    (magenta), or are masked by both `flux + $\chi$-filtering'
    (gray). The horizontal green dashed line at $F=0.94$ ($1-F=0.06$)
    indicates the `flux-filtering' threshold (see left panel of
    Fig.~\ref{fig:PDF_filter}, and the green shaded region indicates
    spectral regions below the threshold (above the $1-F$ threshold)
    that are masked.  \emph{Lower Panels:} Spectrum of $\chi(v)$ (see
    eqn.~\ref{eqn:chi}) computed from the JWST mock spectra in the
    middle panels. Note the triple-peak structure, which naturally
    arises from convolving an absorption doublet with a doublet
    matched-filter due to aliasing when the data/filter overlap with
    the opposite member of the doublet. The magenta dashed line at
    $\chi=10$ indicates the `$\chi$-filtering' threshold (see right panel of Fig.~\ref{fig:PDF_filter}.
    Pixels
    above this threshold are all masked, and we also mask a $\pm
    300~{\rm km~s^{-1}}$ window around any peak identified that has
    $\chi > 10$, as well as at locations $+v_{\ion{Mg}{ii}}$ away to
    account for the redder member of the doublet. We conservatively
    do not attempt to distinguish between real and aliased peaks. The
    magenta shaded regions indicate the regions that are masked by
    this `$\chi$-filtering'.
    \label{fig:mask}}
\end{figure*}

\begin{figure*}
  \includegraphics[trim=7 0 5 0,clip,width=0.49\textwidth]{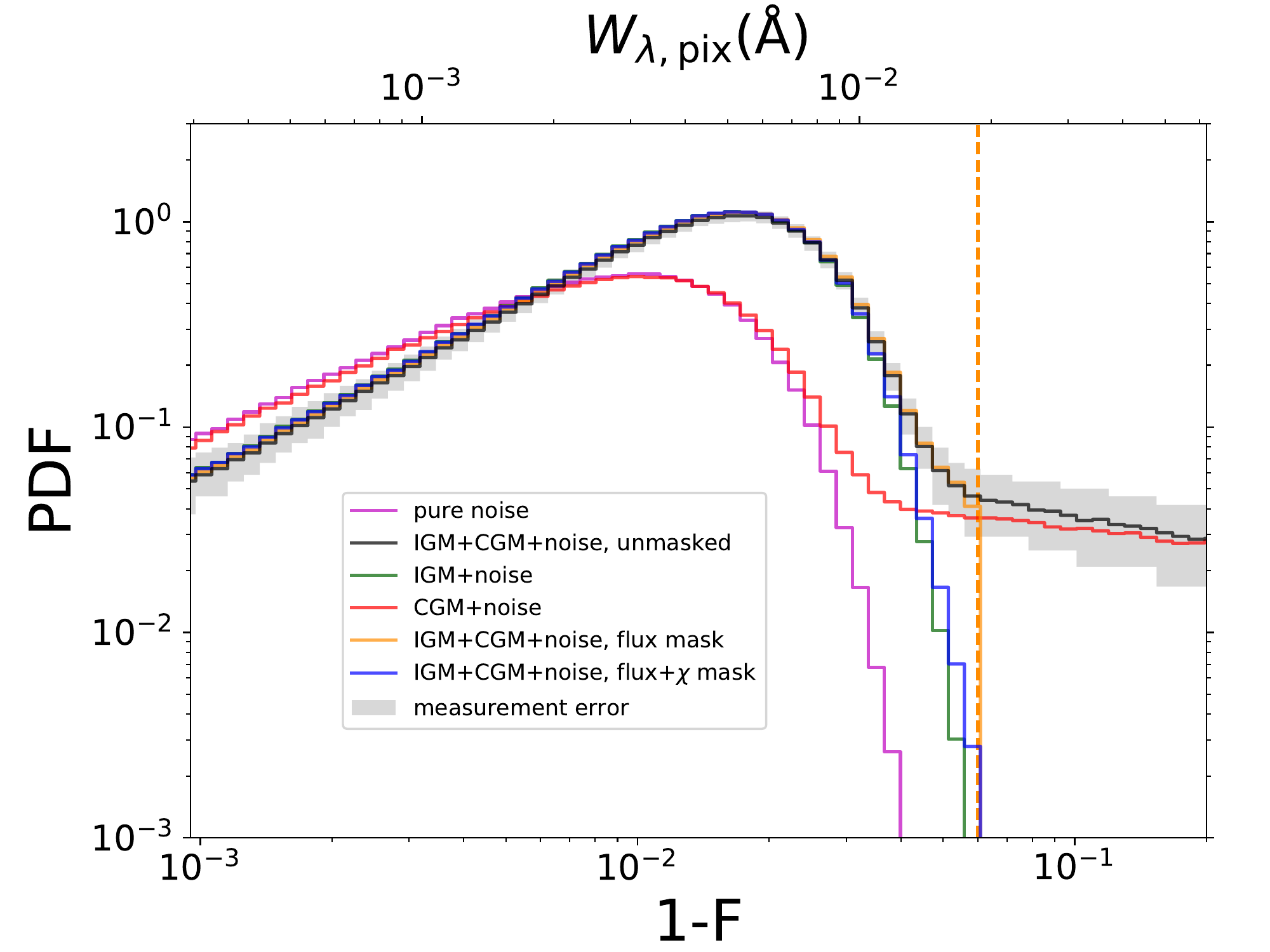}
  \includegraphics[trim=7 0 5 0,clip,width=0.49\textwidth]{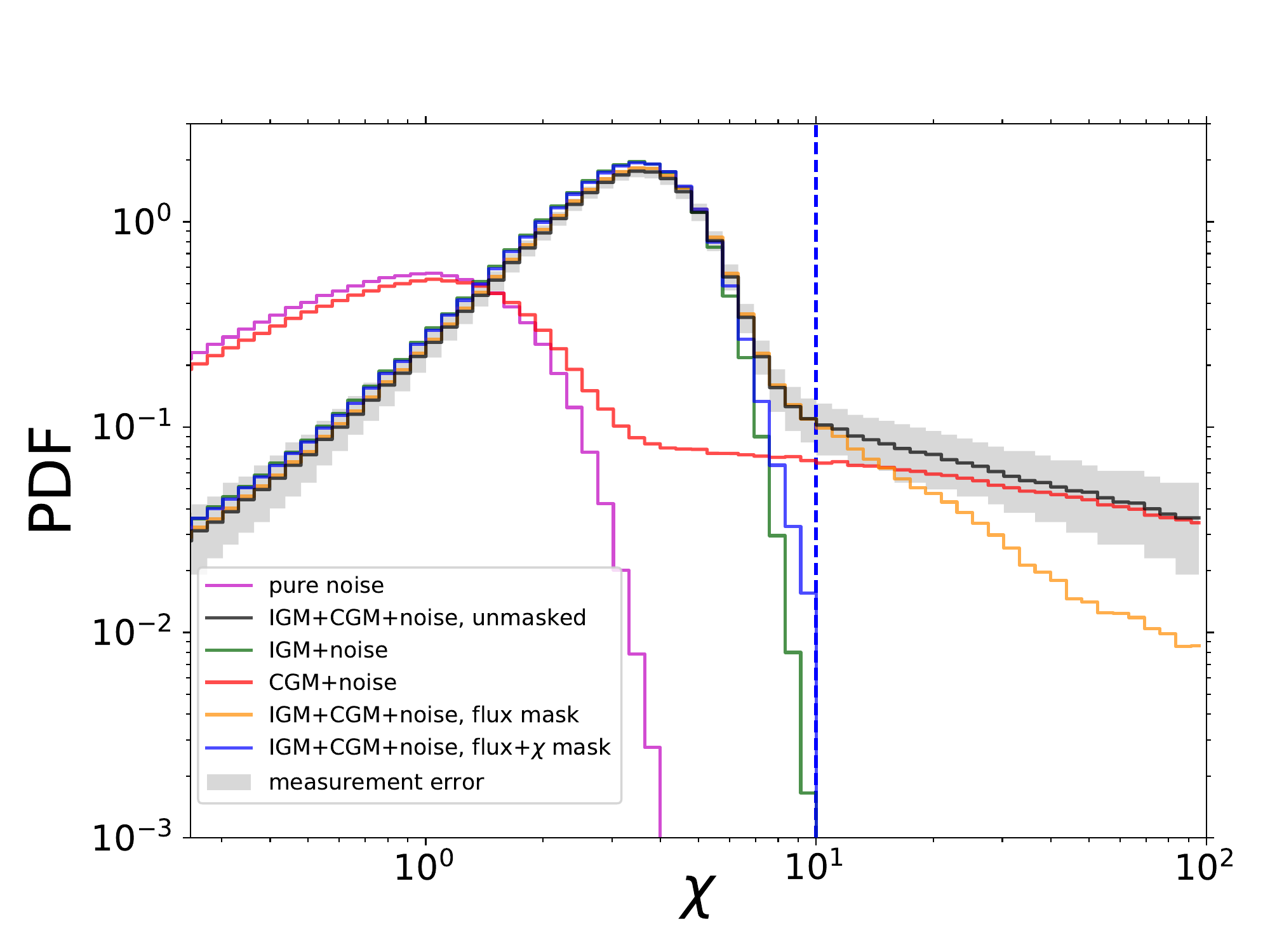}
  \vskip -0.2cm
  \caption{Impact of filtering on the flux and
      significance PDFs. \emph{Left:} The black histogram shows the
    PDF of $1-F$ for spectra which include the IGM \ion{Mg}{ii} forest
    and CGM contamination with noise now added.  The colored
    histograms show the PDFs for various combinations of
    subcomponents: IGM + noise (green), CGM + noise (red), and pure
    noise (magenta). The impact of `flux-filtering' on the PDF,
    whereby all pixels with $1-F > 0.06$ (vertical orange dashed line)
    are masked is illustrated by the orange histogram.  The blue histogram 
    shows the combination of this `flux-filtering' and
    `$\chi$-filtering', whereby pixels with $\chi > 10$ are also
    masked (see right). \emph{Right:} Same as left panel but now
    showing the PDF of the significance field $\chi$ (see
    eqn.~(\ref{eqn:chi}) and Fig.~\ref{fig:mask}). The orange
    histogram now shows the impact of `flux-filtering' on the $\chi$
    PDF, whereas blue shows the combination of `flux + $\chi$-filtering'
    with $\chi > 10$ (vertical blue dashed
    line).\label{fig:PDF_filter}\label{fig:PDF_filter}}
  \vskip -0.2cm
\end{figure*}

We now present a procedure for filtering out the CGM contamination by
identifying and masking pixels impacted by CGM absorption. We first
focus attention on our fiducial IGM model $(\langle x_{\ion{H}{i}}\rangle,[{\rm Mg\slash H}]) = (0.74, -3.7)$, and
later describe how our results generalize to other \ion{Mg}{ii} forest models. To identify
the location of CGM absorbers, we follow standard practice \citep[e.g.][]{ZhuMenard13,Chen17} and convolve our noisy mock spectra with a
matched filter, $W(v)$ corresponding to the transmission profile of a \ion{Mg}{ii} doublet. Extrema in this filtered field
are identified as potential absorber locations. We define the significance field
\begin{equation}
  \chi(v) \equiv \frac{\int [1-F(v^\prime)]W(|v-v^\prime|) dv^\prime}{\sqrt{\int \sigma^2_{F}(v^\prime) W^2(|v-v^\prime|) dv^\prime}}\label{eqn:chi}, 
\end{equation}
where $\sigma^2_{F}$ is the variance of $1-F$ resulting from the
spectrograph noise.  As defined, $\chi$ is essentially a ${\rm S\slash
  N}$ ratio, with the signal being the matched filtered field, and the
noise the one sigma fluctuation of the filtered field that would arise
from noise fluctuations alone.  For $W(v)$ we use $1-e^{-\tau(v)}$,
where $\tau(v)$ is the Voigt
profile describing a \ion{Mg}{ii} doublet with a Gaussian velocity
distribution, assuming a column density of $N_{\ion{Mg}{ii}} =
10^{13.5}~{\rm cm^{-2}}$ and Doppler parameter $b = \sqrt{2}\times
{\rm FWHM}\slash 2.35 = 60.2~{\rm km~s^{-1}}$, where ${\rm
  FWHM}=100~{\rm km~s^{-1}}$ is the resolution of our mock JWST
spectra. This choice for $b$ is sensible because the Doppler parameters
of weak absorbers $b=20~{\rm km~s^{-1}}$ (see \S~\ref{sec:CGM}) are
not resolved by our spectral resolution. Since this combination of $b$
and $N_{\ion{Mg}{ii}}$ puts us on the linear part of the COG (see eqn.~\ref{eqn:Wr}) $W(v) =
1-e^{-\tau(v)}\propto N_{\ion{Mg}{ii}}$ and thus this normalization simply cancels
out of our definition of $\chi$ in eqn.~(\ref{eqn:chi}).

Fig.~\ref{fig:mask} shows simulated \ion{Mg}{ii} forest spectra of the
IGM contaminated by CGM absorbers for two absorption skewers (i.e. top three and bottom three panels).
The upper panel of each plot shows the perfect input spectra (IGM in red, CGM in blue), whereas the middle
panels show forward modeled JWST spectra with finite resolution and ${\rm S\slash N}$. 
The lower panels show $\chi(v)$ computed from these JWST spectra. Notice that
because $W(v)$ has a double Gaussian shape with the two peaks
separated by the \ion{Mg}{ii} doublet separation $v_{\ion{Mg}{ii}}$, the
$\chi$ spectrum of an absorber exhibits a triple-peak structure with
one peak at the true location of the absorber and two `aliased' peaks
at $\pm v_{\ion{Mg}{ii}}$. This aliasing is unavoidable and results when
the $\lambda~2796$ or $\lambda~2804$ part of the filter overlaps with
the opposite member of the doublet in the data.

We implement two distinct filtering procedures by simply masking
spectral regions that are likely to be contaminated by the CGM. The
left panel of Fig.~\ref{fig:PDF_filter} shows the PDF of $1-F$,
analogous to those shown Figs.~\ref{fig:PDF_CGM} and \ref{fig:PDF_IGM}, but where we have now
added noise to the skewers, as illustrated in the middle panels of
Fig.~\ref{fig:mask}. The black histogram in  Fig.~\ref{fig:PDF_filter} shows the full PDF of the \ion{Mg}{ii}
forest plus CGM contamination, whereas the colored histograms show PDFs of
combinations of subcomponents. In particular, the green histogram shows that, for our fiducial
IGM model plus noise, fluctuations with $1-F > 0.06$ (orange dashed
vertical line) lie beyond the exponential cutoff of the IGM PDF. All
of these large fluctuations are caused by CGM absorption (red
histogram). This motivates our `flux-filtering' masking procedure,
whereby pixels with $1-F > 0.06$ are simply masked, corresponding to
the shaded green regions in Fig.~\ref{fig:mask}.

An effective filtering procedure should result in a flux PDF as
close to the pure IGM (green histogram) in the left panel of
Fig.~\ref{fig:PDF_filter} as possible. There it is seen that `flux-filtering' simply
truncates the PDF (orange histogram) above the $1-F=0.06$ threshold
(orange vertical dashed line), but still leaves an appreciable number
of CGM contaminating pixels just below it.  The middle panels of
Fig.~\ref{fig:mask} show that these pixels can be identified with the
wings of CGM absorbers lying just above the $F=0.94$ (below the $1-F=0.06$) threshold (green
horizontal dashed line). Setting the threshold to a lower value of $1-F$ would
mask them, but at the expense of suppressing real IGM signal at $1-F$
values where the IGM and CGM PDFs overlap. Instead, we need to
identify the CGM absorbers and `grow' our mask, which motivates a
second masking procedure, which we refer to as `flux $+ \chi$-filtering'. The
right panel of Fig.~\ref{fig:PDF_filter} shows the PDF of the $\chi$
field (see e.g. middle panels of Fig.~\ref{fig:mask}) for the same
combinations of subcomponents as the left panel.  Analogous to $1-F$, one
observes that IGM fluctuations (green histogram) with $\chi \gtrsim
10$ are exponentially suppressed, and that all of these high-$\chi$
pixels are due to CGM contamination (red histogram). We thus search
for extrema in the $\chi$ field with $\chi > 10$, and mask a $\pm
300~{\rm km~s^{-1}}$ region around each of these peaks at the peak
location, as well as at locations $+v_{\ion{Mg}{ii}}$ away. Our peak
finding is thus conservative: we mask a $600~{\rm km~s^{-1}}$ wide window to ensure
we completely mask all the CGM absorption and we do not attempt to distinguish
between real and aliased peaks.

What we will refer to as `$\chi$-filtering' is the \texttt{OR} of three
distinct boolean bad-pixel masks, i.e. where \texttt{True} corresponds
to a pixel that will be masked in the correlation function
computation. That is
\begin{equation}
\texttt{CHI\_MASK}= \chi > 10~~\texttt{OR}~~\texttt{PEAK}_{2796}~~\texttt{OR}~~\texttt{PEAK}_{2804}, 
\end{equation}
where ${\rm PEAK}_{2796}$ and ${\rm PEAK}_{2804}$ are the masked regions associated with
each peak, and the location $+v_{\ion{Mg}{ii}}$ away, respectively. This mask is illustrated by the magenta shaded regions in the lower panels of Fig.~\ref{fig:mask}.
The final bad-pixel mask for `flux $+ \chi$-filtering' is then
\begin{equation}
  \texttt{FLUX\_CHI\_MASK}= 1-F > 0.06~~\texttt{OR}~~\texttt{CHI\_MASK}, 
\end{equation}
which is depicted by the vertical bars on the spectra in  the middle panels of Fig.~\ref{fig:mask}.
After applying our total flux $+ \chi$-filtering
masks to the entire ensemble of 10,000 skewers for our fiducial model,
we are left with $77\%$ of the pixels being unmasked, indicating
that while our conservative masking does reduce the total
pathlength, the reduction is not severe. 
The PDFs of $1-F$ and $\chi$ for the `flux $+ \chi$-filtered' spectra
are shown as the blue histograms in the left and right panels of
Fig.~\ref{fig:PDF_filter}, respectively. That these PDFs very closely
match those for pure IGM plus noise (green histograms) strongly
suggests that we have achieved of our goal of masking the majority of
the CGM contaminated pixels.

\begin{figure}
  \includegraphics[trim=30 0 2 0,clip,width=0.49\textwidth]{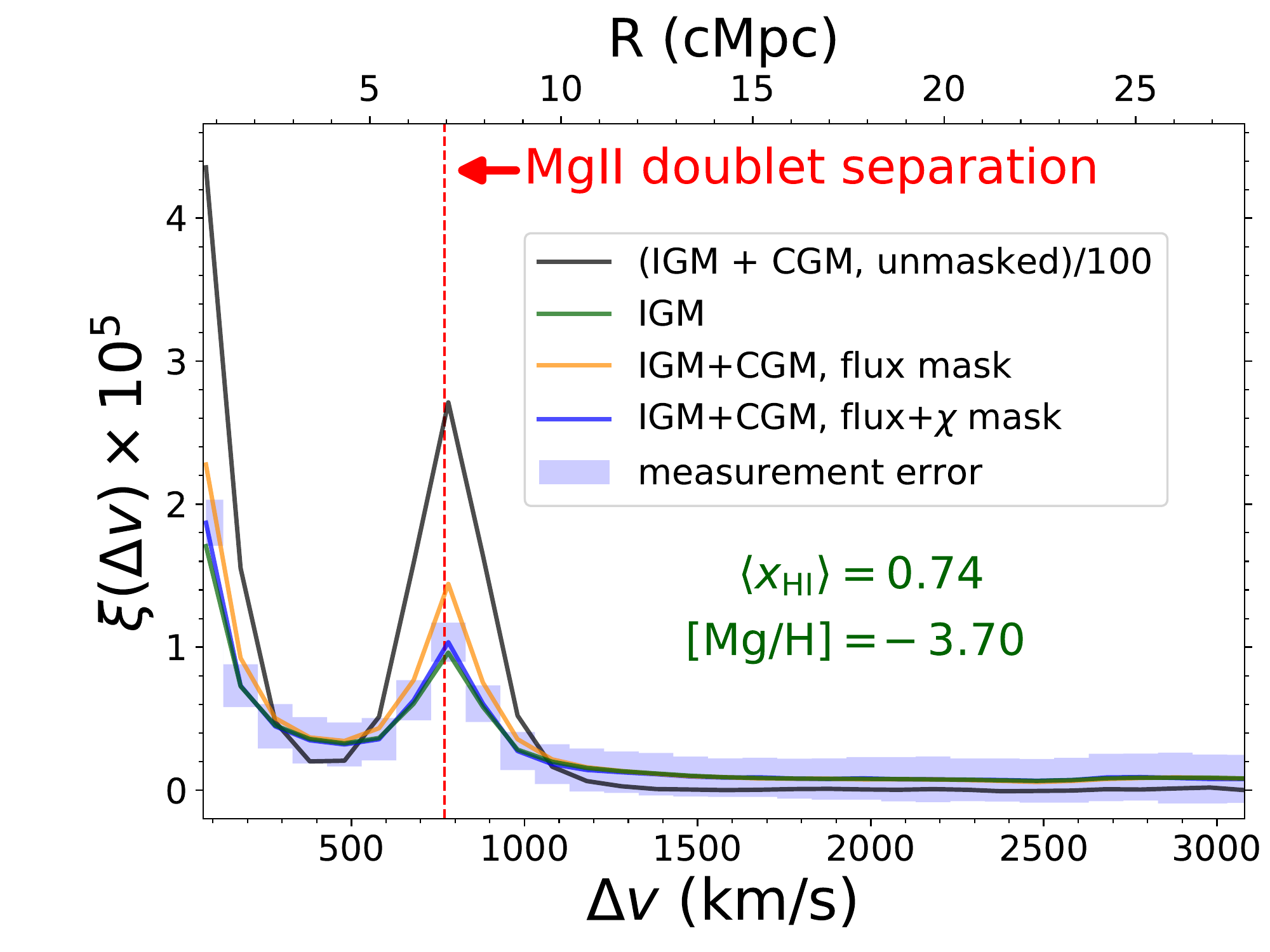}
  \vskip -0.2cm
  \caption{Impact of CGM absorbers on \ion{Mg}{ii} Forest correlation
    function before and after filtering. The black curve shows the
    correlation function divided by $100$ for skewers where CGM
    absorber contamination has been added to the IGM \ion{Mg}{ii}
    forest. For comparison, the green curve shows the \ion{Mg}{ii}
    forest correlation function for the uncontaminated IGM alone. The
    \ion{Mg}{ii} forest model shown here has $(\langle
    x_{\ion{H}{i}}\rangle, [{\rm Mg \slash H}]) = (0.74,-3.7)$. The
    orange line shows that `flux-filtering' greatly suppresses CGM
    contamination resulting in a correlation function very close to
    the IGM alone (green). Even better is the combination of `flux +
    $\chi$-filtering' shown as the blue curve, which results in a
    correlation function that is virtually indistinguishable from the
    pure IGM signal (green) relative to the expected $1\sigma$ errors
    (blue shaded regions) on the correlation function for our fiducial
    JWST dataset.\label{fig:filter_corrfunc}}
    \vskip -0.2cm
\end{figure}

The ultimate validation of this hypothesis comes from the clustering
properties, which formed the basis for our parameter constraints on
reionization and IGM enrichment (see Figs.~\ref{fig:inference} and
\ref{fig:contours}). The impact of CGM contamination and masking on
the correlation function is shown in
Fig.~\ref{fig:filter_corrfunc}. The black curve shows the correlation
function $\xi(\Delta v)$ of the CGM contaminated \ion{Mg}{ii} forest
without masking, which overwhelms the pure IGM signal by over two
orders of magnitude (note the black curve is scaled down by a factor
of 100). Nevertheless, our `flux
$+\chi$-filtering' procedure successfully masks nearly all of the CGM
absorption, and Fig.~\ref{fig:filter_corrfunc} shows that the
resulting clustering signal (blue curve) is virtually
indistinguishable from that of the pure IGM (green curve), especially
in relation to the expected measurement errors for our fiducial JWST
dataset (blue shaded bands).

The outsize influence of CGM absorbers on the \ion{Mg}{ii} forest may
seem counterintuitive given one's experience with the \ion{H}{i}
Ly$\alpha$ forest, where CGM absorbers, i.e. Lyman limit systems and
damped Ly$\alpha$ absorbers, have a small impact on the correlation
function and power spectrum \citep{McDonald05a,Rogers18}. There are
two explanations for this difference.  First, metals are far more
abundant in the CGM than they are in the IGM, enhancing the impact of
CGM contamination on their clustering signal compared to hydrogen. For
example, our fiducial model assumes the IGM is enriched to $[{\rm
    Mg\slash H}]=-3.7$, whereas we know that CGM absorbers at $z\sim
2-4$ have metallicities orders of magnitude higher spanning the range
$-2 \lesssim \log_{10} Z\lesssim -1$ \citep{Fumagalli16}. This large
relative enhancement of metals in the CGM vs IGM also likely holds at
$z\sim 7$.  Second, IGM absorbers in the $z\sim 2-4$ \ion{H}{i}
Ly$\alpha$ forest ($N_{\ion{H}{i}} \sim 10^{12}-10^{14}~{\rm cm^{-2}}$)
are on the linear part of the COG, whereas the most abundant CGM
absorbers have $N_{\ion{H}{i}} \sim 10^{15}-10^{18}~{\rm cm^{-2}}$. Thus
they lie on the
saturated part of the COG limiting their spectral imprint. This is however
not the case for \ion{Mg}{ii} where a large population of CGM contaminants
with $W_{\lambda} \lesssim 0.4\AA$ lie on the linear part of the COG
(see eqn.~\ref{eqn:Wr}).  These absorbers completely dominate the PDF
of $1-F$ for fluctuations greater than a few percent (see left panel
of Fig.~\ref{fig:PDF_IGM}) and, if left unmasked, swamp the IGM
clustering signal.  To build further intuition about why the CGM
absorber matter so much it helps to compare their mean flux decrement
to that resulting from IGM metals. For the CGM model implemented
here, we find that for simulated skewers populated only with CGM
absorbers $1-\langle F\rangle = 0.0115$, which is slightly larger than
the value for pure IGM skewers of $1-\langle F\rangle = 0.00969$ with
$(\langle x_{\ion{H}{i}}\rangle,[{\rm Mg\slash H}])=(1.0,-3.7)$.  This can
be contrasted with the \ion{H}{i} Ly$\alpha$ forest at $z \simeq 3$
where optically thick absorbers with $N_{\ion{H}{i}} > 10^{17.2}~{\rm
  cm^{-2}}$ contribute a negligible amount of transmission $1-\langle
F\rangle \simeq 0.01$ when compared to the mean transmission of $1 -
\langle F\rangle =0.67$ from the lower column density Ly$\alpha$ forest
\citep{Becker2013}. But it still not immediately obvious why CGM
absorbers yield a correlation function more than two orders of
magnitude larger than the IGM \ion{Mg}{ii} forest (see
Fig.~\ref{fig:filter_corrfunc}) given that they contribute comparably
to the IGM in the flux decrement. This outsize effect results from the
fact that the strongest CGM absorbers, although rare, contribute $1 -
F \gtrsim 0.5$ nearly two orders of magnitude larger than the typical
decrement $1 - F \simeq 0.01$ of IGM absorbers (see
e.g. Fig.~\ref{fig:PDF_IGM}), and the correlation function is
sensitive to the square of this enhanced absorption
(see eqn.~\ref{eqn:xi}).

While we have shown that we can effectively suppress the impact of CGM
absorbers on the correlation function for our fiducial model, our
masking procedure involves choosing two thresholds: one for $1-F$
which we took to be $0.06$ and another for $\chi$ chosen to be $10$.
These thresholds were chosen to be just beyond the exponential cutoffs
in the pure IGM + noise PDFs in Fig.~\ref{fig:PDF_filter}, which
prevents overly aggressive masking that could suppress some of the real
IGM \ion{Mg}{ii} forest signal and yield biased parameter estimates.
At face value, given that the choice of these thresholds depends on the
expected fluctuations for a given model, our masking procedure appears
to be model dependent. However, we argue that this actually is not the
case, since the location of these thresholds can be determined by
simply inspecting the PDFs in Fig.~\ref{fig:PDF_filter} for the real
data. The gray shaded regions in the left and right panels show the
expected $1\sigma$ measurement error on the $1-F$ and $\chi$ PDFs
respectively, for a mock JWST dataset. Guided by the generic
expectation, elucidated in Figs.~\ref{fig:PDF_CGM}, \ref{fig:PDF_IGM}
and \ref{fig:PDF_filter}, that the IGM PDF will peak at a
characteristic value and sharply cutoff towards higher $1-F$ (or
$\chi$), whereas the CGM PDF will be flat at these large $1-F$
($\chi$), the masking thresholds can be chosen by simply inspecting
the $1-F$ and $\chi$ PDFs estimated from the data, and the relative
size of the shaded error bars indicate that one would have ample
signal-to-noise on these PDFs to do so.

\section{Summary and Conclusions}
\label{sec:summary}

We proposed a novel experiment to detect the weak forest of
low-ionization \ion{Mg}{ii} absorbers in quasar spectra that will be
present if the IGM is both significantly neutral and sufficiently
enriched with metals.  In contrast to the traditional approach of
searching for discrete absorption systems, we advocated treating this
forest of metal absorption as a continuous cosmological random field
and measuring its two-point correlation function and PDF, leveraging
techniques from precision cosmology.  To quantify the efficacy of approach 
method, we simulated the \ion{Mg}{ii} forest for the first time by
combining a large cosmological hydrodynamical simulation of the
pre-reionization IGM with a semi-numerical computation of the global
reionization topology, assuming a simple enrichment model where the
IGM is uniformly suffused with metals.  We studied the behavior of the
\ion{Mg}{ii} forest correlation function, $\xi(\Delta v)$, and find
that it exhibits the following properties: 1) a steep rise towards
small velocity lags (small-scales) resulting from the clumpy
small-scale structure of the pre-reionization IGM, 2) a conspicuous
peak at a $\Delta v_{\ion{Mg}{ii}} = 768\,{\rm km\,s^{-1}}$ arising from
the doublet nature of the \ion{Mg}{ii} transition, 3) a power-law
shape at intermediate to large velocity lags induced by
the topology of neutral regions during reionization, 
which is highly
sensitive to their volume averaged filling fraction $\langle x_{\ion{H}{i}}\rangle$, 4) an overall amplitude which scales as the square of
the Mg abundance $[{\rm Mg}\slash {\rm H}]$.

We perform statistical inference for a correlation measurement based
on a realistic mock dataset of 10 JWST spectra and find that one can
simultaneously determine the Mg abundance $[{\rm Mg}\slash {\rm H}]$,
with a $1\sigma$ precision of 0.02~dex and measure the global neutral
fraction $\langle x_{\ion{H}{i}}\rangle$ to 5\%, for a fiducial model
with $\langle x_{\ion{H}{i}}\rangle = 0.74$, and $[{\rm Mg}\slash {\rm
    H}] = -3.7$. Contrary to the naive expectation that this
enrichment level, $[{\rm Mg\slash H}]$, should be degenerate with the
global neutral fraction, $\langle x_{\ion{H}{i}}\rangle$, we find that
they can be uniquely constrained owing to the distinct dependence of
the correlation function shape on each parameter.  Alternatively, if
the IGM is pristine, then a null-detection of the \ion{Mg}{ii} forest
would place a stringent upper limit on the metallicity of the
pre-reionization IGM of $[{\rm Mg}\slash {\rm H}] < -4.4$ at 95\%
credibility, assuming an independent constraints on $\langle
x_{\ion{H}{i}}\rangle > 0.5$ from another reionization probe.

We investigated the degree to which concentrations of metals in the
CGM around galaxies could potentially contaminate a \ion{Mg}{ii}
forest signal arising from the IGM. CGM absorbers with a line density
and equivalent width distribution consistent with current
observational constraints were injected into our mock spectra.  We
analyzed the flux PDF for the models of interest, and find that the
PDF for IGM absorption exhibits a broad peak around $1-F\simeq
0.01$, and a sharp exponential cutoff for $1-F$ fluctuations a factor
of a few larger. In contrast, CGM absorbers give rise to a flat flux
PDF producing one to two orders of magnitude lower probability at the
$1-F$ where the IGM flux PDF peaks, but overwhelmingly dominates the
absorption statistics at the larger $1-F$ where IGM fluctuations are
exponentially suppressed. Exploiting the distinct shapes of the flux
PDF for IGM and CGM absorption, we present a strategy for masking the
CGM contamination, and show that the difference between the
correlation function $\xi(\Delta v)$ recovered from masked
data and the uncontaminated IGM correlation
function is negligible compared to the statistical errors.

Low frequency radio observations of the 21cm line have been touted as
the premier probe of reionization because of their potential to
measure $x_{\ion{H}{i}}(z)$ and characterize the topology of reionization
\citep[see e.g.][]{Pritchard12}. Our analysis illustrates that the 
\ion{Mg}{ii} forest has tremendous potential to constrain 
reionization if the pre-reionization IGM is significantly
enriched, so it is interesting to compare and contrast the two probes
of the EoR. A rapidly developing area of 21cm cosmology are
interferometric measurements with arrays like HERA \citep{HERA} and
LOFAR \citep{LOFAR} which hope to map fluctuations in 21cm emission
during the EoR. These studies aim to constrain reionization by
measuring the 3D 21cm emission power spectrum, whose amplitude
and shape depends on the timing and topology of reionization
\citep{Madau97,FOB06,Barkana09}. Our approach to the \ion{Mg}{ii} forest is
very similar in spirit to these 21cm studies. We can also constrain
the timing and topology of reionization by measuring a clustering
signal, in our case the 1D correlation function of \ion{Mg}{ii}
absorption toward quasar sightlines.  The primary disadvantage of the  \ion{Mg}{ii} forest is that it probes a metal line, not a primordial transition, and
it is unknown whether the pre-reionization IGM is enriched. Of course given the
major challenge facing all 21cm experiments -- that of teasing out a
miniscule signal buried beneath foregrounds and instrumental
systematics that are five orders of magnitude larger \citep{Cheng18}
-- they have yet to deliver competitive constraints on the
Universe's reionization history \citep{Mertens20,Trott20}. In contrast, detecting the
\ion{Mg}{ii} forest in absorption is far simpler from a technical
perspective, since the noise arising from the sky background and
detector read noise is white, and systematics are expected to be
negligible.

Of course, the most direct 21cm analog of the \ion{Mg}{ii} forest 
is the so-called 21cm forest, which is the prospect of observing
neutral hydrogen in 21cm absorption toward a bright background
radio source \citep{Furlanetto06_forest,Carilli02,Ciardi13}.  This technique
has not yet yielded any constraints owing to both the limited
sensitivity of current instrumentation and the lack of sufficiently
bright radio sources residing in the EoR. A significant concern is the
extremely low expected 21cm optical depth,
\begin{equation}
  \tau_{\rm 21cm} =
0.002\left(\frac{x_{\ion{H}{i}}}{1.0}\right)\left(\frac{T_S}{100~{\rm
    K}}\right)^{-1}\left(\frac{1 + z}{8.5}\right)^{3\slash 2}, 
\end{equation}
which can be directly compared
to that for \ion{Mg}{ii} in eqn.~(\ref{eqn:mgiitau0}).
While the
enrichment level of the pre-reionization IGM is unknown and could be
extremely small, the 21cm spin 
temperature $T_{\rm S}$, determined by the temperature of the
pre-reionization IGM, is also highly uncertain.  Indeed, the vast majority of models of early IGM
thermal evolution predict that a metagalactic X-ray background sourced
by early black holes photoelectrically heated the IGM to $T_S \sim
1000\,{\rm K}$ \citep[e.g.][but see \citet{Fialkov14}]{Furlanetto06}. It is thus most likely
that the 21cm forest optical depth is $\tau_{\rm 21cm}\simeq 2\times
10^{-4}$ which will be exceedingly difficult to detect. This
comparison highlights the identical dependencies of the 21cm forest
and \ion{Mg}{ii} forest optical depths. The metallicity $Z$ can be
simply substituted with inverse spin temperature $T_{\rm S}^{-1}$, and
hence a spin temperature of $T_{\rm S} \simeq 1000\,{\rm K}$ gives the
same optical depth in hydrogen as $Z\slash \simeq 10^{-5}Z_{\odot}$
would for \ion{Mg}. This analogy between the 21cm forest and the
\ion{Mg}{ii} forest is essentially perfect\footnote{The only
  differences arise from the different amounts of thermal
  broadening because the atomic weight of \ion{Mg} is 24 times that of
  hydrogen (see eqn.~\ref{eqn:bval})}, and the formalism, modeling,
and inference procedure we have developed here can be directly applied
to the 21cm forest, which will be a subject of a future paper \citep[see also][]{Thyagarajan20}. 

The foregoing discussion further highlights that the Achilles' heel of
the \ion{Mg}{ii} forest technique is the unknown enrichment level of
the pre-reionization IGM.  Theory has yet to provide much guidance on
this question because cosmological hydrodynamical simulations that
track enrichment and properly model reionization via radiative
transfer are extremely challenging, and thus only a handful of studies
have addressed this question to date
\citep{Oppenheimer09,Pallottini14,Jaacks18,Jaacks19,Doughty19}.  We
opted for an extremely simplistic model of a uniformly enriched IGM
for the sensitivity analyses performed here, but our correlation
function approach can be easily generalized to accommodate more complex
distributions of metals and/or or applied directly to hydrodynamical
simulations.

Indeed, empirical constraints are likely to provide the first answers
to the question of early IGM enrichment.  Most obvious would be a lack
of detection of the \ion{Mg}{ii} forest towards quasars at
sufficiently high-redshift that the IGM is known to be highly neutral,
which we argued would yield a 95\% upper limit of $[{\rm Mg}\slash
  {\rm H}] < -4.4$ for our fiducial JWST dataset. A complementary
approach would be to measure the enrichment level of the
\emph{ionized} IGM as a function of redshift approaching the EoR,
using ions like \ion{C}{iv} or \ion{Si}{iv}.  To date the highest
redshift IGM metallicity measurement comes from \citet{Simcoe11a} who
measured ${\rm C\slash H} \simeq -3.55$ at $z\sim 4.3$ via Voigt
profile fitting of \ion{C}{iv} absorbers. As previously discussed, at
higher redshifts $4.5 < z < 7$, both the reduction in sensitivity as
transitions redshift into the near-IR, as well as the rapidly
increasing Ly$\alpha$ forest opacity render the line-fitting of
individual absorbers hopeless. But a fruitful direction for progress
would be to apply the correlation function techniques presented here
to these intermediate redshift forests of absorbers in the ionized
IGM, both for metal-metal auto-correlations and metal-hydrogen
cross-correlations. Such auto and cross correlation measurements would
enable IGM metallicity measurements in the critical $z \sim 5-6$
window after reionization. By comparing the clustering of
high-ionization lines (e.g. \ion{C}{iv} and \ion{Si}{iv}) to
low-ionization lines (e.g. \ion{Mg}{ii}, \ion{O}{i}, \ion{Si}{ii},
\ion{C}{ii}) through the EoR, one could observe the high-ionization
lines disappear as the low-ionization lines appear, providing the smoking-gun
for the phase transition. Interpreting these metal-line forests for
bluer transitions than \ion{Mg}{ii} will be more complicated, because
of foreground metal-line contamination, which is why we have initially
focused on \ion{Mg}{ii}.  But similar to the approach adopted for the
Ly$\alpha$ forest \citep{McDonald06,Nathalie15}
foreground metal-line
contamination can be quantified using lower redshift quasars \citep{Blomqvist18}
and statistically subtracted. This would open up the possibility of
characterizing the clustering of a host of different metal absorption
lines, and might even enable a measurement of relative abundance
patterns in the high-redshift IGM prior to reionization, which could shed light on PopIII
enrichment \citep{Kulkarni13,Kulkarni14}.

In this study we assumed a mock dataset of 10 JWST spectra with ${\rm
  FWHM}=100~{\rm km~s^{-1}}$ and ${\rm S\slash N}=100$ probing the the
redshift range $z = 6.9-7.5$. At present, there are five $z > 7$
quasars known bright enough $m_{1450}\sim 20.5$ for this experiment,
including two at $z > 7.5$
\citep{Mortlock11,Wang18z7,Yang19,Banados18,Yang20b}. Four\footnote{The
  fifth J0038$-$1527 is a broad absorption line quasar, which
  precludes using it to measure an IGM damping wing.} of them exhibit
compelling evidence for an IGM damping wing indicating that a
significantly neutral IGM
lies
in the foreground of these quasars
\citep{Mortlock11,Greig17b,Banados18,Davies18b,Greig19,Wang20a,Yang20b}. While
the full set of 10 $z > 7.5$ quasars does not yet exist, Euclid will
discover over 100 quasars with $7.0 < z < 7.5$, and $\sim$ 25 quasars
beyond the current record of $z = 7.5$, including $\sim$ 8 beyond $z =
8.0$ \citep{Euclid19}. The brightest of these will provide the ideal
targets for searching for the \ion{Mg}{ii} forest with JWST, enabling
a new powerful probe of the the reionization and enrichment history of
the Universe.

\section*{Acknowledgements}
We acknowledge helpful conversations with the ENIGMA group at UC Santa Barbara. 
JFH acknowledges support from the National Science Foundation under
Grant No. 1816006.  FW acknowledge the support provided by NASA
through the NASA Hubble Fellowship grant \#HST-HF2-51448.001-A awarded
by the Space Telescope Science Institute, which is operated by the
Association of Universities for Research in Astronomy, Incorporated,
under NASA contract NAS5-26555.  We are grateful to PRACE for awarding us
access to JUWELS hosted by GCS@FZJ, Germany.




\bibliographystyle{mnras}

\begin{thebibliography}{}
\makeatletter
\relax
\def\mn@urlcharsother{\let\do\@makeother \do\$\do\&\do\#\do\^\do\_\do\%\do\~}
\def\mn@doi{\begingroup\mn@urlcharsother \@ifnextchar [ {\mn@doi@}
  {\mn@doi@[]}}
\def\mn@doi@[#1]#2{\def\@tempa{#1}\ifx\@tempa\@empty \href
  {http://dx.doi.org/#2} {doi:#2}\else \href {http://dx.doi.org/#2} {#1}\fi
  \endgroup}
\def\mn@eprint#1#2{\mn@eprint@#1:#2::\@nil}
\def\mn@eprint@arXiv#1{\href {http://arxiv.org/abs/#1} {{\tt arXiv:#1}}}
\def\mn@eprint@dblp#1{\href {http://dblp.uni-trier.de/rec/bibtex/#1.xml}
  {dblp:#1}}
\def\mn@eprint@#1:#2:#3:#4\@nil{\def\@tempa {#1}\def\@tempb {#2}\def\@tempc
  {#3}\ifx \@tempc \@empty \let \@tempc \@tempb \let \@tempb \@tempa \fi \ifx
  \@tempb \@empty \def\@tempb {arXiv}\fi \@ifundefined
  {mn@eprint@\@tempb}{\@tempb:\@tempc}{\expandafter \expandafter \csname
  mn@eprint@\@tempb\endcsname \expandafter{\@tempc}}}

\bibitem[\protect\citeauthoryear{{Adelberger}, {Steidel}, {Shapley}  \&
  {Pettini}}{{Adelberger} et~al.}{2003}]{adel03}
{Adelberger} K.~L.,  {Steidel} C.~C.,  {Shapley} A.~E.,   {Pettini} M.,  2003,
  \mn@doi [\apj] {10.1086/345660}, \href
  {http://adsabs.harvard.edu/cgi-bin/nph-bib\_query?bibcode=2003ApJ...584...45A\&db\_key=AST}
  {584, 45}

\bibitem[\protect\citeauthoryear{{Adelberger}, {Shapley}, {Steidel}, {Pettini},
  {Erb}  \& {Reddy}}{{Adelberger} et~al.}{2005}]{ass+05}
{Adelberger} K.~L.,  {Shapley} A.~E.,  {Steidel} C.~C.,  {Pettini} M.,  {Erb}
  D.~K.,   {Reddy} N.~A.,  2005, \mn@doi [\apj] {10.1086/431753}, \href
  {http://adsabs.harvard.edu/cgi-bin/nph-bib\_query?bibcode=2005ApJ...629..636A\&db\_key=AST}
  {629, 636}

\bibitem[\protect\citeauthoryear{{Aguirre}, {Schaye}, {Kim}, {Theuns}, {Rauch}
  \& {Sargent}}{{Aguirre} et~al.}{2004}]{Aguirre04}
{Aguirre} A.,  {Schaye} J.,  {Kim} T.-S.,  {Theuns} T.,  {Rauch} M.,
  {Sargent} W. L.~W.,  2004, \mn@doi [The Astrophysical Journal]
  {10.1086/380961}, \href
  {https://ui.adsabs.harvard.edu/abs/2004ApJ...602...38A} {602, 38}

\bibitem[\protect\citeauthoryear{{Aguirre}, {Dow-Hygelund}, {Schaye}  \&
  {Theuns}}{{Aguirre} et~al.}{2008}]{Aguirre08}
{Aguirre} A.,  {Dow-Hygelund} C.,  {Schaye} J.,   {Theuns} T.,  2008, \mn@doi
  [The Astrophysical Journal] {10.1086/592554}, \href
  {https://ui.adsabs.harvard.edu/abs/2008ApJ...689..851A} {689, 851}

\bibitem[\protect\citeauthoryear{{Almgren}, {Bell}, {Lijewski}, {Luki{\'c}}  \&
  {Van Andel}}{{Almgren} et~al.}{2013}]{Almgren13}
{Almgren} A.~S.,  {Bell} J.~B.,  {Lijewski} M.~J.,  {Luki{\'c}} Z.,   {Van
  Andel} E.,  2013, \mn@doi [\apj] {10.1088/0004-637X/765/1/39}, \href
  {http://adsabs.harvard.edu/abs/2013ApJ...765...39A} {765, 39}

\bibitem[\protect\citeauthoryear{{Asplund}, {Grevesse}, {Sauval}  \&
  {Scott}}{{Asplund} et~al.}{2009}]{Asplung09}
{Asplund} M.,  {Grevesse} N.,  {Sauval} A.~J.,   {Scott} P.,  2009, \mn@doi
  [\araa] {10.1146/annurev.astro.46.060407.145222}, \href
  {https://ui.adsabs.harvard.edu/abs/2009ARA&A..47..481A} {47, 481}

\bibitem[\protect\citeauthoryear{{Ba{\~n}ados} et~al.,}{{Ba{\~n}ados}
  et~al.}{2018}]{Banados18}
{Ba{\~n}ados} E.,  et~al., 2018, \mn@doi [\nat] {10.1038/nature25180}, \href
  {https://ui.adsabs.harvard.edu/abs/2018Natur.553..473B} {553, 473}

\bibitem[\protect\citeauthoryear{{Barkana}}{{Barkana}}{2009}]{Barkana09}
{Barkana} R.,  2009, \mn@doi [\mnras] {10.1111/j.1365-2966.2009.14929.x}, \href
  {https://ui.adsabs.harvard.edu/abs/2009MNRAS.397.1454B} {397, 1454}

\bibitem[\protect\citeauthoryear{{Becker}, {Sargent}, {Rauch}  \&
  {Simcoe}}{{Becker} et~al.}{2006}]{Becker06}
{Becker} G.~D.,  {Sargent} W. L.~W.,  {Rauch} M.,   {Simcoe} R.~A.,  2006,
  \mn@doi [The Astrophysical Journal] {10.1086/500079}, \href
  {https://ui.adsabs.harvard.edu/abs/2006ApJ...640...69B} {640, 69}

\bibitem[\protect\citeauthoryear{{Becker}, {Rauch}  \& {Sargent}}{{Becker}
  et~al.}{2009}]{Becker09}
{Becker} G.~D.,  {Rauch} M.,   {Sargent} W. L.~W.,  2009, \mn@doi [The
  Astrophysical Journal] {10.1088/0004-637X/698/2/1010}, \href
  {https://ui.adsabs.harvard.edu/abs/2009ApJ...698.1010B} {698, 1010}

\bibitem[\protect\citeauthoryear{{Becker}, {Sargent}, {Rauch}  \&
  {Calverley}}{{Becker} et~al.}{2011}]{Becker11a}
{Becker} G.~D.,  {Sargent} W. L.~W.,  {Rauch} M.,   {Calverley} A.~P.,  2011,
  \mn@doi [The Astrophysical Journal] {10.1088/0004-637X/735/2/93}, \href
  {https://ui.adsabs.harvard.edu/abs/2011ApJ...735...93B} {735, 93}

\bibitem[\protect\citeauthoryear{{Becker}, {Hewett}, {Worseck}  \&
  {Prochaska}}{{Becker} et~al.}{2013}]{Becker2013}
{Becker} G.~D.,  {Hewett} P.~C.,  {Worseck} G.,   {Prochaska} J.~X.,  2013,
  \mn@doi [\mnras] {10.1093/mnras/stt031}, \href
  {http://adsabs.harvard.edu/abs/2013MNRAS.430.2067B} {430, 2067}

\bibitem[\protect\citeauthoryear{{Becker} et~al.,}{{Becker}
  et~al.}{2019}]{Becker19}
{Becker} G.~D.,  et~al., 2019, arXiv e-prints, \href
  {https://ui.adsabs.harvard.edu/abs/2019arXiv190702983B} {p. arXiv:1907.02983}

\bibitem[\protect\citeauthoryear{{Bergeron}, {Aracil}, {Petitjean}  \&
  {Pichon}}{{Bergeron} et~al.}{2002}]{Bergeron02}
{Bergeron} J.,  {Aracil} B.,  {Petitjean} P.,   {Pichon} C.,  2002, \mn@doi
  [Astronomy and Astrophysics] {10.1051/0004-6361:20021611}, \href
  {https://ui.adsabs.harvard.edu/abs/2002A&A...396L..11B} {396, L11}

\bibitem[\protect\citeauthoryear{{Blomqvist} et~al.,}{{Blomqvist}
  et~al.}{2018}]{Blomqvist18}
{Blomqvist} M.,  et~al., 2018, \mn@doi [Journal of Cosmology and Astro-Particle
  Physics] {10.1088/1475-7516/2018/05/029}, \href
  {https://ui.adsabs.harvard.edu/abs/2018JCAP...05..029B} {2018, 029}

\bibitem[\protect\citeauthoryear{{Boksenberg} \& {Sargent}}{{Boksenberg} \&
  {Sargent}}{2015}]{Boksenberg15}
{Boksenberg} A.,  {Sargent} W. L.~W.,  2015, \mn@doi [\apjs]
  {10.1088/0067-0049/218/1/7}, \href
  {https://ui.adsabs.harvard.edu/abs/2015ApJS..218....7B} {218, 7}

\bibitem[\protect\citeauthoryear{{Booth}, {Schaye}, {Delgado}  \& {Dalla
  Vecchia}}{{Booth} et~al.}{2012}]{Booth12}
{Booth} C.~M.,  {Schaye} J.,  {Delgado} J.~D.,   {Dalla Vecchia} C.,  2012,
  \mn@doi [\mnras] {10.1111/j.1365-2966.2011.20047.x}, \href
  {https://ui.adsabs.harvard.edu/abs/2012MNRAS.420.1053B} {420, 1053}

\bibitem[\protect\citeauthoryear{{Bosman}, {Becker}, {Haehnelt}, {Hewett},
  {McMahon}, {Mortlock}, {Simpson}  \& {Venemans}}{{Bosman}
  et~al.}{2017}]{Bosman17}
{Bosman} S. E.~I.,  {Becker} G.~D.,  {Haehnelt} M.~G.,  {Hewett} P.~C.,
  {McMahon} R.~G.,  {Mortlock} D.~J.,  {Simpson} C.,   {Venemans} B.~P.,  2017,
  \mn@doi [\mnras] {10.1093/mnras/stx1305}, \href
  {https://ui.adsabs.harvard.edu/abs/2017MNRAS.470.1919B} {470, 1919}

\bibitem[\protect\citeauthoryear{{Bouch{\'e}}, {Murphy}, {P{\'e}roux}, {Csabai}
   \& {Wild}}{{Bouch{\'e}} et~al.}{2006}]{BoucheLRG06}
{Bouch{\'e}} N.,  {Murphy} M.~T.,  {P{\'e}roux} C.,  {Csabai} I.,   {Wild} V.,
  2006, \mn@doi [\mnras] {10.1111/j.1365-2966.2006.10685.x}, \href
  {https://ui.adsabs.harvard.edu/abs/2006MNRAS.371..495B} {371, 495}

\bibitem[\protect\citeauthoryear{{Busca} et~al.,}{{Busca}
  et~al.}{2013}]{Busca13}
{Busca} N.~G.,  et~al., 2013, \mn@doi [Astronomy and Astrophysics]
  {10.1051/0004-6361/201220724}, \href
  {https://ui.adsabs.harvard.edu/abs/2013A&A...552A..96B} {552, A96}

\bibitem[\protect\citeauthoryear{{Calverley}, {Becker}, {Haehnelt}  \&
  {Bolton}}{{Calverley} et~al.}{2011}]{Calverley11}
{Calverley} A.~P.,  {Becker} G.~D.,  {Haehnelt} M.~G.,   {Bolton} J.~S.,  2011,
  \mn@doi [\mnras] {10.1111/j.1365-2966.2010.18072.x}, \href
  {http://adsabs.harvard.edu/abs/2011MNRAS.412.2543C} {412, 2543}

\bibitem[\protect\citeauthoryear{{Carilli}, {Gnedin}  \& {Owen}}{{Carilli}
  et~al.}{2002}]{Carilli02}
{Carilli} C.~L.,  {Gnedin} N.~Y.,   {Owen} F.,  2002, \mn@doi [\apj]
  {10.1086/342179}, \href
  {https://ui.adsabs.harvard.edu/abs/2002ApJ...577...22C} {577, 22}

\bibitem[\protect\citeauthoryear{{Chen} et~al.,}{{Chen} et~al.}{2017}]{Chen17}
{Chen} S.-F.~S.,  et~al., 2017, \mn@doi [The Astrophysical Journal]
  {10.3847/1538-4357/aa9707}, \href
  {https://ui.adsabs.harvard.edu/abs/2017ApJ...850..188C} {850, 188}

\bibitem[\protect\citeauthoryear{{Cheng} et~al.,}{{Cheng}
  et~al.}{2018}]{Cheng18}
{Cheng} C.,  et~al., 2018, \mn@doi [\apj] {10.3847/1538-4357/aae833}, \href
  {https://ui.adsabs.harvard.edu/abs/2018ApJ...868...26C} {868, 26}

\bibitem[\protect\citeauthoryear{{Churchill} \& {Vogt}}{{Churchill} \&
  {Vogt}}{2001}]{Churchill01}
{Churchill} C.~W.,  {Vogt} S.~S.,  2001, \mn@doi [\aj] {10.1086/321174}, \href
  {https://ui.adsabs.harvard.edu/abs/2001AJ....122..679C} {122, 679}

\bibitem[\protect\citeauthoryear{{Churchill}, {Rigby}, {Charlton}  \&
  {Vogt}}{{Churchill} et~al.}{1999}]{Churchill99}
{Churchill} C.~W.,  {Rigby} J.~R.,  {Charlton} J.~C.,   {Vogt} S.~S.,  1999,
  \mn@doi [\apjs] {10.1086/313168}, \href
  {https://ui.adsabs.harvard.edu/abs/1999ApJS..120...51C} {120, 51}

\bibitem[\protect\citeauthoryear{{Churchill}, {Vogt}  \&
  {Charlton}}{{Churchill} et~al.}{2003}]{Churchill03}
{Churchill} C.~W.,  {Vogt} S.~S.,   {Charlton} J.~C.,  2003, \mn@doi [\aj]
  {10.1086/345513}, \href
  {https://ui.adsabs.harvard.edu/abs/2003AJ....125...98C} {125, 98}

\bibitem[\protect\citeauthoryear{{Ciardi} et~al.,}{{Ciardi}
  et~al.}{2013}]{Ciardi13}
{Ciardi} B.,  et~al., 2013, \mn@doi [\mnras] {10.1093/mnras/sts156}, \href
  {https://ui.adsabs.harvard.edu/abs/2013MNRAS.428.1755C} {428, 1755}

\bibitem[\protect\citeauthoryear{{Coc}, {Uzan}  \& {Vangioni}}{{Coc}
  et~al.}{2014}]{Coc14}
{Coc} A.,  {Uzan} J.-P.,   {Vangioni} E.,  2014, \mn@doi [\jcap]
  {10.1088/1475-7516/2014/10/050}, \href
  {https://ui.adsabs.harvard.edu/abs/2014JCAP...10..050C} {2014, 050}

\bibitem[\protect\citeauthoryear{{Cooke}, {Pettini}  \& {Steidel}}{{Cooke}
  et~al.}{2017}]{Cooke17}
{Cooke} R.~J.,  {Pettini} M.,   {Steidel} C.~C.,  2017, \mn@doi [\mnras]
  {10.1093/mnras/stx037}, \href
  {https://ui.adsabs.harvard.edu/abs/2017MNRAS.467..802C} {467, 802}

\bibitem[\protect\citeauthoryear{Coppolani, Petitjean, Stoehr
  et~al.}{Coppolani et~al.}{2006}]{Coppolani06}
Coppolani F.,  Petitjean P.,  Stoehr F.,   et~al., 2006, MNRAS, 370, 1804

\bibitem[\protect\citeauthoryear{{Crighton}, {O'Meara}  \& {Murphy}}{{Crighton}
  et~al.}{2016}]{Crighton16}
{Crighton} N. H.~M.,  {O'Meara} J.~M.,   {Murphy} M.~T.,  2016, \mn@doi
  [\mnras] {10.1093/mnrasl/slv191}, \href
  {https://ui.adsabs.harvard.edu/abs/2016MNRAS.457L..44C} {457, L44}

\bibitem[\protect\citeauthoryear{{D'Aloisio}, {McQuinn}, {Davies}  \&
  {Furlanetto}}{{D'Aloisio} et~al.}{2018}]{Anson18}
{D'Aloisio} A.,  {McQuinn} M.,  {Davies} F.~B.,   {Furlanetto} S.~R.,  2018,
  \mn@doi [\mnras] {10.1093/mnras/stx2341}, \href
  {https://ui.adsabs.harvard.edu/abs/2018MNRAS.473..560D} {473, 560}

\bibitem[\protect\citeauthoryear{{D'Aloisio}, {McQuinn}, {Trac}, {Cain}  \&
  {Mesinger}}{{D'Aloisio} et~al.}{2020}]{Anson20}
{D'Aloisio} A.,  {McQuinn} M.,  {Trac} H.,  {Cain} C.,   {Mesinger} A.,  2020,
  arXiv e-prints, \href {https://ui.adsabs.harvard.edu/abs/2020arXiv200202467D}
  {p. arXiv:2002.02467}

\bibitem[\protect\citeauthoryear{{D'Odorico}, {Calura}, {Cristiani}  \&
  {Viel}}{{D'Odorico} et~al.}{2010}]{Dodorico10}
{D'Odorico} V.,  {Calura} F.,  {Cristiani} S.,   {Viel} M.,  2010, \mn@doi
  [\mnras] {10.1111/j.1365-2966.2009.15856.x}, \href
  {https://ui.adsabs.harvard.edu/abs/2010MNRAS.401.2715D} {401, 2715}

\bibitem[\protect\citeauthoryear{{D'Odorico} et~al.,}{{D'Odorico}
  et~al.}{2013}]{Dodorico13a}
{D'Odorico} V.,  et~al., 2013, \mn@doi [Monthly Notices of the Royal
  Astronomical Society] {10.1093/mnras/stt1365}, \href
  {https://ui.adsabs.harvard.edu/abs/2013MNRAS.435.1198D} {435, 1198}

\bibitem[\protect\citeauthoryear{{D'Odorico} et~al.,}{{D'Odorico}
  et~al.}{2016}]{Dodorico16}
{D'Odorico} V.,  et~al., 2016, \mn@doi [\mnras] {10.1093/mnras/stw2161}, \href
  {https://ui.adsabs.harvard.edu/abs/2016MNRAS.463.2690D} {463, 2690}

\bibitem[\protect\citeauthoryear{{Davies} \& {Furlanetto}}{{Davies} \&
  {Furlanetto}}{2016}]{davies16b}
{Davies} F.~B.,  {Furlanetto} S.~R.,  2016, \mn@doi [\mnras]
  {10.1093/mnras/stw931}, \href
  {http://adsabs.harvard.edu/abs/2016MNRAS.460.1328D} {460, 1328}

\bibitem[\protect\citeauthoryear{{Davies} \& {Hennawi}}{{Davies} \&
  {Hennawi}}{2020}]{Davies20_smallscale}
{Davies} F.~B.,  {Hennawi} J.~F.,  2020, in prep

\bibitem[\protect\citeauthoryear{{Davies}, {Hennawi}, {Eilers}  \&
  {Luki{\'c}}}{{Davies} et~al.}{2018a}]{Davies18ABC}
{Davies} F.~B.,  {Hennawi} J.~F.,  {Eilers} A.-C.,   {Luki{\'c}} Z.,  2018a,
  \mn@doi [\apj] {10.3847/1538-4357/aaaf70}, \href
  {https://ui.adsabs.harvard.edu/abs/2018ApJ...855..106D} {855, 106}

\bibitem[\protect\citeauthoryear{{Davies} et~al.,}{{Davies}
  et~al.}{2018b}]{Davies18b}
{Davies} F.~B.,  et~al., 2018b, \mn@doi [\apj] {10.3847/1538-4357/aad6dc},
  \href {https://ui.adsabs.harvard.edu/abs/2018ApJ...864..142D} {864, 142}

\bibitem[\protect\citeauthoryear{{DeBoer} et~al.,}{{DeBoer}
  et~al.}{2017}]{HERA}
{DeBoer} D.~R.,  et~al., 2017, \mn@doi [\pasp]
  {10.1088/1538-3873/129/974/045001}, \href
  {https://ui.adsabs.harvard.edu/abs/2017PASP..129d5001D} {129, 045001}

\bibitem[\protect\citeauthoryear{{Doughty} \& {Finlator}}{{Doughty} \&
  {Finlator}}{2019}]{Doughty19}
{Doughty} C.,  {Finlator} K.,  2019, \mn@doi [\mnras] {10.1093/mnras/stz2331},
  \href {https://ui.adsabs.harvard.edu/abs/2019MNRAS.tmp.2245D} {p.~2245}

\bibitem[\protect\citeauthoryear{{Ellison}, {Songaila}, {Schaye}  \&
  {Pettini}}{{Ellison} et~al.}{2000}]{Ellison00}
{Ellison} S.~L.,  {Songaila} A.,  {Schaye} J.,   {Pettini} M.,  2000, \mn@doi
  [The Astronomical Journal] {10.1086/301511}, \href
  {https://ui.adsabs.harvard.edu/abs/2000AJ....120.1175E} {120, 1175}

\bibitem[\protect\citeauthoryear{{Euclid} et~al.,}{{Euclid}
  et~al.}{2019}]{Euclid19}
{Euclid} et~al., 2019, arXiv e-prints, \href
  {https://ui.adsabs.harvard.edu/abs/2019arXiv190804310E} {p. arXiv:1908.04310}

\bibitem[\protect\citeauthoryear{{Ferland} et~al.,}{{Ferland}
  et~al.}{2017}]{Cloudy17}
{Ferland} G.~J.,  et~al., 2017, \rmxaa, \href
  {https://ui.adsabs.harvard.edu/abs/2017RMxAA..53..385F} {53, 385}

\bibitem[\protect\citeauthoryear{{Ferrara}}{{Ferrara}}{2016}]{Ferrara16}
{Ferrara} A.,  2016, in {Mesinger} A.,  ed.,  Astrophysics and Space Science
  Library Vol. 423, Understanding the Epoch of Cosmic Reionization: Challenges
  and Progress. p.~163 (\mn@eprint {arXiv} {1511.01120}),
  \mn@doi{10.1007/978-3-319-21957-8_6}

\bibitem[\protect\citeauthoryear{{Fialkov}, {Barkana}  \& {Visbal}}{{Fialkov}
  et~al.}{2014}]{Fialkov14}
{Fialkov} A.,  {Barkana} R.,   {Visbal} E.,  2014, \mn@doi [\nat]
  {10.1038/nature12999}, \href
  {https://ui.adsabs.harvard.edu/abs/2014Natur.506..197F} {506, 197}

\bibitem[\protect\citeauthoryear{{Frebel} \& {Norris}}{{Frebel} \&
  {Norris}}{2015}]{Frebel15}
{Frebel} A.,  {Norris} J.~E.,  2015, \mn@doi [\araa]
  {10.1146/annurev-astro-082214-122423}, \href
  {https://ui.adsabs.harvard.edu/abs/2015ARA&A..53..631F} {53, 631}

\bibitem[\protect\citeauthoryear{{Fumagalli}, {O'Meara}  \&
  {Prochaska}}{{Fumagalli} et~al.}{2011}]{Fumagalli_sci11}
{Fumagalli} M.,  {O'Meara} J.~M.,   {Prochaska} J.~X.,  2011, \mn@doi [Science]
  {10.1126/science.1213581}, \href
  {https://ui.adsabs.harvard.edu/abs/2011Sci...334.1245F} {334, 1245}

\bibitem[\protect\citeauthoryear{{Fumagalli}, {O'Meara}  \&
  {Prochaska}}{{Fumagalli} et~al.}{2016}]{Fumagalli16}
{Fumagalli} M.,  {O'Meara} J.~M.,   {Prochaska} J.~X.,  2016, \mn@doi [\mnras]
  {10.1093/mnras/stv2616}, \href
  {https://ui.adsabs.harvard.edu/abs/2016MNRAS.455.4100F} {455, 4100}

\bibitem[\protect\citeauthoryear{{Furlanetto}}{{Furlanetto}}{2006a}]{Furlanetto06_forest}
{Furlanetto} S.~R.,  2006a, \mn@doi [\mnras]
  {10.1111/j.1365-2966.2006.10603.x}, \href
  {https://ui.adsabs.harvard.edu/abs/2006MNRAS.370.1867F} {370, 1867}

\bibitem[\protect\citeauthoryear{{Furlanetto}}{{Furlanetto}}{2006b}]{Furlanetto06}
{Furlanetto} S.~R.,  2006b, \mn@doi [\mnras]
  {10.1111/j.1365-2966.2006.10725.x}, \href
  {https://ui.adsabs.harvard.edu/abs/2006MNRAS.371..867F} {371, 867}

\bibitem[\protect\citeauthoryear{{Furlanetto}, {Oh}  \& {Briggs}}{{Furlanetto}
  et~al.}{2006}]{FOB06}
{Furlanetto} S.~R.,  {Oh} S.~P.,   {Briggs} F.~H.,  2006, \mn@doi [\physrep]
  {10.1016/j.physrep.2006.08.002}, \href
  {https://ui.adsabs.harvard.edu/abs/2006PhR...433..181F} {433, 181}

\bibitem[\protect\citeauthoryear{{Gauthier}, {Chen}, {Cooksey}, {Simcoe},
  {Seyffert}  \& {O'Meara}}{{Gauthier} et~al.}{2014}]{Gauthier14}
{Gauthier} J.-R.,  {Chen} H.-W.,  {Cooksey} K.~L.,  {Simcoe} R.~A.,  {Seyffert}
  E.~N.,   {O'Meara} J.~M.,  2014, \mn@doi [\mnras] {10.1093/mnras/stt2443},
  \href {https://ui.adsabs.harvard.edu/abs/2014MNRAS.439..342G} {439, 342}

\bibitem[\protect\citeauthoryear{{Gnedin} \& {Hui}}{{Gnedin} \&
  {Hui}}{1998}]{GH98}
{Gnedin} N.~Y.,  {Hui} L.,  1998, \mn@doi [\mnras]
  {10.1046/j.1365-8711.1998.01249.x}, \href
  {http://adsabs.harvard.edu/abs/1998MNRAS.296...44G} {296, 44}

\bibitem[\protect\citeauthoryear{{Gnedin} \& {Ostriker}}{{Gnedin} \&
  {Ostriker}}{1997}]{GnedinOstriker97}
{Gnedin} N.~Y.,  {Ostriker} J.~P.,  1997, \mn@doi [The Astrophysical Journal]
  {10.1086/304548}, \href
  {https://ui.adsabs.harvard.edu/abs/1997ApJ...486..581G} {486, 581}

\bibitem[\protect\citeauthoryear{{Greig}, {Mesinger}, {Haiman}  \&
  {Simcoe}}{{Greig} et~al.}{2017}]{Greig17b}
{Greig} B.,  {Mesinger} A.,  {Haiman} Z.,   {Simcoe} R.~A.,  2017, \mn@doi
  [\mnras] {10.1093/mnras/stw3351}, \href
  {http://adsabs.harvard.edu/abs/2017MNRAS.466.4239G} {466, 4239}

\bibitem[\protect\citeauthoryear{{Greig}, {Mesinger}  \& {Ba{\~n}ados}}{{Greig}
  et~al.}{2019}]{Greig19}
{Greig} B.,  {Mesinger} A.,   {Ba{\~n}ados} E.,  2019, \mn@doi [\mnras]
  {10.1093/mnras/stz230}, \href
  {https://ui.adsabs.harvard.edu/abs/2019MNRAS.484.5094G} {484, 5094}

\bibitem[\protect\citeauthoryear{{Haardt} \& {Madau}}{{Haardt} \&
  {Madau}}{2012}]{HM12}
{Haardt} F.,  {Madau} P.,  2012, \mn@doi [\apj] {10.1088/0004-637X/746/2/125},
  \href {http://adsabs.harvard.edu/abs/2012ApJ...746..125H} {746, 125}

\bibitem[\protect\citeauthoryear{{Hahn} \& {Abel}}{{Hahn} \&
  {Abel}}{2011}]{Hahn11}
{Hahn} O.,  {Abel} T.,  2011, \mn@doi [\mnras]
  {10.1111/j.1365-2966.2011.18820.x}, \href
  {https://ui.adsabs.harvard.edu/abs/2011MNRAS.415.2101H} {415, 2101}

\bibitem[\protect\citeauthoryear{{Hoag} et~al.,}{{Hoag} et~al.}{2019}]{Hoag19}
{Hoag} A.,  et~al., 2019, \mn@doi [\apj] {10.3847/1538-4357/ab1de7}, \href
  {https://ui.adsabs.harvard.edu/abs/2019ApJ...878...12H} {878, 12}

\bibitem[\protect\citeauthoryear{{Howlett}, {Lewis}, {Hall}  \&
  {Challinor}}{{Howlett} et~al.}{2012}]{Howlett12}
{Howlett} C.,  {Lewis} A.,  {Hall} A.,   {Challinor} A.,  2012, \mn@doi [\jcap]
  {10.1088/1475-7516/2012/04/027}, \href
  {https://ui.adsabs.harvard.edu/abs/2012JCAP...04..027H} {2012, 027}

\bibitem[\protect\citeauthoryear{{Ir{\v{s}}i{\v{c}}}
  et~al.,}{{Ir{\v{s}}i{\v{c}}} et~al.}{2017}]{Irsic17}
{Ir{\v{s}}i{\v{c}}} V.,  et~al., 2017, \mn@doi [\prd]
  {10.1103/PhysRevD.96.023522}, \href
  {https://ui.adsabs.harvard.edu/abs/2017PhRvD..96b3522I} {96, 023522}

\bibitem[\protect\citeauthoryear{{Jaacks}, {Thompson}, {Finkelstein}  \&
  {Bromm}}{{Jaacks} et~al.}{2018}]{Jaacks18}
{Jaacks} J.,  {Thompson} R.,  {Finkelstein} S.~L.,   {Bromm} V.,  2018, \mn@doi
  [\mnras] {10.1093/mnras/sty062}, \href
  {https://ui.adsabs.harvard.edu/abs/2018MNRAS.475.4396J} {475, 4396}

\bibitem[\protect\citeauthoryear{{Jaacks}, {Finkelstein}  \& {Bromm}}{{Jaacks}
  et~al.}{2019}]{Jaacks19}
{Jaacks} J.,  {Finkelstein} S.~L.,   {Bromm} V.,  2019, \mn@doi [\mnras]
  {10.1093/mnras/stz1529}, \href
  {https://ui.adsabs.harvard.edu/abs/2019MNRAS.488.2202J} {488, 2202}

\bibitem[\protect\citeauthoryear{{Kacprzak} \& {Churchill}}{{Kacprzak} \&
  {Churchill}}{2011}]{Kacp11}
{Kacprzak} G.~G.,  {Churchill} C.~W.,  2011, \mn@doi [\apjl]
  {10.1088/2041-8205/743/2/L34}, \href
  {https://ui.adsabs.harvard.edu/abs/2011ApJ...743L..34K} {743, L34}

\bibitem[\protect\citeauthoryear{{Keating}, {Haehnelt}, {Becker}  \&
  {Bolton}}{{Keating} et~al.}{2014}]{Keating14}
{Keating} L.~C.,  {Haehnelt} M.~G.,  {Becker} G.~D.,   {Bolton} J.~S.,  2014,
  \mn@doi [Monthly Notices of the Royal Astronomical Society]
  {10.1093/mnras/stt2324}, \href
  {https://ui.adsabs.harvard.edu/abs/2014MNRAS.438.1820K} {438, 1820}

\bibitem[\protect\citeauthoryear{{Keating}, {Weinberger}, {Kulkarni},
  {Haehnelt}, {Chardin}  \& {Aubert}}{{Keating} et~al.}{2020}]{Keating20}
{Keating} L.~C.,  {Weinberger} L.~H.,  {Kulkarni} G.,  {Haehnelt} M.~G.,
  {Chardin} J.,   {Aubert} D.,  2020, \mn@doi [\mnras] {10.1093/mnras/stz3083},
  \href {https://ui.adsabs.harvard.edu/abs/2020MNRAS.491.1736K} {491, 1736}

\bibitem[\protect\citeauthoryear{{Kirihara}, {Hasegawa}, {Umemura}, {Mori}  \&
  {Ishiyama}}{{Kirihara} et~al.}{2020}]{Kirihara20}
{Kirihara} T.,  {Hasegawa} K.,  {Umemura} M.,  {Mori} M.,   {Ishiyama} T.,
  2020, \mn@doi [\mnras] {10.1093/mnras/stz3376}, \href
  {https://ui.adsabs.harvard.edu/abs/2020MNRAS.491.4387K} {491, 4387}

\bibitem[\protect\citeauthoryear{{Kulkarni}, {Rollinde}, {Hennawi}  \&
  {Vangioni}}{{Kulkarni} et~al.}{2013}]{Kulkarni13}
{Kulkarni} G.,  {Rollinde} E.,  {Hennawi} J.~F.,   {Vangioni} E.,  2013,
  \mn@doi [\apj] {10.1088/0004-637X/772/2/93}, \href
  {http://adsabs.harvard.edu/abs/2013ApJ...772...93K} {772, 93}

\bibitem[\protect\citeauthoryear{{Kulkarni}, {Hennawi}, {Rollinde}  \&
  {Vangioni}}{{Kulkarni} et~al.}{2014}]{Kulkarni14}
{Kulkarni} G.,  {Hennawi} J.~F.,  {Rollinde} E.,   {Vangioni} E.,  2014,
  \mn@doi [\apj] {10.1088/0004-637X/787/1/64}, \href
  {http://adsabs.harvard.edu/abs/2014ApJ...787...64K} {787, 64}

\bibitem[\protect\citeauthoryear{{Kulkarni}, {Hennawi}, {O{\~n}orbe}, {Rorai}
  \& {Springel}}{{Kulkarni} et~al.}{2015}]{Kulkarni15}
{Kulkarni} G.,  {Hennawi} J.~F.,  {O{\~n}orbe} J.,  {Rorai} A.,   {Springel}
  V.,  2015, \mn@doi [\apj] {10.1088/0004-637X/812/1/30}, \href
  {http://adsabs.harvard.edu/abs/2015ApJ...812...30K} {812, 30}

\bibitem[\protect\citeauthoryear{{Kulkarni}, {Keating}, {Haehnelt}, {Bosman},
  {Puchwein}, {Chardin}  \& {Aubert}}{{Kulkarni} et~al.}{2019}]{Kulkarni19b}
{Kulkarni} G.,  {Keating} L.~C.,  {Haehnelt} M.~G.,  {Bosman} S. E.~I.,
  {Puchwein} E.,  {Chardin} J.,   {Aubert} D.,  2019, \mn@doi [\mnras]
  {10.1093/mnrasl/slz025}, \href
  {https://ui.adsabs.harvard.edu/abs/2019MNRAS.485L..24K} {485, L24}

\bibitem[\protect\citeauthoryear{{Lacey} \& {Cole}}{{Lacey} \&
  {Cole}}{1993}]{LaceyCole93}
{Lacey} C.,  {Cole} S.,  1993, \mnras, \href
  {http://adsabs.harvard.edu/abs/1993MNRAS.262..627L} {262, 627}

\bibitem[\protect\citeauthoryear{{Lee} et~al.,}{{Lee} et~al.}{2015}]{Lee15}
{Lee} K.-G.,  et~al., 2015, \mn@doi [The Astrophysical Journal]
  {10.1088/0004-637X/799/2/196}, \href
  {https://ui.adsabs.harvard.edu/abs/2015ApJ...799..196L} {799, 196}

\bibitem[\protect\citeauthoryear{{Lewis}, {Challinor}  \& {Lasenby}}{{Lewis}
  et~al.}{2000}]{CAMB}
{Lewis} A.,  {Challinor} A.,   {Lasenby} A.,  2000, \mn@doi [\apj]
  {10.1086/309179}, \href
  {https://ui.adsabs.harvard.edu/abs/2000ApJ...538..473L} {538, 473}

\bibitem[\protect\citeauthoryear{{Luki{\'c}}, {Stark}, {Nugent}, {White},
  {Meiksin}  \& {Almgren}}{{Luki{\'c}} et~al.}{2015}]{Lukic15}
{Luki{\'c}} Z.,  {Stark} C.~W.,  {Nugent} P.,  {White} M.,  {Meiksin} A.~A.,
  {Almgren} A.,  2015, \mn@doi [\mnras] {10.1093/mnras/stu2377}, \href
  {http://adsabs.harvard.edu/abs/2015MNRAS.446.3697L} {446, 3697}

\bibitem[\protect\citeauthoryear{{Lundgren} et~al.,}{{Lundgren}
  et~al.}{2009}]{Lundgren09}
{Lundgren} B.~F.,  et~al., 2009, \mn@doi [\apj] {10.1088/0004-637X/698/1/819},
  \href {https://ui.adsabs.harvard.edu/abs/2009ApJ...698..819L} {698, 819}

\bibitem[\protect\citeauthoryear{{Madau} \& {Fragos}}{{Madau} \&
  {Fragos}}{2017}]{Madau17}
{Madau} P.,  {Fragos} T.,  2017, \mn@doi [\apj] {10.3847/1538-4357/aa6af9},
  \href {https://ui.adsabs.harvard.edu/abs/2017ApJ...840...39M} {840, 39}

\bibitem[\protect\citeauthoryear{{Madau} \& {Shull}}{{Madau} \&
  {Shull}}{1996}]{MadauShull96}
{Madau} P.,  {Shull} J.~M.,  1996, \mn@doi [The Astrophysical Journal]
  {10.1086/176751}, \href
  {https://ui.adsabs.harvard.edu/abs/1996ApJ...457..551M} {457, 551}

\bibitem[\protect\citeauthoryear{{Madau}, {Meiksin}  \& {Rees}}{{Madau}
  et~al.}{1997}]{Madau97}
{Madau} P.,  {Meiksin} A.,   {Rees} M.~J.,  1997, \mn@doi [\apj]
  {10.1086/303549}, \href
  {https://ui.adsabs.harvard.edu/abs/1997ApJ...475..429M} {475, 429}

\bibitem[\protect\citeauthoryear{{Madau}, {Ferrara}  \& {Rees}}{{Madau}
  et~al.}{2001}]{MFR01}
{Madau} P.,  {Ferrara} A.,   {Rees} M.~J.,  2001, \mn@doi [The Astrophysical
  Journal] {10.1086/321474}, \href
  {https://ui.adsabs.harvard.edu/abs/2001ApJ...555...92M} {555, 92}

\bibitem[\protect\citeauthoryear{{Madau}, {Rees}, {Volonteri}, {Haardt}  \&
  {Oh}}{{Madau} et~al.}{2004}]{Madau04}
{Madau} P.,  {Rees} M.~J.,  {Volonteri} M.,  {Haardt} F.,   {Oh} S.~P.,  2004,
  \mn@doi [\apj] {10.1086/381935}, \href
  {https://ui.adsabs.harvard.edu/abs/2004ApJ...604..484M} {604, 484}

\bibitem[\protect\citeauthoryear{{Martin}, {Scannapieco}, {Ellison}, {Hennawi},
  {Djorgovski}  \& {Fournier}}{{Martin} et~al.}{2010}]{Martin10}
{Martin} C.~L.,  {Scannapieco} E.,  {Ellison} S.~L.,  {Hennawi} J.~F.,
  {Djorgovski} S.~G.,   {Fournier} A.~P.,  2010, \mn@doi [\apj]
  {10.1088/0004-637X/721/1/174}, \href
  {https://ui.adsabs.harvard.edu/abs/2010ApJ...721..174M} {721, 174}

\bibitem[\protect\citeauthoryear{{Mas-Ribas}, {Riemer-S{\o}rensen}, {Hennawi},
  {Miralda-Escud{\'e}}, {O'Meara}, {P{\'e}rez-R{\`a}fols}, {Murphy}  \&
  {Webb}}{{Mas-Ribas} et~al.}{2018}]{MasRibas18}
{Mas-Ribas} L.,  {Riemer-S{\o}rensen} S.,  {Hennawi} J.~F.,
  {Miralda-Escud{\'e}} J.,  {O'Meara} J.~M.,  {P{\'e}rez-R{\`a}fols} I.,
  {Murphy} M.~T.,   {Webb} J.~K.,  2018, \mn@doi [\apj]
  {10.3847/1538-4357/aac81a}, \href
  {https://ui.adsabs.harvard.edu/abs/2018ApJ...862...50M} {862, 50}

\bibitem[\protect\citeauthoryear{{Mason}, {Treu}, {Dijkstra}, {Mesinger},
  {Trenti}, {Pentericci}, {de Barros}  \& {Vanzella}}{{Mason}
  et~al.}{2018}]{Mason18}
{Mason} C.~A.,  {Treu} T.,  {Dijkstra} M.,  {Mesinger} A.,  {Trenti} M.,
  {Pentericci} L.,  {de Barros} S.,   {Vanzella} E.,  2018, \mn@doi [\apj]
  {10.3847/1538-4357/aab0a7}, \href
  {https://ui.adsabs.harvard.edu/abs/2018ApJ...856....2M} {856, 2}

\bibitem[\protect\citeauthoryear{{Mason} et~al.,}{{Mason}
  et~al.}{2019}]{Mason19}
{Mason} C.~A.,  et~al., 2019, \mn@doi [\mnras] {10.1093/mnras/stz632}, \href
  {https://ui.adsabs.harvard.edu/abs/2019MNRAS.485.3947M} {485, 3947}

\bibitem[\protect\citeauthoryear{{Mathes}, {Churchill}  \& {Murphy}}{{Mathes}
  et~al.}{2017}]{Vulture17}
{Mathes} N.~L.,  {Churchill} C.~W.,   {Murphy} M.~T.,  2017, arXiv e-prints,
  \href {https://ui.adsabs.harvard.edu/abs/2017arXiv170105624M} {p.
  arXiv:1701.05624}

\bibitem[\protect\citeauthoryear{{McDonald}, {Miralda-Escud{\'e}}, {Rauch},
  {Sargent}, {Barlow}  \& {Cen}}{{McDonald} et~al.}{2001}]{McDonald01}
{McDonald} P.,  {Miralda-Escud{\'e}} J.,  {Rauch} M.,  {Sargent} W. L.~W.,
  {Barlow} T.~A.,   {Cen} R.,  2001, \mn@doi [The Astrophysical Journal]
  {10.1086/323426}, \href
  {https://ui.adsabs.harvard.edu/abs/2001ApJ...562...52M} {562, 52}

\bibitem[\protect\citeauthoryear{{McDonald}, {Seljak}, {Cen}, {Bode}  \&
  {Ostriker}}{{McDonald} et~al.}{2005a}]{McDonald05a}
{McDonald} P.,  {Seljak} U.,  {Cen} R.,  {Bode} P.,   {Ostriker} J.~P.,  2005a,
  \mn@doi [Monthly Notices of the Royal Astronomical Society]
  {10.1111/j.1365-2966.2005.09141.x}, \href
  {https://ui.adsabs.harvard.edu/abs/2005MNRAS.360.1471M} {360, 1471}

\bibitem[\protect\citeauthoryear{{McDonald} et~al.,}{{McDonald}
  et~al.}{2005b}]{McDonald05}
{McDonald} P.,  et~al., 2005b, \mn@doi [The Astrophysical Journal]
  {10.1086/497563}, \href
  {https://ui.adsabs.harvard.edu/abs/2005ApJ...635..761M} {635, 761}

\bibitem[\protect\citeauthoryear{{McDonald} et~al.,}{{McDonald}
  et~al.}{2006}]{McDonald06}
{McDonald} P.,  et~al., 2006, \mn@doi [The Astrophysical Journal Supplement
  Series] {10.1086/444361}, \href
  {https://ui.adsabs.harvard.edu/abs/2006ApJS..163...80M} {163, 80}

\bibitem[\protect\citeauthoryear{{Mertens} et~al.,}{{Mertens}
  et~al.}{2020}]{Mertens20}
{Mertens} F.~G.,  et~al., 2020, \mn@doi [\mnras] {10.1093/mnras/staa327}, \href
  {https://ui.adsabs.harvard.edu/abs/2020MNRAS.493.1662M} {493, 1662}

\bibitem[\protect\citeauthoryear{{Mesinger}, {Furlanetto}  \& {Cen}}{{Mesinger}
  et~al.}{2011}]{Mesinger11}
{Mesinger} A.,  {Furlanetto} S.,   {Cen} R.,  2011, \mn@doi [\mnras]
  {10.1111/j.1365-2966.2010.17731.x}, \href
  {http://adsabs.harvard.edu/abs/2011MNRAS.411..955M} {411, 955}

\bibitem[\protect\citeauthoryear{{Miralda-Escud{\'e}} \&
  {Rees}}{{Miralda-Escud{\'e}} \& {Rees}}{1997}]{MR97}
{Miralda-Escud{\'e}} J.,  {Rees} M.~J.,  1997, \mn@doi [The Astrophysical
  Journal] {10.1086/310550}, \href
  {https://ui.adsabs.harvard.edu/abs/1997ApJ...478L..57M} {478, L57}

\bibitem[\protect\citeauthoryear{{Mortlock} et~al.,}{{Mortlock}
  et~al.}{2011}]{Mortlock11}
{Mortlock} D.~J.,  et~al., 2011, \mn@doi [\nat] {10.1038/nature10159}, \href
  {http://adsabs.harvard.edu/abs/2011Natur.474..616M} {474, 616}

\bibitem[\protect\citeauthoryear{{Narayanan}, {Misawa}, {Charlton}  \&
  {Kim}}{{Narayanan} et~al.}{2007}]{Narayanan07}
{Narayanan} A.,  {Misawa} T.,  {Charlton} J.~C.,   {Kim} T.-S.,  2007, \mn@doi
  [\apj] {10.1086/512852}, \href
  {https://ui.adsabs.harvard.edu/abs/2007ApJ...660.1093N} {660, 1093}

\bibitem[\protect\citeauthoryear{{Nasir} \& {D'Aloisio}}{{Nasir} \&
  {D'Aloisio}}{2020}]{Nasir20}
{Nasir} F.,  {D'Aloisio} A.,  2020, \mn@doi [\mnras] {10.1093/mnras/staa894},
  \href {https://ui.adsabs.harvard.edu/abs/2020MNRAS.494.3080N} {494, 3080}

\bibitem[\protect\citeauthoryear{{Nestor}, {Turnshek}  \& {Rao}}{{Nestor}
  et~al.}{2005}]{ntr05}
{Nestor} D.~B.,  {Turnshek} D.~A.,   {Rao} S.~M.,  2005, \mn@doi [\apj]
  {10.1086/427547}, \href {http://adsabs.harvard.edu/abs/2005ApJ...628..637N}
  {628, 637}

\bibitem[\protect\citeauthoryear{{O{\~n}orbe}, {Hennawi}  \&
  {Luki{\'c}}}{{O{\~n}orbe} et~al.}{2017}]{Onorbe17a}
{O{\~n}orbe} J.,  {Hennawi} J.~F.,   {Luki{\'c}} Z.,  2017, \mn@doi [\apj]
  {10.3847/1538-4357/aa6031}, \href
  {http://adsabs.harvard.edu/abs/2017ApJ...837..106O} {837, 106}

\bibitem[\protect\citeauthoryear{{O{\~n}orbe}, {Davies}, {Luki{\'c}}, {},
  {Hennawi}  \& {Sorini}}{{O{\~n}orbe} et~al.}{2019}]{Onorbe19}
{O{\~n}orbe} J.,  {Davies} F.~B.,  {Luki{\'c}} {} Z.,  {Hennawi} J.~F.,
  {Sorini} D.,  2019, \mn@doi [\mnras] {10.1093/mnras/stz984}, \href
  {https://ui.adsabs.harvard.edu/abs/2019MNRAS.486.4075O} {486, 4075}

\bibitem[\protect\citeauthoryear{{Oh}}{{Oh}}{2002}]{Oh02}
{Oh} S.~P.,  2002, \mn@doi [Monthly Notices of the Royal Astronomical Society]
  {10.1046/j.1365-8711.2002.05859.x}, \href
  {https://ui.adsabs.harvard.edu/abs/2002MNRAS.336.1021O} {336, 1021}

\bibitem[\protect\citeauthoryear{{Oppenheimer}, {Dav{\'e}}  \&
  {Finlator}}{{Oppenheimer} et~al.}{2009}]{Oppenheimer09}
{Oppenheimer} B.~D.,  {Dav{\'e}} R.,   {Finlator} K.,  2009, \mn@doi [Monthly
  Notices of the Royal Astronomical Society]
  {10.1111/j.1365-2966.2009.14771.x}, \href
  {https://ui.adsabs.harvard.edu/abs/2009MNRAS.396..729O} {396, 729}

\bibitem[\protect\citeauthoryear{{Palanque-Delabrouille}
  et~al.,}{{Palanque-Delabrouille} et~al.}{2015}]{Nathalie15}
{Palanque-Delabrouille} N.,  et~al., 2015, \mn@doi [Journal of Cosmology and
  Astro-Particle Physics] {10.1088/1475-7516/2015/11/011}, \href
  {https://ui.adsabs.harvard.edu/abs/2015JCAP...11..011P} {2015, 011}

\bibitem[\protect\citeauthoryear{{Pallottini}, {Ferrara}, {Gallerani},
  {Salvadori}  \& {D'Odorico}}{{Pallottini} et~al.}{2014}]{Pallottini14}
{Pallottini} A.,  {Ferrara} A.,  {Gallerani} S.,  {Salvadori} S.,   {D'Odorico}
  V.,  2014, \mn@doi [Monthly Notices of the Royal Astronomical Society]
  {10.1093/mnras/stu451}, \href
  {https://ui.adsabs.harvard.edu/abs/2014MNRAS.440.2498P} {440, 2498}

\bibitem[\protect\citeauthoryear{{Pawlik}, {Rahmati}, {Schaye}, {Jeon}  \&
  {Dalla Vecchia}}{{Pawlik} et~al.}{2017}]{Pawlik17}
{Pawlik} A.~H.,  {Rahmati} A.,  {Schaye} J.,  {Jeon} M.,   {Dalla Vecchia} C.,
  2017, \mn@doi [\mnras] {10.1093/mnras/stw2869}, \href
  {https://ui.adsabs.harvard.edu/abs/2017MNRAS.466..960P} {466, 960}

\bibitem[\protect\citeauthoryear{{Petitjean} \& {Bergeron}}{{Petitjean} \&
  {Bergeron}}{1990}]{Petitjean90}
{Petitjean} P.,  {Bergeron} J.,  1990, \aap, \href
  {https://ui.adsabs.harvard.edu/abs/1990A&A...231..309P} {231, 309}

\bibitem[\protect\citeauthoryear{{Petitjean} \& {Bergeron}}{{Petitjean} \&
  {Bergeron}}{1994}]{Petitjean94}
{Petitjean} P.,  {Bergeron} J.,  1994, \aap, \href
  {https://ui.adsabs.harvard.edu/abs/1994A&A...283..759P} {283, 759}

\bibitem[\protect\citeauthoryear{{Pichon}, {Scannapieco}, {Aracil},
  {Petitjean}, {Aubert}, {Bergeron}  \& {Colombi}}{{Pichon}
  et~al.}{2003}]{Pichon03}
{Pichon} C.,  {Scannapieco} E.,  {Aracil} B.,  {Petitjean} P.,  {Aubert} D.,
  {Bergeron} J.,   {Colombi} S.,  2003, \mn@doi [\apjl] {10.1086/380087}, \href
  {https://ui.adsabs.harvard.edu/abs/2003ApJ...597L..97P} {597, L97}

\bibitem[\protect\citeauthoryear{{Pieri}}{{Pieri}}{2014}]{Pieri14}
{Pieri} M.~M.,  2014, \mn@doi [Monthly Notices of the Royal Astronomical
  Society] {10.1093/mnrasl/slu142}, \href
  {https://ui.adsabs.harvard.edu/abs/2014MNRAS.445L.104P} {445, L104}

\bibitem[\protect\citeauthoryear{{Planck} et~al.,}{{Planck}
  et~al.}{2018}]{Planck18}
{Planck} et~al., 2018, arXiv e-prints, \href
  {https://ui.adsabs.harvard.edu/abs/2018arXiv180706209P} {p. arXiv:1807.06209}

\bibitem[\protect\citeauthoryear{{Pritchard} \& {Loeb}}{{Pritchard} \&
  {Loeb}}{2012}]{Pritchard12}
{Pritchard} J.~R.,  {Loeb} A.,  2012, \mn@doi [Reports on Progress in Physics]
  {10.1088/0034-4885/75/8/086901}, \href
  {https://ui.adsabs.harvard.edu/abs/2012RPPh...75h6901P} {75, 086901}

\bibitem[\protect\citeauthoryear{{Prochaska} et~al.,}{{Prochaska}
  et~al.}{2013}]{QPQ6}
{Prochaska} J.~X.,  et~al., 2013, \mn@doi [\apj] {10.1088/0004-637X/776/2/136},
  \href {http://adsabs.harvard.edu/abs/2013ApJ...776..136P} {776, 136}

\bibitem[\protect\citeauthoryear{{Quashnock} \& {Stein}}{{Quashnock} \&
  {Stein}}{1999}]{Quashnock99}
{Quashnock} J.~M.,  {Stein} M.~L.,  1999, \mn@doi [\apj] {10.1086/307067},
  \href {https://ui.adsabs.harvard.edu/abs/1999ApJ...515..506Q} {515, 506}

\bibitem[\protect\citeauthoryear{{Quashnock} \& {Vanden Berk}}{{Quashnock} \&
  {Vanden Berk}}{1998}]{QuashnockVB98}
{Quashnock} J.~M.,  {Vanden Berk} D.~E.,  1998, \mn@doi [\apj]
  {10.1086/305705}, \href
  {https://ui.adsabs.harvard.edu/abs/1998ApJ...500...28Q} {500, 28}

\bibitem[\protect\citeauthoryear{{Rauch}, {Sargent}, {Womble}  \&
  {Barlow}}{{Rauch} et~al.}{1996}]{Rauch96}
{Rauch} M.,  {Sargent} W.~L.~W.,  {Womble} D.~S.,   {Barlow} T.~A.,  1996,
  \mn@doi [\apjl] {10.1086/310187}, \href
  {https://ui.adsabs.harvard.edu/abs/1996ApJ...467L...5R} {467, L5}

\bibitem[\protect\citeauthoryear{{Ricotti} \& {Ostriker}}{{Ricotti} \&
  {Ostriker}}{2004}]{Ricotti04a}
{Ricotti} M.,  {Ostriker} J.~P.,  2004, \mn@doi [Monthly Notices of the Royal
  Astronomical Society] {10.1111/j.1365-2966.2004.07662.x}, \href
  {https://ui.adsabs.harvard.edu/abs/2004MNRAS.350..539R} {350, 539}

\bibitem[\protect\citeauthoryear{{Robert}, {Murphy}, {O'Meara}, {Crighton}  \&
  {Fumagalli}}{{Robert} et~al.}{2019}]{Robert19}
{Robert} P.~F.,  {Murphy} M.~T.,  {O'Meara} J.~M.,  {Crighton} N. H.~M.,
  {Fumagalli} M.,  2019, \mn@doi [\mnras] {10.1093/mnras/sty3287}, \href
  {https://ui.adsabs.harvard.edu/abs/2019MNRAS.483.2736R} {483, 2736}

\bibitem[\protect\citeauthoryear{{Rogers}, {Bird}, {Peiris}, {Pontzen},
  {Font-Ribera}  \& {Leistedt}}{{Rogers} et~al.}{2018}]{Rogers18}
{Rogers} K.~K.,  {Bird} S.,  {Peiris} H.~V.,  {Pontzen} A.,  {Font-Ribera} A.,
   {Leistedt} B.,  2018, \mn@doi [\mnras] {10.1093/mnras/stx2942}, \href
  {https://ui.adsabs.harvard.edu/abs/2018MNRAS.474.3032R} {474, 3032}

\bibitem[\protect\citeauthoryear{{Rorai} et~al.,}{{Rorai}
  et~al.}{2017}]{Rorai17}
{Rorai} A.,  et~al., 2017, \mn@doi [Science] {10.1126/science.aaf9346}, \href
  {http://adsabs.harvard.edu/abs/2017Sci...356..418R} {356, 418}

\bibitem[\protect\citeauthoryear{{Ryan-Weber}, {Pettini}, {Madau}  \&
  {Zych}}{{Ryan-Weber} et~al.}{2009}]{RyanWeber09}
{Ryan-Weber} E.~V.,  {Pettini} M.,  {Madau} P.,   {Zych} B.~J.,  2009, \mn@doi
  [Monthly Notices of the Royal Astronomical Society]
  {10.1111/j.1365-2966.2009.14618.x}, \href
  {https://ui.adsabs.harvard.edu/abs/2009MNRAS.395.1476R} {395, 1476}

\bibitem[\protect\citeauthoryear{{Sargent}, {Young}, {Boksenberg}  \&
  {Tytler}}{{Sargent} et~al.}{1980}]{Sargent80}
{Sargent} W.~L.~W.,  {Young} P.~J.,  {Boksenberg} A.,   {Tytler} D.,  1980,
  \mn@doi [\apjs] {10.1086/190644}, \href
  {https://ui.adsabs.harvard.edu/abs/1980ApJS...42...41S} {42, 41}

\bibitem[\protect\citeauthoryear{{Sargent}, {Boksenberg}  \&
  {Steidel}}{{Sargent} et~al.}{1988}]{Sargent88}
{Sargent} W. L.~W.,  {Boksenberg} A.,   {Steidel} C.~C.,  1988, \mn@doi [\apjs]
  {10.1086/191300}, \href
  {https://ui.adsabs.harvard.edu/abs/1988ApJS...68..539S} {68, 539}

\bibitem[\protect\citeauthoryear{{Scannapieco}, {Pichon}, {Aracil},
  {Petitjean}, {Thacker}, {Pogosyan}, {Bergeron}  \& {Couchman}}{{Scannapieco}
  et~al.}{2006}]{Scannapieco06}
{Scannapieco} E.,  {Pichon} C.,  {Aracil} B.,  {Petitjean} P.,  {Thacker}
  R.~J.,  {Pogosyan} D.,  {Bergeron} J.,   {Couchman} H.~M.~P.,  2006, \mn@doi
  [\mnras] {10.1111/j.1365-2966.2005.09753.x}, \href
  {https://ui.adsabs.harvard.edu/abs/2006MNRAS.365..615S} {365, 615}

\bibitem[\protect\citeauthoryear{{Schaye}, {Aguirre}, {Kim}, {Theuns}, {Rauch}
  \& {Sargent}}{{Schaye} et~al.}{2003}]{Schaye03}
{Schaye} J.,  {Aguirre} A.,  {Kim} T.-S.,  {Theuns} T.,  {Rauch} M.,
  {Sargent} W. L.~W.,  2003, \mn@doi [The Astrophysical Journal]
  {10.1086/378044}, \href
  {https://ui.adsabs.harvard.edu/abs/2003ApJ...596..768S} {596, 768}

\bibitem[\protect\citeauthoryear{{Schechter}}{{Schechter}}{1976}]{Schechter76}
{Schechter} P.,  1976, \mn@doi [\apj] {10.1086/154079}, \href
  {https://ui.adsabs.harvard.edu/abs/1976ApJ...203..297S} {203, 297}

\bibitem[\protect\citeauthoryear{{Seljak}, {McDonald}  \& {Makarov}}{{Seljak}
  et~al.}{2003}]{Seljak03}
{Seljak} U.,  {McDonald} P.,   {Makarov} A.,  2003, \mn@doi [Monthly Notices of
  the Royal Astronomical Society] {10.1046/j.1365-8711.2003.06809.x}, \href
  {https://ui.adsabs.harvard.edu/abs/2003MNRAS.342L..79S} {342, L79}

\bibitem[\protect\citeauthoryear{{Simcoe}}{{Simcoe}}{2011}]{Simcoe11a}
{Simcoe} R.~A.,  2011, \mn@doi [The Astrophysical Journal]
  {10.1088/0004-637X/738/2/159}, \href
  {https://ui.adsabs.harvard.edu/abs/2011ApJ...738..159S} {738, 159}

\bibitem[\protect\citeauthoryear{{Simcoe} et~al.,}{{Simcoe}
  et~al.}{2011}]{Simcoe11b}
{Simcoe} R.~A.,  et~al., 2011, \mn@doi [The Astrophysical Journal]
  {10.1088/0004-637X/743/1/21}, \href
  {https://ui.adsabs.harvard.edu/abs/2011ApJ...743...21S} {743, 21}

\bibitem[\protect\citeauthoryear{{Simcoe}, {Sullivan}, {Cooksey}, {Kao},
  {Matejek}  \& {Burgasser}}{{Simcoe} et~al.}{2012}]{Simcoe12}
{Simcoe} R.~A.,  {Sullivan} P.~W.,  {Cooksey} K.~L.,  {Kao} M.~M.,  {Matejek}
  M.~S.,   {Burgasser} A.~J.,  2012, \mn@doi [\nat] {10.1038/nature11612},
  \href {http://adsabs.harvard.edu/abs/2012Natur.492...79S} {492, 79}

\bibitem[\protect\citeauthoryear{{Slosar} et~al.,}{{Slosar}
  et~al.}{2013}]{Slosar13}
{Slosar} A.,  et~al., 2013, \mn@doi [Journal of Cosmology and Astro-Particle
  Physics] {10.1088/1475-7516/2013/04/026}, \href
  {https://ui.adsabs.harvard.edu/abs/2013JCAP...04..026S} {2013, 026}

\bibitem[\protect\citeauthoryear{{Songaila}}{{Songaila}}{2005}]{Songaila05}
{Songaila} A.,  2005, \mn@doi [\aj] {10.1086/491704}, \href
  {https://ui.adsabs.harvard.edu/abs/2005AJ....130.1996S} {130, 1996}

\bibitem[\protect\citeauthoryear{{Steidel} \& {Sargent}}{{Steidel} \&
  {Sargent}}{1992}]{SteidelSargent92}
{Steidel} C.~C.,  {Sargent} W. L.~W.,  1992, \mn@doi [\apjs] {10.1086/191660},
  \href {https://ui.adsabs.harvard.edu/abs/1992ApJS...80....1S} {80, 1}

\bibitem[\protect\citeauthoryear{{Thyagarajan}}{{Thyagarajan}}{2020}]{Thyagarajan20}
{Thyagarajan} N.,  2020, arXiv e-prints, \href
  {https://ui.adsabs.harvard.edu/abs/2020arXiv200610070T} {p. arXiv:2006.10070}

\bibitem[\protect\citeauthoryear{{Trott} et~al.,}{{Trott}
  et~al.}{2020}]{Trott20}
{Trott} C.~M.,  et~al., 2020, \mn@doi [\mnras] {10.1093/mnras/staa414}, \href
  {https://ui.adsabs.harvard.edu/abs/2020MNRAS.493.4711T} {493, 4711}

\bibitem[\protect\citeauthoryear{{Tytler} et~al.,}{{Tytler}
  et~al.}{2009}]{Tytler09}
{Tytler} D.,  et~al., 2009, \mn@doi [\mnras]
  {10.1111/j.1365-2966.2008.14159.x}, \href
  {https://ui.adsabs.harvard.edu/abs/2009MNRAS.392.1539T} {392, 1539}

\bibitem[\protect\citeauthoryear{{Viel}, {Becker}, {Bolton}  \&
  {Haehnelt}}{{Viel} et~al.}{2013}]{Viel13}
{Viel} M.,  {Becker} G.~D.,  {Bolton} J.~S.,   {Haehnelt} M.~G.,  2013, \mn@doi
  [\prd] {10.1103/PhysRevD.88.043502}, \href
  {http://adsabs.harvard.edu/abs/2013PhRvD..88d3502V} {88, 043502}

\bibitem[\protect\citeauthoryear{{Vikas} et~al.,}{{Vikas}
  et~al.}{2013}]{Vikas13}
{Vikas} S.,  et~al., 2013, \mn@doi [\apj] {10.1088/0004-637X/768/1/38}, \href
  {https://ui.adsabs.harvard.edu/abs/2013ApJ...768...38V} {768, 38}

\bibitem[\protect\citeauthoryear{{Walther}, {Hennawi}, {Hiss}, {O{\~n}orbe},
  {Lee}, {Rorai}  \& {O'Meara}}{{Walther} et~al.}{2018}]{Walther17}
{Walther} M.,  {Hennawi} J.~F.,  {Hiss} H.,  {O{\~n}orbe} J.,  {Lee} K.-G.,
  {Rorai} A.,   {O'Meara} J.,  2018, \mn@doi [\apj] {10.3847/1538-4357/aa9c81},
  \href {https://ui.adsabs.harvard.edu/abs/2018ApJ...852...22W} {852, 22}

\bibitem[\protect\citeauthoryear{{Walther}, {O{\~n}orbe}, {Hennawi}  \&
  {Luki{\'c}}}{{Walther} et~al.}{2019}]{Walther19}
{Walther} M.,  {O{\~n}orbe} J.,  {Hennawi} J.~F.,   {Luki{\'c}} Z.,  2019,
  \mn@doi [The Astrophysical Journal] {10.3847/1538-4357/aafad1}, \href
  {https://ui.adsabs.harvard.edu/abs/2019ApJ...872...13W} {872, 13}

\bibitem[\protect\citeauthoryear{{Wang} et~al.,}{{Wang}
  et~al.}{2018}]{Wang18z7}
{Wang} F.,  et~al., 2018, \mn@doi [\apjl] {10.3847/2041-8213/aaf1d2}, \href
  {https://ui.adsabs.harvard.edu/abs/2018ApJ...869L...9W} {869, L9}

\bibitem[\protect\citeauthoryear{{Wang} et~al.,}{{Wang} et~al.}{2020}]{Wang20a}
{Wang} F.,  et~al., 2020, \mn@doi [\apj] {10.3847/1538-4357/ab8c45}, \href
  {https://ui.adsabs.harvard.edu/abs/2020ApJ...896...23W} {896, 23}

\bibitem[\protect\citeauthoryear{{Wyithe} \& {Bolton}}{{Wyithe} \&
  {Bolton}}{2011}]{Wyithe11}
{Wyithe} J.~S.~B.,  {Bolton} J.~S.,  2011, \mn@doi [\mnras]
  {10.1111/j.1365-2966.2010.18030.x}, \href
  {http://adsabs.harvard.edu/abs/2011MNRAS.412.1926W} {412, 1926}

\bibitem[\protect\citeauthoryear{{Xu}, {Ahn}, {Norman}, {Wise}  \&
  {O'Shea}}{{Xu} et~al.}{2016}]{Xu16}
{Xu} H.,  {Ahn} K.,  {Norman} M.~L.,  {Wise} J.~H.,   {O'Shea} B.~W.,  2016,
  \mn@doi [\apjl] {10.3847/2041-8205/832/1/L5}, \href
  {https://ui.adsabs.harvard.edu/abs/2016ApJ...832L...5X} {832, L5}

\bibitem[\protect\citeauthoryear{{Yang} et~al.,}{{Yang} et~al.}{2019}]{Yang19}
{Yang} J.,  et~al., 2019, \mn@doi [\aj] {10.3847/1538-3881/ab1be1}, \href
  {https://ui.adsabs.harvard.edu/abs/2019AJ....157..236Y} {157, 236}

\bibitem[\protect\citeauthoryear{{Yang} et~al.,}{{Yang} et~al.}{2020}]{Yang20b}
{Yang} J.,  et~al., 2020, \mn@doi [\apjl] {10.3847/2041-8213/ab9c26}, \href
  {https://ui.adsabs.harvard.edu/abs/2020ApJ...897L..14Y} {897, L14}

\bibitem[\protect\citeauthoryear{{Zel'Dovich}}{{Zel'Dovich}}{1970}]{Zeldovich70}
{Zel'Dovich} Y.~B.,  1970, \aap, \href
  {https://ui.adsabs.harvard.edu/abs/1970A&A.....5...84Z} {500, 13}

\bibitem[\protect\citeauthoryear{{Zhu} \& {M{\'e}nard}}{{Zhu} \&
  {M{\'e}nard}}{2013}]{ZhuMenard13}
{Zhu} G.,  {M{\'e}nard} B.,  2013, \mn@doi [The Astrophysical Journal]
  {10.1088/0004-637X/773/1/16}, \href
  {https://ui.adsabs.harvard.edu/abs/2013ApJ...773...16Z} {773, 16}

\bibitem[\protect\citeauthoryear{{du Mas des Bourboux} et~al.,}{{du Mas des
  Bourboux} et~al.}{2019}]{Bourboux19}
{du Mas des Bourboux} H.,  et~al., 2019, \mn@doi [The Astrophysical Journal]
  {10.3847/1538-4357/ab1d49}, \href
  {https://ui.adsabs.harvard.edu/abs/2019ApJ...878...47D} {878, 47}

\bibitem[\protect\citeauthoryear{{van Haarlem} et~al.,}{{van Haarlem}
  et~al.}{2013}]{LOFAR}
{van Haarlem} M.~P.,  et~al., 2013, \mn@doi [\aap]
  {10.1051/0004-6361/201220873}, \href
  {https://ui.adsabs.harvard.edu/abs/2013A&A...556A...2V} {556, A2}

\makeatother
\end{thebibliography}




\bsp	
\label{lastpage}
\end{document}